\documentclass[aps,pra,twocolumn,superscriptaddress,nofootinbib]{revtex4-2}
\usepackage[T1]{fontenc}
\usepackage{xcolor}
\usepackage{graphicx}
\usepackage{booktabs}

\usepackage{float} 
\usepackage{mathtools}

\usepackage{amsmath}
\usepackage{yhmath}
\usepackage{amssymb}
\usepackage[export]{adjustbox}
\usepackage{dcolumn}
\usepackage{bm}
\usepackage{hyperref}
\usepackage{enumitem}

\hypersetup{linktocpage,colorlinks,citecolor={blue},pdfdisplaydoctitle=true,pdfpagemode=UseOutlines,bookmarksnumbered=true}

\usepackage{mathrsfs,dsfont}
\usepackage{xcolor}

\usepackage{orcidlink}
\usepackage{bbold}
\usepackage{gensymb}
\usepackage{csquotes}

\usepackage{braket}
\usepackage{cancel}

\newcommand{\bb}{{\textbf{b}}}

\graphicspath{{Figures/PNG/}{Figures/PDF/}{Figures/EPS/}{Figures/TEX/}{Figures/}}

\begin{document}

\title{
Refinements of the Eigenstate Thermalization Hypothesis \\ under Local Rotational Invariance via Free Probability}

\author{Elisa Vallini}
\affiliation{Institut f\"ur Theoretische Physik, Universit\"at zu K\"oln, Z\"ulpicher Straße 77, 50937 K\"oln, Germany}

\author{Laura Foini}
\affiliation{Institut de Physique Théorique, CNRS, CEA, Université Paris-Saclay, 91191 Gif-sur-Yvette, France}

\author{Silvia Pappalardi}
\affiliation{Institut f\"ur Theoretische Physik, Universit\"at zu K\"oln, Z\"ulpicher Straße 77, 50937 K\"oln, Germany}

\date{\today}

\begin{abstract}
The Eigenstate Thermalization Hypothesis (ETH) was developed as a framework for understanding how the principles of statistical mechanics emerge in the long-time limit of isolated quantum many-body systems. Since then, ETH has shifted the attention towards the study of matrix elements of physical observables in the energy eigenbasis. 
In this work, we revisit recent developments leading to the formulation of full ETH, a generalization of the original ETH ansatz that accounts for multi-point correlation functions. Using tools from free probability, we explore the implications of local rotational invariance, a property that emerges from the statistical invariance of observables under random basis transformations induced by small perturbations of the Hamiltonian. This approach allows us to make quantitative predictions and derive an analytical characterization of subleading corrections to matrix-element correlations, thereby refining the ETH ansatz.
Moreover, our analysis links the statistical properties of matrix elements under random basis changes to the empirical averages over energy windows that are usually considered when dealing with a single instance of the ensemble.
We validate our analytical predictions through comparison with numerical simulations in non-integrable Floquet systems.
\end{abstract}

\maketitle

\section{Introduction}
The Eigenstate Thermalization Hypothesis (ETH) was introduced as a natural extension of ideas from quantum chaos, typicality, and random matrix theory (RMT) to explain thermalization in isolated quantum systems \cite{PhysRevA.30.504,PhysRevA.34.591,srednicki1994chaos, srednicki1999approach,deutsch1991quantum, reimann2015eigenstate, dalessio2016from}. 
In the quantum chaos community, it is common to model deterministic systems as effectively random. In this spirit, the Bohigas--Giannoni--Schmidt conjecture established that quantum chaotic systems exhibit eigenvalue statistics consistent with RMT \cite{bohigas1984characterization}. Beyond spectral properties, quantum chaos also manifests in the statistics of eigenstates. According to Berry's random wave conjecture, eigenfunctions of chaotic systems \(H\ket{E_i}=E_i\ket{E_i}\) behave as superpositions of plane waves with Gaussian-distributed amplitudes and random phases \cite{berry1977regular, voros1979semi}. 

Building on these foundations, ETH focuses on the matrix elements of few-body observables in the energy eigenbasis, $A_{ij}=\braket{E_i|A|E_j}$. It has been proposed that these matrix elements can be regarded as random variables: the average of the diagonal elements vary smoothly with energy and reproduce thermal expectation values, while the off-diagonal elements have zero mean and an energy-dependent variance \cite{srednicki1994chaos, srednicki1999approach}. These small fluctuations can be viewed as arising from a fictitious ensemble of infinitesimally perturbed systems, effectively introducing randomness through the arbitrary mixing of nearby eigenstates, without altering the relevant physical properties at the scales of interest in statistical mechanics \cite{deutsch1991quantum}.
ETH connects the statistical properties of the matrix elements to the thermal behavior of $A$, thereby explaining thermalization in isolated chaotic systems. 
Randomness emerges as the underlying mechanism: subject to the constraint of preserving a minimal structure in energy, 
the chaotic eigenstates are otherwise as random as possible,
encoding the microcanonical ensemble \cite{srednicki1999approach, goldstein2006canonical, popescu2006entanglement, rigol2008thermalization, vonNeumann2010proof, goldstein2010approach, dymarsky2018subsystem}.

Recent developments have extended the standard ETH framework by studying higher-order correlations among matrix elements \cite{foini2019eigenstate}. These correlations are now known to play a crucial role in determining quantities such as out-of-time-order correlators (OTOCs), and have been investigated in different contexts, ranging from condensed matter \cite{chan2019eigenstate,murthy2019bounds,alves2025probes,richter2020eigenstate, hahn2024eigenstate, foini2025out} to high energy physics \cite{belin2021random,belin2022non,chandra2022semiclassical,jafferis2023jt,hosur2016chaos,kawamoto2025strategy,saad2019late,jafferis2023matrix}. \\ The \textit{full ETH} \cite{foini2019eigenstate} states the following expression, for products with distinct indices $i_1 \ne \dots \ne i_n$:
\begin{subequations}
\label{eth}
\begin{align}
&\overline{A_{i_1 i_2}\dots A_{i_n i_1}}  =  \\& \qquad  =D_{E_+}^{1-n}\,  F_{e_+}^{(n)}(\omega_{_1},\dots, \omega_n) +\mathcal{O}\left(D_{E_+}^{2-n}\right)\ , \notag
\end{align}
where $F^{(n)}$ are some smooth functions of the energy differences $\omega_\alpha=E_{i_\alpha}-E_{i_{\alpha+1}}$ and of the mean energy density $e_+=E_+/L = (E_{i_1}+\dots+E_{i_n})/nL$, with $L$ system size. They are commonly known as \emph{ETH smooth function} and are related to the Fourier transforms of higher-order connected thermal correlation functions. $D_{E_+} = e^{S(E_+)}$ is the density of states at the average energy, with $S$ macroscopic entropy.
When indices are repeated, the average of the products factorizes, at the leading order in the density of states $D_{E_+}$, as: 
\begin{align}\label{Eq_intro_cacti}
    &\overline{A_{i_1 i_2} \dots A_{i_{m-1} i_1} A_{i_1 i_{m+1}} \dots A_{i_ni_1} } \\ &\  \qquad \qquad \approx \overline{A_{i_1 i_2} \dots A_{i_{m-1} i_1}}\ \ \overline{ A_{i_1 i_{m+1}} \dots A_{i_n i_1} }  \  . \notag
\end{align}
\end{subequations}
 
Remarkably, it has been recognized that this ansatz can be reformulated within the mathematical framework of \textit{free probability theory} \cite{pappalardi2022eigenstate, pappalardi2025full}.
The latter generalizes classical probability to non-commuting variables by replacing the notion of statistical independence with \textit{freeness} \cite{voiculescu1991limit, voiculescu1992free, speicher1997free}. Specifically, it introduces powerful tools for understanding the statistical properties of large
matrices, such as those that arise in chaotic quantum
systems. Recasting ETH in this language has led to several conceptual developments \cite{camargo2025quantum, alves2025probes, fritzsch2025free, fritzsch2025microcanonical, fritzsch2025freecumulants, pathak2025full}, including the intriguing proposal that, in ergodic systems, local observables at different times become freely independent in the long-time limit \cite{jindal2024generalized, fava2025designs, cipolloni2022thermalisation, chen2024free, vallini2024longtime, dowling2025free}.

In this work we further develop the \textit{principle of local rotational invariance}, which originally motivated the formulation of the full ETH \cite{foini2019eigenstate2}. On the one hand, we clarify in detail how the full ETH emerges from local rotational invariance; on the other hand, pursuing this goal also provides a quantitative characterization of the subleading corrections to Eq.~\eqref{eth}, which were only partially discussed in \cite{foini2019eigenstate}.

Rotational invariance can be realized at different levels. In its simplest implementation, basis transformations are taken to be global: the probability distribution of the Hamiltonian remains unchanged under arbitrary unitary transformations\footnote{In this work we focus mostly on random unitary rotations, that is, when the system belongs to the unitary symmetry class. The orthogonal case is only briefly mentioned in Apps.~\ref{App_HCIZ} and \ref{app_num_GOE}.}, leading to completely random eigenvectors uniformly distributed on the unit sphere \cite{wigner1955characteristic, mehta2004random}. Consequently, the statistical properties of $A$ remain invariant under random global unitary rotations.
In realistic Hamiltonians, however, randomness emerges only locally in energy
: eigenstates exhibit random-like fluctuations on small energy scales while preserving smooth variations and physical structure across the spectrum \cite{brenes2020lowfrequency, wang2022eigenstate, bertoni2024typical, wang2024emergence, wang2025eigenstatefewbody, fuellgraf2025scaling}. This observation motivated refined RMT models, such as banded random matrix ensembles, where randomness is restricted to nearby energy levels \cite{casati1990scaling, fyodorov1991statistical}. Within these models, the statistical properties of observables remain invariant only under \textit{local} unitary transformations, providing a more faithful statistical description of chaotic many-body systems. 
The transition from global to local rotational invariance can thus be viewed as a refinement of the ETH paradigm \cite{foini2019eigenstate}: the global case represents the featureless random limit, while the local version incorporates energy structure. This hierarchical picture provides a controlled approach to study the internal structure of ETH and to explicitly evaluate its first subleading corrections, which naturally emerge in terms of \textit{local free cumulants} \cite{bernard2024structured}, as we aim to show in this work. These corrections are compatible with the postulates of structured random matrices introduced in \cite{bernard2024structured} beyond the leading order.

\subsection*{Motivation and main results}

Before entering the technical development, we summarize here the physical idea behind our approach and our main results.\\

\begin{figure*}[t]
\centering
\includegraphics[width=1.03\linewidth]{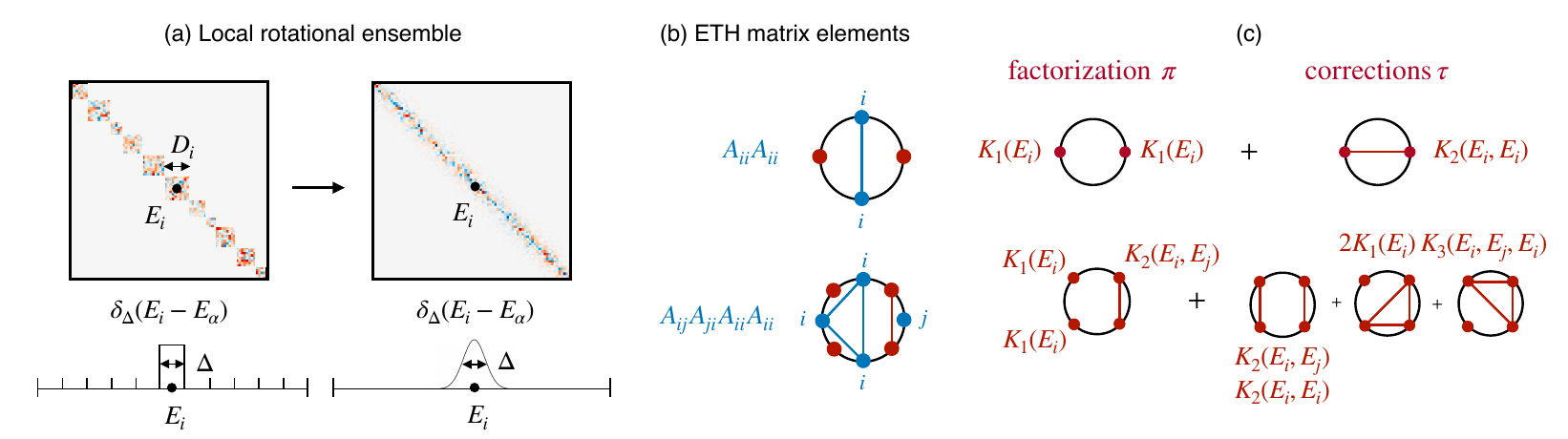}
\caption{\emph{Pictorial summary.}
(a) The local rotational ensemble $U_{\rm loc}$: a random unitary with block structure (left) or with a smoothed/banded structure in energy (right). Only the $D_i$ states within an energy window of width $\Delta$ around each mesoscopic energy $E_i$ are mixed by a matrix of the ensemble. This window is described by the function $\delta_\Delta(E_i-E_\alpha)$, which may take the form of a sharp box or a smooth profile.\\
(b) Products of matrix elements, such as $A_{ii}A_{ii}$ and $A_{ij}A_{ji}A_{ii}A_{ii}$, are represented by blue diagrams on $n$ points, where repeated indices are connected by blue lines. This uniquely fixes $\pi$, identified by the associated diagram on the red edges.\\
(c) Leading factorization and subleading corrections of correlations with repeated indices [cf. Eq.~\eqref{main_summary}]. The leading contribution is determined by $\pi$ and given by the product over its blocks of smooth functions evaluated at the energies labeling the nodes. Subleading corrections are obtained from diagrams $\tau$ formed by merging two blocks of $\pi$.\\
Top row: $k=2$ factorization in Eq.~\eqref{ex2b_summary}. Bottom row: $k=4$ factorization in Eq.~\eqref{ex4_summary}.}
\label{fig_summary}
\end{figure*}

\paragraph{Local rotational invariance and the ensemble $U_{\rm loc}$.}
Our starting point is the physical observation that the observable statistical properties that are relevant for thermalization should not change if we slightly perturb the Hamiltonian in a way that only reshuffles eigenstates that are close in energy. Indeed, mesoscopic physics is expected to be insensitive to such small perturbations, which mix only nearby eigenvectors while leaving the smooth, coarse-grained structure of the spectrum intact. This expectation is encoded in the \emph{principle of local rotational invariance}~\cite{foini2019eigenstate2}: an observable $A$ in the energy eigenbase shares its statistical properties with a version $A^{U_{}} = U_{\rm loc}^\dagger A\, U_{\rm loc}$ rotated by a random unitary that acts \emph{only locally} in energy,
\begin{equation}
A \sim A^{U_{}}\ , \qquad\quad U_{\rm loc} = \bigoplus_{k} U^{(I_k)}\ .
\label{intro_loc}
\end{equation}
Each $U^{(I_k)}$ is an independent Haar-random unitary that mixes only the states inside a small energy window of width $\Delta$, centered on a mesoscopic energy $E_i$. Such a window is described by a function $\delta_\Delta(E_i-E_\alpha)$, either a sharp box or a smoothed profile peaked around $E_i$, as illustrated in Fig.~\ref{fig_summary}(a), so that the number of states it contains is $D_i = \sum_\alpha \delta_\Delta(E_i - E_\alpha)\simeq e^{S(E_i)}$. Averaging over $U_{\rm loc}$ reshuffles the matrix elements of $A$ only locally in energy, while preserving a hierarchy of correlations.
\\

\paragraph{Main results.}
In this work, we compute averages of products of matrix elements, such as those appearing in Eqs.~\eqref{eth}. Local rotational invariance allows these quantities to be computed as ensemble averages over the corresponding unitaries \eqref{intro_loc}. Carrying out this calculation yields the full ETH ansatz, showing that local rotational invariance provides the underlying principle governing the structure of the correlations encoded in ETH. The explicit evaluation further leads to the following results:
\begin{enumerate}
    \item the definition of the \emph{smooth ETH functions in terms of global matrix quantities}, which make the energy dependence of matrix elements explicit;
    \item the definition of the \emph{empirical averages} entering the ETH ansatz. While the derivation relies on ensemble averages over random rotations, these averages ultimately admit a precise formulation in terms of matrix elements of a single deterministic operator.
    \item the identification of \emph{subleading corrections} to the ETH factorization;
\end{enumerate}
These are our main results.\\

\paragraph*{The simplest example: the standard ETH. }
In this paragraph, we illustrate these results on the simplest examples at first and second order, 
which correspond to the \emph{standard ETH ansatz}~\cite{srednicki1999approach, dalessio2016from}. 

When the indices are distinct, the average over the local rotational invariant ensemble gives \emph{smooth functions of the energies} (result 1.) as:
\begin{subequations}
\label{ex_summary}
\begin{align}
\label{ex1_summary}
\overline{A^U_{ii}}(E_i) & = \mathrm{Tr}(\hat\sigma_{E_i} A) \equiv K_1(E_i)\ ,\\
\label{ex2_summary}
\overline{A^U_{ij}A^U_{ji}}(E_i,E_j) & \simeq \mathrm{Tr}(\hat\sigma_{E_i} A\,\hat\sigma_{E_j} A) - [\mathrm{Tr}(\hat\sigma_{E_i} A)]^2\, \sigma(E_i-E_j) \notag
\\ & \equiv K_2(E_i,E_j)\ ,
\end{align}
\end{subequations}
where 
$\displaystyle \hat\sigma_{E}=\sum_\alpha \tfrac{\delta_\Delta(E-E_\alpha)}{D_E}\ket{\alpha}\bra{\alpha}$ is a normalized microcanonical projector peaked around the energy $E$, and $\sigma(E_i-E_j)=\delta_\Delta(E_i-E_j)/D_{E_i}$. In this way we obtain energy-resolved moments of $A$, i.e. traces projected onto the energy windows, which define the smooth functions of the energy $K_1(E)= F^{(1)}_e$ and $K_2(E_1, E_2)= F_{e^+}^{(2)}(\omega)/D_{E_+}$ [using the notation in Eqs.\eqref{eth}].

Furthermore, we note that, for $i \neq j$, we can rewrite Eq.\eqref{ex2_summary} as
\begin{equation}\label{Eq_Aij_results}
\overline{A^U_{ij}A^U_{ij}} \simeq \sum_{\alpha\ne \beta} A_{\alpha \beta} \frac{\delta_{\Delta}(E_i-E_\alpha)}{D_i} A_{\beta \alpha}\frac{\delta_{\Delta}(E_j-E_\beta)}{D_j}\ .
\end{equation}
As we discuss below, this shows that ETH averages can be interpreted as averages over small energy windows of a single
realization of the operator, the \textit{empirical averages} entering the ETH ansatz (result 2.).

Finally, when some indices are repeated, the average of the matrix elements \emph{factorizes} into products of these smooth functions, plus \emph{a subleading correction} (result 3.); the simplest instance is the diagonal two-point function
\begin{equation}
\label{ex2b_summary}
\overline{A^U_{ii}A^U_{ii}}(E_i)
\;\simeq\;
\overline{A^U_{ii}}(E_i)\overline{A^U_{ii}}(E_i)
\;+\;
\overline{A^U_{ij}A^U_{ji}}(E_i,E_i)\ ,
\end{equation}
where the first term is the factorization and the second, of relative order $1/D$, is the subleading correction.
As we explain better below, the latter encodes a nontrivial \emph{gluing} of leading contributions, mechanism that generates higher-order correlations, although constrained to equal-energy sectors.
Equation~\eqref{ex2b_summary} is the well-known statement that the variance of the diagonal matrix elements is fixed by the off-diagonal ones at zero frequency \cite{dalessio2016from}.
Indeed, the subleading contribution is already present in the standard ETH, but missing in the usual higher-order formulations. One of the goals of this paper is to extend these results to arbitrary orders, thereby filling a structural gap in the full ETH framework. \\

\paragraph*{The general result. }
We now show how the results above hold at arbitrary order. \\
For an all-distinct cyclic product (a simple loop), the average is a single smooth function of the energies,
\begin{equation}
\label{simple_summary}
\overline{A^U_{i_1 i_2}\, A^U_{i_2 i_3}\dots A^U_{i_n i_1}} \;\simeq\; K_n(E_{i_1},\dots,E_{i_n})\ ,
\end{equation}
which extends Eqs.~\eqref{ex1_summary}--\eqref{ex2_summary} to arbitrary $n$. We will be able to write the smooth functions $K_n$ in terms of moments of $A$ resolved in energies by microcanonical projectors $\hat \sigma_E$, and to show that these expressions are \emph{local free cumulants} \cite{bernard2024structured}, directly connected to operator valued free probability. These provide a natural definition of the ETH smooth functions, ${F_{e_+}^{(n)}()=\lim_{N\to\infty} D_{E^+}^{\,n-1}\,K_n()}$ in terms of global matrix quantities.

A key outcome is that the ensemble averages coincide with \emph{empirical averages} of a single, fixed operator over small energy windows. Using $A\sim A^U$ and self-averaging, the diagrams with distinct indices read
\begin{align}
\label{emp_summary}
\overline{A^U_{i_1 i_2}\dots A^U_{i_n i_1}}
\;\simeq
&\sum_{\alpha_1\neq\dots\neq\alpha_n}
A_{\alpha_1\alpha_2}\frac{\delta_\Delta(E_{i_1}\!-\!E_{\alpha_1})}{D_{i_1}}\dots \nonumber\\
&\quad\cdots
A_{\alpha_n\alpha_1}\frac{\delta_\Delta(E_{i_n}\!-\!E_{\alpha_n})}{D_{i_n}}\ ,
\end{align}
thus generalizing Eq.~\eqref{Eq_Aij_results}. The right-hand side involves only the matrix elements of the fixed operator $A$, summed over normalized energy windows around $E_{i_1},\dots,E_{i_n}$: the ensemble randomness is replaced by a \emph{smoothing over small energy windows} of a single deterministic matrix. Nicely, this provides the standard recipe used to extract ETH functions numerically, and it extends to repeated indices by smoothing only over the distinct ones.

The factorization at repeated indices, together with its subleading corrections, also admits a simple interpretation as a generalization of the mechanism already highlighted in Eq.~\eqref{ex2b_summary}.
At the fourth order, for instance,
\begin{align}
\label{ex4_summary}
&\overline{A^U_{ij}A^U_{ji}A^U_{ii}A^U_{ii}}
\;\simeq\;
\overline{A^U_{ij}A^U_{ji}}(E_i,E_j)\,\overline{A^U_{ii}}^{\,2}(E_i) \nonumber\\
&\quad+\,\overline{A^U_{ij}A^U_{ji}}(E_i,E_j)\,\overline{A^U_{il}A^U_{li}}(E_i,E_i)\nonumber\\
&\quad+\,2\,\overline{A^U_{ii}}(E_i)\,\overline{A^U_{ij}A^U_{jk}A^U_{ki}}(E_i,E_j,E_i)\ ,
\end{align}
that is, a leading factorization followed by subleading corrections built from products with one more distinct index, evaluated at equal mesoscopic energies.\\
This mechanism can be understood pictorially from Fig.~\ref{fig_summary}(b--c). Firstly, we associate to each product of matrix elements a diagram $\pi$, as illustrated in Fig.~\ref{fig_summary}(b) for the examples of Eq.~\eqref{ex2b_summary} and Eq.~\eqref{ex4_summary}. The matrix elements are represented by blue dots, joined by a blue line whenever their indices coincide, while the diagram $\pi$ is read off from the red edges. \\ We focus on products for which this diagram is non-crossing and show that the average reads 
\begin{equation}
\label{main_summary}
\overline{[A^U_{i_1 i_2}\dots A^U_{i_n i_{1}}]_\pi}
\;\simeq\;
\underbrace{K_\pi(\{E\})}_{\text{factorization}}
\;+\;
\underbrace{\sum_{\tau} K_\tau(\{E\})}_{\text{subleading}}\ .
\end{equation}
The leading term $K_\pi$ is read directly from the diagram $\pi$ [Fig.~\ref{fig_summary}(c), left]: it is given by the product of smooth functions $K_n$, with one factor for each block of $\pi$, evaluated at the energies associated with the dots in that block. The subleading corrections arise from diagrams $\tau$ obtained by merging two blocks of $\pi$ [Fig.~\ref{fig_summary}(c), right]; each $K_\tau$ is again a product of smooth functions. For instance, at $k=2$ [top row of Fig.~\ref{fig_summary}(b--c)] the diagram of $A^U_{ii}A^U_{ii}$ has two blocks, yielding the factorization $[K_1(E_i)]^2=\overline{A^U_{ii}}^2(E_i)$. Merging them gives the single correction $K_2(E_i,E_i)=\overline{A^U_{ij}A^U_{ji}}(E_i,E_i)$, which reproduces Eq.~\eqref{ex2b_summary}. In the body of the work, we will show that this procedure has a precise mathematical formulation and that $\pi$ and $\tau$ are naturally organized on the lattice of non-crossing partitions, appearing in free probability.\\

\paragraph{Strategy of the paper.}
To build intuition, we begin with the simplest case of \emph{global} rotational invariance, i.e.\ a single global Haar rotation acting on all matrix elements. This corresponds to taking a single block of size $D$, equal to the full Hilbert space dimension, in $U_{\rm loc}$. This toy model already captures the factorization properties and the structure of the subleading corrections, but is blind to any energy dependence: in this case the local free cumulants reduce to the normalized free cumulants of the observable $A$, $K_n()=\kappa_n/D^{n-1}$, independent of energy. We then refine this construction to local rotational invariance, which restores energy dependence and, crucially, yields the smooth energy dependence of the ETH functions.

\subsection*{Paper's structure}
The rest of the manuscript is organized as follows. 
In Sec.~\ref{sec_change} we present in detail the idea of global and local rotational invariance; we briefly review Weingarten calculus for the calculation of averages of products of matrix elements, and key concepts of free probability.
The next four sections contain the bulk of our new results:
\begin{itemize}[leftmargin=1.2em, itemsep=0.5em, topsep=0.5em]
    \item Sec.~\ref{sec_global}: we start by the simplest toy model of global rotational invariance;
we derive a closed formula for the leading and subleading contributions; those can be read in terms of free cumulants on the lattice of non-crossing partitions. 
\item Sec.~\ref{sec_local}: we consider the second toy model with local rotational invariance, which allows us to improve the closed formula from the global case by incorporating the energy dependence; we translate the corresponding quantities from standard free cumulants to local free cumulants, related to operator-valued free probability.
\item Sec.~\ref{sec_ETH}: these findings lead us to define empirical averages over small energy windows for the fixed matrix $A$ and to directly link the smooth functions of ETH and local free cumulants. In this way, we gain back ETH and the first subleading correction to the factorization of the diagrams. 
\item Sec.~\ref{sec_numerical} provides numerical tests conducted in a many-body non-integrable Floquet system, demonstrating strong agreement with our theoretical predictions.
\end{itemize}
In Sec.~\ref{sec_conclusions} we conclude and discuss the implications of our work,
with an outlook on future directions suggested by the principle of local rotational invariance and possible applications of the subleading corrections.\\
There are several appendices. In App.~\ref{app_FB} we introduce additional concepts from free probability that complement the ones in the main text. We then discuss the
detailed calculations through Weingarten calculus in Apps.~\ref{app_GRIderivation} and \ref{app_LRIderivation}, respectively for global and local rotational invariance. In App.~\ref{app_linkFB} we link the results obtained from local rotational invariance to operator-valued free probability. An alternative approach to free probability, based on the Harish-Chandra–Itzykson–Zuber integral, is discussed in App.~\ref{App_HCIZ}, while App.~\ref{app_num} presents additional numerical analyses.
In App.~\ref{app_corr_fun} we show how the subleading corrections discussed in this paper determine the late-time plateau of \emph{thermal free cumulants}~\cite{pappalardi2022eigenstate}, providing a quantitative characterization of long-time freeness in chaotic dynamics.

\section{Random Basis Transformations}
\label{sec_change}

\begin{figure}[t]
\includegraphics[width=0.49 \textwidth]{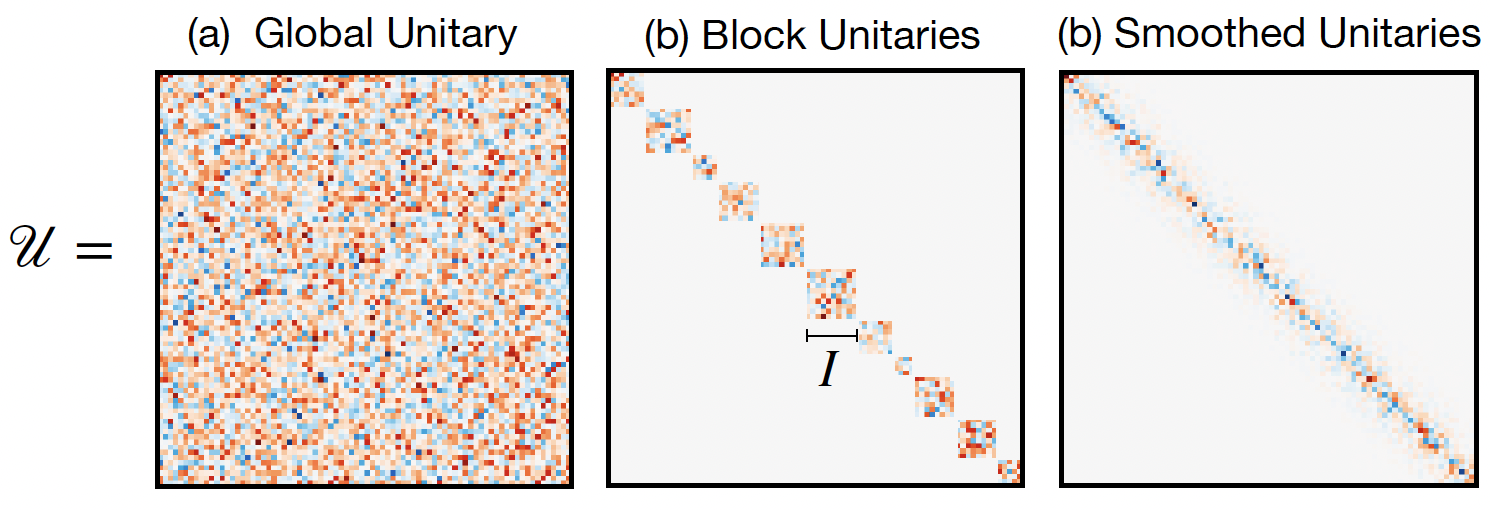}
\caption{Illustrative examples of the random matrix $\mathcal{U}$ in Eq.~\eqref{ASS}, employed to rotate the observable $A$. Different refined toy models for deriving ETH are shown. In (a) the matrix $\mathcal{U}$ is a global Haar unitary matrix, so that $A$ exhibits consequently global rotational invariant. Local rotational invariance is achieved by incorporating a local structure, as in (b) and (c): in (b) the matrix $\mathcal{U}$ is composed of disjointed blocks of independent Haar unitaries, each acting only on a local interval of states. The last more refined toy model in (c) is a smoothed version of the previous one, see Sec.~\ref{sec_change} for discussion.}
\label{fig_unitary}
\end{figure}

In this work we study the matrix elements of some observable $A$ in a randomly rotated basis by studying correlations between the entries of $A^{\mathcal{U}} = \mathcal{U}^\dagger A \mathcal{U}$ with $\mathcal{U}$ a unitary matrix drawn from an ensemble that we will specify.
The basic assumption is to look at a single matrix $A$ in the basis of the Hamiltonian and assume that it is characterized by \emph{global or local rotational invariance}.
With this we mean that $A$ shares the statistical properties with a globally or locally rotated version:
\begin{equation}
    \label{ASS}
    A \sim A^{\mathcal{U}} = \mathcal{U}^\dagger A \mathcal{U}\ ,
\end{equation}
where $\mathcal{U}$ can be a Haar distributed random unitary matrix, as the one in Fig. \ref{fig_unitary}(a), or a banded random unitary characterized by a local structure in energy, such as the block diagonal matrix in Fig. \ref{fig_unitary}(b) or its smoothed version in Fig. \ref{fig_unitary}(c). Moments of $A$ are clearly equal to those of $A^{\mathcal{U}}$ as they depend only on the spectral properties of the matrix.
In the context of local rotational invariance we will encounter generalized moments, and we will assume that in the large size limit these moments of $A$ equal those of $A^{\mathcal{U}}$, and that in particular these quantities are self-averaging, namely that a single realization of the matrix equals the result on average.

Random matrix theory and free probability provide a framework for evaluating matrix elements' $n-$th moments
$\overline{A^{\mathcal{U}}_{i_1j_1}\dots A^{\mathcal{U}}_{i_n j_n}}$, which are the interest of ETH. We explicitly keep the dependence on $\mathcal{U}$ to emphasize that we are averaging over an ensemble of random matrices, assuming $A\sim A^\mathcal{U}$. 
The empirical averages performed on the matrix elements of a fixed observable $A$ will be discussed in Sec.~\ref{sec_ETH}. 
We now define the different toy models that we will consider.
\bigskip

\medskip 
In the first toy model $\mathcal{U} = U$ is a structureless random unitary, and the observable $A^U$ possesses global rotational invariance.
Therefore, $A_{ij}^U = \sum_{\bar i,\bar j} U^*_{\bar ii}A_{\bar i\, \bar j} U_{\bar jj}$ with $U$ sampled by the Haar ensemble.
The matrix elements' $n-$th moment follows as
\begin{align}
\label{mat_el_prod}
    &\overline{A^U_{i_1j_1}\dots A^U_{i_n j_n}} = \\
    &= \sum_{\substack{\bar{i}_1 \dots \bar{i}_n \\ \bar{j}_1 \dots \bar{j}_n}} \overline{U^*_{\bar{i}_1i_1}\dots U^*_{\bar{i}_ni_n}U_{\bar{j}_1j_1}\dots U_{\bar{j}_nj_n}}
    A_{\bar{i}_1\bar{j}_1}\dots A_{\bar{i}_n\bar{j}_n} \, .
    \notag
\end{align}
The statistical properties of the Haar ensemble are known exactly and can be expressed compactly using the Weingarten calculus, which follows from the Schur–Weyl duality and Schur’s lemma, allowing unitary invariants to be written in terms of permutation operators as \cite{collins2006integration}:  
\begin{align}
\label{Haar_av_1}
    &\overline{U^*_{\bar{i}_1i_1}\dots U^*_{\bar{i}_ni_n}U_{\bar{j}_1j_1}\dots U_{\bar{j}_nj_n}}=
    \\
    &\sum_{\alpha,\beta \in S_n } 
    \text{Wg}_{\alpha,\beta}(D)\,
    \delta_{i_{\alpha(1)},j_1}\dots \delta_{i_{\alpha(n)},j_n}
    \delta_{\bar{i}_{\beta(1)},\bar{j}_1}\dots \delta_{\bar{i}_{\beta(n)},\bar{j}_n} \, ,
    \notag
\end{align}
with $\text{Wg}_{\alpha,\beta}(D)=\text{Wg}(\alpha^{-1}\beta,D)$ Weingarten matrix for $\alpha, \beta$ permutations in the symmetric group of $n$ elements $S_n$. $D$ is the dimension of the matrix $A$ i.e. the Hilbert space dimension.  Further discussion of the Weingarten function will be given in App.~\ref{app_GRIderivation}.

A different approach based on an explicit formula for the generating function of the matrix elements, the (low rank limit of the) Harish-Chandra-Itzykson-Zuber integral, is discussed in App.~\ref{App_HCIZ}.

\bigskip

The second toy model takes into account that $A$ remains invariant under local basis transformations i.e. reshufflings of its elements restricted to local neighborhoods in energy. We consider $\mathcal{U} = U_{\rm loc}$ to have local structure in the energy of the Hamiltonian, i.e. $ H\ket{E_i} = E_i \ket {E_i}$.
We divide the total energy range in $M$ disjoint intervals, $I_1\dots I_M$. 
The size of each interval, in terms of energy, is denoted by $\Delta$, and intuitively all states in the same block are associated to the same \emph{mesoscopic} energy. 
Considering a generic interval $I$, whichever $i\in I$ can be taken as representative of the associated mesoscopic energy that we therefore label by the state, as $E_i$. 
The number of states of $I$, that we call $D_i$, is defined as
\begin{equation}
    \label{D_I}
    D_i \equiv \sum_{\alpha\in I}1 = \sum_{\alpha}\delta_\Delta(E_i-E_\alpha)
    \simeq e^{S(E_i)}
    \ ,
\end{equation}
where the $\delta_\Delta(E_i-E_\alpha)$ selects states within the block $I$ to which $i$ belongs, as a sharp box, and $e^{S(E_i)}$ is the density of states at energy $E_i$, with $S$ the macroscopic entropy.
One can therefore define a matrix $U^{(I)}$ for each interval $I$, as a block-sparse matrix with a single random unitary block of size $D_i\times D_i$, supported on the indices $I$. Globally, as depicted in Fig.~\ref{fig_unitary}(b),
\begin{equation}
    U_{\rm loc} = \bigoplus_{k=1}^M U^{(I_k)}
\end{equation}
where $U^{(I_k)}$ are independent. As a consequence, the locally rotated matrix is 
    \begin{equation}
        \label{A_U_local}
        A^U_{ij} =  (U^\dagger_{\rm loc} A U_{\rm loc})_{ij} = \sum_{\substack{{\bar i }\in I \\ \bar j \in J}} U^{(I)\,*}_{\bar i i}  A_{\bar i\, \bar j } U^{(J)}_{\bar j j}\ ,
    \end{equation} 
where now $U^{(I)}$ and $U^{(J)}$ are independent Haar unitaries for the blocks $I$ and $J$ respectively: the rotation is made locally around the energies of the states $i$ and $j$. 
The matrix elements' $n-$th moments for locally rotated matrices are 
\begin{align}
\label{mat_el_prod_local}
    & \overline{A^U_{i_1j_1}\dots A^U_{i_n j_n}} = \\
    & \sum_{\substack{\bar{i}_1\in I_1 \dots \bar{i}_n\in I_n \\ \bar{j}_1\in J_1 \dots \bar{j}_n\in J_n}} \overline{U^{(I_1)\,*}_{\bar{i}_1i_1}\dots U^{(I_n)\,*}_{\bar{i}_ni_n}U^{(J_1)}_{\bar{j}_1j_1}\dots U^{(J_n)}_{\bar{j}_nj_n}}
    A_{\bar{i}_1\bar{j}_1}\dots A_{\bar{i}_n\bar{j}_n} \, .
    \notag
\end{align}
Differently from the case of global rotational invariance, one must account for the fact that the matrices are random unitaries only on the associated set of indices. Therefore, Eq.~\eqref{Haar_av_1} cannot be applied directly; however, Weingarten calculus can still be used by appropriately coupling matrices on different intervals.
One clearly retrieves the global Haar case for a single interval $M=1$ with a flat $\delta_\Delta()=1$ in Eq.~\eqref{D_I}.

This minimal toy model, although convenient for analytical computations, exhibits intrinsic discontinuities, since indices $i$ and $j$ that are close in energy may still belong to distinct blocks $I$ and $J$. Nevertheless, it has a natural counterpart that can be schematically related to the banded unitary shown in Fig.~\ref{fig_unitary}(c). Indeed, its formulation allows one to generalize the minimal model into a smoothed version by considering $\delta_\Delta()$ as peaked around $E_i$ for each state $i$ with a characteristic width $\Delta$.
Moving from the sharp to the smoothed function yields a refined version of the local rotational invariance model. 

In the next sections, Eq.~\eqref{mat_el_prod} and \eqref{mat_el_prod_local} will be evaluated explicitly through Weingarten calculus, and products of matrix elements will be given in terms of free cumulants.
Since these concepts are essential to explain our results, we briefly review in the next paragraph the basic elements of free probability theory.

\begin{figure*}[t]
\centering
\includegraphics[width=0.5 \textwidth]{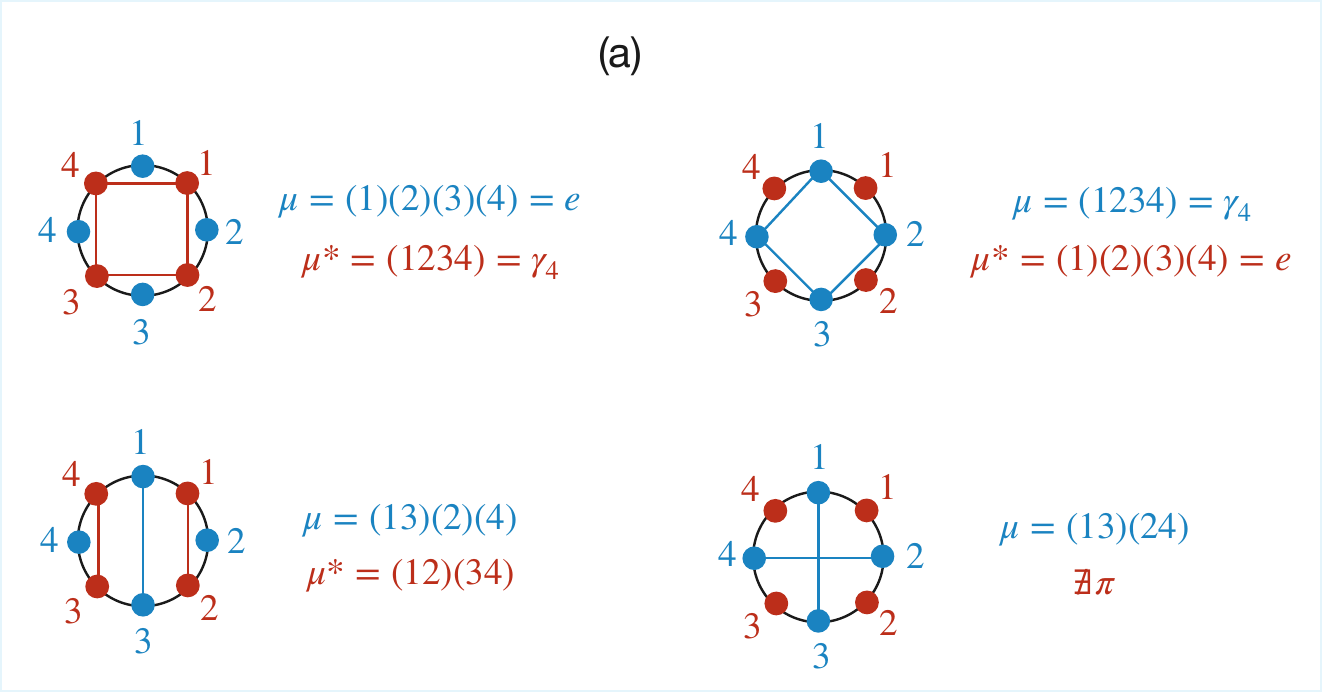}
\hspace{1cm}
\includegraphics[width=0.43 \textwidth]{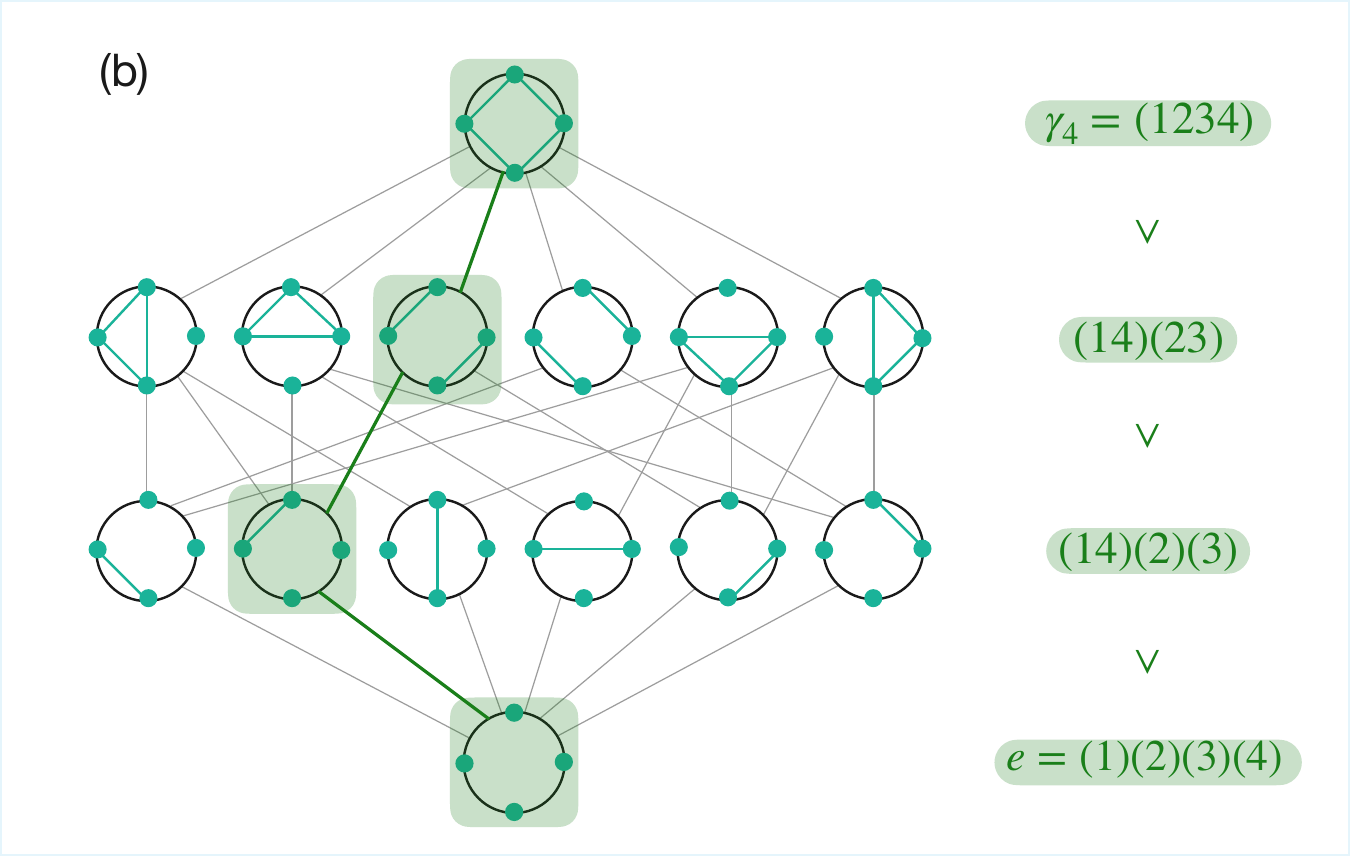}
\caption{(a) Graphical representation of some partitions of four elements (in blue) together with their respective Kreweras complements (in red). The unique crossing partition for $n=4$ is shown in the lower right corner, and has no Kreweras complement. (b) $NC(4)$ (sub)lattice of non-crossing partitions; lines between rows are connecting partitions at a unit distance. At the bottom, the discrete partition (finest), at the top the trivial one (coarsest). Correspondingly, $NC(4)$ can be seen as the geodesic interval $[e,\gamma_4]$, subset of the Cayley graph, for which permutations from one row to the other are related via a single transposition. Each set of permutations following a path from $e$ to $\gamma_4$ defines a single geodesic in the interval; an explicit example is highlighted in green, and on the right.}
\label{fig_free_prob}
\end{figure*}

\subsection{Key concepts of Free Probability}
\label{sec_free_prob}

Free probability is
an extension of traditional probability theory to non-commuting
variables, such as random matrices. Originating in planar field theory \cite{brezin1978planar, cvitanovic1981planar, cvitanovic1982planar}, free
probability has recently played a role in many branches of many-body physics, such
as quantum information theory \cite{collins2016random}, tensor networks \cite{collins2010random, kutlerflam2021distinguishing, kudler2022negativity,cheng2024random}, disordered systems \cite{movassagh2010isotropic,movassagh2011density, chen2012error, hruza2023coherent, bauer2023bernoulli, bernard2023exact}, and gravity \cite{berkooz2019towards, penington2022replica, wang2023beyond, wu2024non, chandrasekaran2023large}. 

This section provides a concise introduction to the aspects of Free Probability theory required to understand the main results of this paper, focusing on its combinatorial formulation. Additional material is presented in App.~\ref{app_FB}, where we provide a more standard mathematical introduction to free probability and a brief overview of operator-valued free probability. The latter extends free probability to variables with additional structure, such as block random matrices, and as we show below, some of the structures emerging from our analysis admit a natural interpretation within this framework.\\
For a more comprehensive treatment of free probability theory, see \cite{speicher2016free, mingo2017free, xia2019simple}.  

In Fig.~\ref{fig_free_prob} some examples of the main concepts are illustrated.

\paragraph{Partitions.}
In order to describe the combinatorics of free probability, it is useful to introduce partitions. 
A partition of a set of $n$ elements is a collection of non-empty, disjoint subsets whose union equals the original set; we indicate it as a sequence of blocks $B_m$, $\mu = \{B_1,\dots B_{\#\mu}\}$, with $\# \mu$ the total number of blocks in the partition. Each partition is usually represented graphically by $n$ points on a circle, ordered clockwise, with lines connecting the elements in the same block. Crossing partitions are the ones for which \textit{blocks cross}, i.e. lines in the respective diagram intersect. Non-crossing partitions are all the others, are denoted by $NC(n)$, and their number is given by the Catalan number $C_n=\binom{2n}{n}/(n+1)$.
Two special partitions are the discrete partition $e$ ($n$ blocks of 1 element) and the trivial one $\gamma_n$ (1 block of $n$ elements).

\paragraph{Lattice.}
Partitions are partially ordered by refinement: for two partitions 
$\mu$ and $\nu$, we write $\nu \le \mu$ if every block of $\nu$ is contained in a block of $\mu$. Under this order, the set of all partitions forms a lattice which can be visualized as a hierarchy of partitions arranged from the coarsest at the top, which is the trivial partition, to the finest at the bottom, which is the discrete partition. Not all partitions are comparable, which makes the refinement only a partial order: incomparable partitions appear on the same level of the lattice. 
When considering only non-crossing partitions, a similar structure can be drawn, forming a sublattice of the lattice of all partitions under the same refinement order.
The sublattice of $NC(n)$ exhibits a top-bottom symmetry which is a special feature of non-crossing partitions and does not hold in the lattice of all partitions. The symmetry consists in exchanging the coarsest and finest partitions and reversing the order of all others; this leaves the lattice combinatorially unchanged. This inversion symmetry can be made precise, defining the Kreweras complement.

\paragraph{Dual partition or Kreweras complement.}
The Kreweras complement of a non-crossing partition $\mu$ is obtained diagrammatically by placing a second copy of the set $\{1, \dots, n\}$ around the same circle and taking the maximal non-crossing partition on the new points without crossing any lines of $\mu$. Intuitively, $\mu^*$ completes $\mu$ so that together they form the maximal non-crossing partition of the doubled set. It is unique.  Crossing partitions do not have a dual complement.  
The Kreweras complement defines an order-reversing bijection on $NC(n)$; that is, if $\nu\le \mu$ then $\nu^*\ge \mu^*$. In other words, the coarsest partitions become the finest under the Kreweras complement, the finest become the coarsest, and intermediate levels are mirrored. As a consequence, the Kreweras complement realizes an isomorphism between $NC(n)$ and the order-reversed lattice. $\mu^*$ is also called the dual partition of $\mu$, and $NC(n)$ is said to be self-dual.

\paragraph{Free Cumulants.}
The combinatorics of partitions plays a central role in probability theory when defining 
cumulants recursively from the moments of a random variable $A$, through the 
moments-cumulants formula. In the classical case all partitions 
contribute, but in free probability theory
only 
non-crossing partitions appear. The moments-free cumulants formula then reads
\begin{equation}
\langle A^n \rangle = \sum_{\mu \in NC(n)} \kappa_\mu, \qquad \quad \kappa_\mu = \prod_{m=1}^{\#\mu} \kappa_{l_m} \ ,
\label{free_cum_NC}
\end{equation}
where $\kappa_\mu$ is defined implicitly for each $\mu \in NC(n)$. Each $\kappa_\mu$ can 
be expressed as a product of free cumulants associated with the blocks of $\mu$: $l_m$ denotes the length of the $m$-th block $B_m$ of $\mu$. 
As an example, the first free cumulants are defined implicitly by
\begin{align}
\notag
\langle A \rangle &= \kappa_1 \\ \label{ex_momcum} 
\langle A^2 \rangle &= \kappa_2 + \kappa_1^2 \\ \notag
\langle A^3\rangle & =  \kappa_3 + 3\kappa_1\kappa_2 + \kappa_1^3 \\\notag
\langle A^4 \rangle& = \kappa_4 + 4\kappa_1\kappa_3 + 2\kappa_2^2 + 6\kappa_1^2\kappa_2+\kappa_1^4 \ .
\end{align}
This formulation highlights the combinatorial structure underlying free cumulants: 
each cumulant naturally corresponds to a non-crossing partition of the indices, 
and the $n$-th moment is reconstructed by summing over the contributions associated 
with all non-crossing partitions of size $n$.
The moment functional $\langle \bullet \rangle$ can be chosen quite generally; in RMT, a common and natural choice is the normalized trace $\langle\bullet\rangle=\text{Tr}(\bullet)/D$.

\paragraph{Isomorphism partitions-permutations.}
This correspondence is discussed in more detail in App.~\ref{app_permutations}, while here we only outline the main idea. Every permutation in the symmetric group $\alpha \in S_n$ induces a partition $\mu_\alpha$ by grouping together the elements belonging to the same cycle: $\mu_{\alpha} = \{\text{cycles of } \alpha\}$. 
The converse is not true: a partition does not specify a unique permutation, because one must choose an order of the elements within each block to define a cycle. 

When fixing an order in the blocks, which for us will be the normal ordering, and excluding the crossing partitions, the correspondence becomes bijective. Therefore, one can establish an isomorphism between non-crossing partitions and a specific subset of permutations, the ones lying on the geodesic interval from the identity to the normal-ordered maximal cycle, where the geodesic is defined with respect to the Cayley distance. This statement is made more precise in App.~\ref{app_permutations}, where also the correspondence between the lattice structure and the Cayley graph is highlighted, after introducing a partial order on the permutations.
Acknowledging this isomorphism, we will identify a non-crossing partition with the associated permutation. In particular, we will call them with the same letter, and represent blocks and cycles in the same way. The interpretation depends on the context: when viewed as a map, the notation indicates the action of a permutation, otherwise, it represents the blocks of a partition. We indicate as $e=(1)\dots(n)$ the identity permutation i.e. the discrete partition, and with $\gamma_n=(1\dots n)$ the maximal normal-ordered cycle i.e. the trivial partition.

\section{Global Rotational Invariance}
\label{sec_global}
Our goal is to evaluate the products in Eq.~\eqref{mat_el_prod}, and derive a formula that provides a combinatorial prescription for computing the leading order contribution and the first subleading, in the Hilbert space dimension $D$. 
In this section, we consider the case in which the observable $A^U$ possesses global rotational invariance. The results presented here are derived in App.~\ref{app_GRIderivation} and are obtained by inserting Eq.~\eqref{Haar_av_1} into Eq.~\eqref{mat_el_prod} and performing a systematic expansion in powers of $1/D$. Our analysis focuses on specific products of the form $\overline{A^U_{i_1i_2} A^U_{i_2 i_3}\dots A^U_{i_ni_1}}$, which we refer to as \textit{ETH-cycles}\footnote{These averages of products of matrix elements are the only ones formulated by the full ETH (see Eq.~\eqref{eth}); the interest towards those quantities is motivated by the evaluation of thermal correlation functions. Indeed, when calculating thermal moments, they naturally appear:
\begin{equation}
\langle A^n\rangle_{\beta}= \frac{\text{Tr}(A^ne^{-\beta H})}{\text{Tr}(e^{-\beta H})}=  \frac{1}{Z}\sum_{i_1,\dots, i_n} A_{i_1i_2} A_{i_2 i_3}\dots A_{i_ni_1}e^{-\beta E_{i_1}} \ ,
\label{ther_corr_fun}
\end{equation}
with $Z = \sum_i e^{-\beta E_i}$ partition function at temperature $T=\beta^{-1}$.}
For convenience, we define the ordered set of indices $\textbf{i} :=\{i_1,\dots,i_n\}$, where different $i_j$ may or may not coincide, for different $j$.
By allowing all possible repetitions in \textbf{i}, we encompass all the relevant products of matrix elements that we aim to study. Each can be uniquely associated with a partition of the set $\{1 \dots n\}$, where the numbers indicate the position of indices in $\textbf{i}$, meaning the subindices: we allocate in the same block the positions of equal indices in the set $\textbf{i}$. Specifically, let $B_m$ denote the $m-$th block of the partition, containing $l_m$ elements. All indices corresponding to that block are equal, i.e. $i_{B_m(1)} = \dots = i_{B_m(l_m)}$. The total number of distinct indices in the product is then given by the number of blocks in the partition.
For instance, $A_{ij}A_{ji}A_{ii}A_{ii}$ has $\textbf{i}=\{i,j,i,i\}$ and it is associated to the 2-blocks partition $(134)(2)$. Indeed, $i_1=i_3=i_4\equiv i \ne i_2\equiv j$, with respect to the generic ETH-cycle $A_{i_1i_2}A_{i_2i_3}A_{i_3i_4}A_{i_4i_1}$.

 Based on which indices are repeated, we can have:
\begin{itemize}[leftmargin=1.2em, itemsep=0.6em, topsep=0.6em]
    \item \emph{Simple loops}, if they are all different. These are defined by the $n-$blocks partitions; 
    \item \emph{Cacti}, if there are repeated indices and the associated partition is non-crossing;
    \item \emph{Non-cacti}, if there are repeated indices and the associated partition is crossing. For instance, this is the case for $A_{ij}A_{ji}A_{ij}A_{ji}$ with partition $(13)(24)$.  We will exclude non-cacti diagrams from our discussion\footnote{Those products of matrix elements are usually excluded by ETH discussions, because they give a negligible contribution in the calculation of thermal correlation functions (see Eq.~\eqref{ther_corr_fun}), for entropic reasons. This also explains why free probability, rather than classical probability theory, is relevant for quantum dynamics.}. 
\end{itemize} 
Each ETH-cycle is associated with a diagram that can be identified with the one of the partition so defined \cite{pappalardi2022eigenstate}. For instance, some examples for $n=4$ are
\begin{equation}
    \raisebox{-0.3\height}{\includegraphics[width=0.9\linewidth]{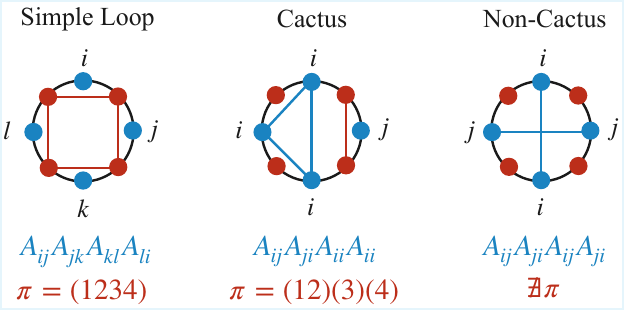}} \ .
\notag
\end{equation}
In this diagrammatic representation, we highlight the original partition of the indices in blue and its dual non-crossing partition (when it exists) in red.
In fact to write the closed formula for leading and subleading contributions, it is convenient to classify the diagrams in terms of the Kreweras complement of the partition introduced before, which we denote generically by $\pi$. While the partition introduced previously ensured that indices labelled by numbers in the same block were equal, for $\pi$ this is not the case; it is more convenient to interpret it as a permutation. 
Concretely, $\pi$ can be seen as a permutation of the set of subindices $\{1,\dots,n\}$ of \textbf{i}; while $\textbf{i}=\{i_1,\dots,i_n\}$ is the set of the indices in the first position in the matrix elements, $\pi(\textbf{i}):=\{i_{\pi(1)},\dots,i_{\pi(n)}\}$ is the set of indices in the second position in the matrix elements.
This allows us to express a product of matrix elements, excluding non-cacti, in the form
$
A_{i_1 i_{\pi(1)}} \dots A_{i_n i_{\pi(n)}}.
$
For example, $A_{ij}A_{ji}A_{ii}A_{ii}$ has Kreweras complement $\pi = (1 2)(3)(4)$ on $\textbf{i}=\{i,j,i,i\}$, which reflects the fact that the first two indices form a cycle and are exchanged in the first two matrix elements: $i_{\pi(1)}=i_2=j$ and $i_{\pi(2)} = i_1=i$, while $i_{\pi(3)} = i_3=i$ and $i_{\pi(4)} = i_4=i$. 
For cacti, the number of blocks $\#\pi$ of the dual partition is the number of \textit{leaves} in the diagram, meaning the number of parts in which the circle is divided when \textit{pinching} equal indices.

\bigskip

With the tools introduced so far, we can now explain the formula for leading and (first) subleading contributions of $\overline{A^U_{i_1i_{\pi(1)}}\dots A^U_{i_ni_{\pi(n)}}}$. This is given here as an ansatz, which it is
proven in App.~\ref{app_GRIderivation}:
\begin{align}
    &\overline{A^U_{i_1i_{\pi(1)}}\dots A^U_{i_ni_{\pi(n)}}} = \frac{\kappa_{\pi}}{D^{n-\#\pi}} 
    \label{close_formula_RMT}
    + \\ &+ \sum_{\substack{\tau\in NC(n)\,  \text{with}\\ \tau > \pi, \  \#\tau= \#\pi-1}}\frac{\kappa_{\tau} }{D^{n-\#\tau}} +\mathcal{O}\left( \frac{1}{D^{n-\#\pi+2}}\right) \ , \nonumber
\end{align}
where  
$\kappa_\mu=\prod_{m=1}^{\#\mu} \kappa^{}_{l_m}$ 
is a product of free cumulants, one for each block, being $\#\mu$ the number of blocks of $\mu$, cf. Eq.~\eqref{free_cum_NC}. These free cumulants are the ones associated to moments given by the normalized traces of powers of the observable $A$. 
We emphasize that, in Eq. 
(\ref{close_formula_RMT}), $\pi$ does not have to be thought as a generic permutation, but as the one associated to the Kreweras complement of the non-crossing partition describing simple loops and cacti. 

Let us now discuss the implications of Eq.~\eqref{close_formula_RMT}.
The leading term contribution is given by the product of free cumulants associated to the blocks of $\pi$, rescaled by the Hilbert space dimension to a power dependent on the number of blocks. This constitutes the simplest ``toy model'' of ETH (see for instance \cite{dalessio2016from} for $n=2$), in which products of matrix elements have expectation values which do not depend on the indices, the constant free cumulants of Eq.~\eqref{ex_momcum}.
The leading term was obtained in a different context and with the methods of App.~\ref{App_HCIZ} in \cite{maillard2019high}.
Remarkably, even the subleading correction admits a compact expression in terms of free cumulants, the ones associated to partitions $\tau$ bigger than $\pi$ at distance one on the lattice of non-crossing partitions. Therefore, the subleading contributions can be read immediately from the lattice $NC(n)$, locating $\pi$, and considering all the partitions on the row above, connected with a line to it. 
The partitions $\tau$ may, depending on $\pi$, be unique, non-unique, or may not exist. 

Let us illustrate these results by explicit examples.
For simple loops, the partition $\tau$ at distance one does not exist. In fact, $\pi$ is the trivial and the coarsest partition possible: it is located at the top of the lattice, and there are no partitions above. Hence, the first order correction is zero\footnote{From the derivation in App.~\ref{app_GRIderivation}, it becomes clear that higher-order contributions are considerably more difficult to compute, which goes beyond the scope of this work. Therefore, we do not provide subleading corrections for simple loops.}.
Its  associated free cumulant is $\kappa_{\pi}=\kappa_n$. For instance, for $\pi=(1234)$, partition at the top of $NC(4)$ in Fig.~\ref{fig_free_prob}, the formula reads:
\begin{subequations}
\begin{equation} 
\label{ex_ijkl}
\overline{A^U_{ij}A^U_{jk}A^U_{kl}A^U_{li}} = \frac{\kappa_4}{D^3}  + \mathcal{O}\left( \frac{1}{D^5}\right) \ .
\end{equation}
As a consequence, one can identify a free cumulant with a diagram with all distinct indices as:
\begin{equation}
    \raisebox{-0.3\height}{\includegraphics[width=0.75\linewidth]{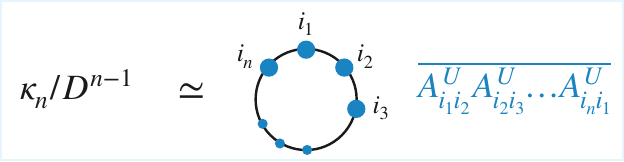}} \ .
\label{simple_diagram}
\end{equation}
Throughout this work, the symbol $\simeq$ denotes the leading and next-to-leading behavior, while all corrections ${o}(D^{-2})$ smaller by two orders of magnitude are neglected.\\

For the case of cacti, the formula gives the factorization of the matrix elements' $n-$th moments in terms of lower order moments with different indices only, each of them given by a single free cumulant $\kappa$, and each of them representing a block of the initial diagram. This is the usual factorization encoded by full ETH: one should divide the statistical average every time a repeated index is present. 
The subleading contribution is a correction to the standard factorization. This can be understood more clearly in the following examples.

For $\pi=(12)(3)(4)$, Kreweras complement of $\overline{A^U_{ij}A^U_{ji}A^U_{ii}A^U_{ii}}$, the leading order is given by $\kappa_2\kappa_1^2/D$, which corresponds to the factorization $\overline{A^U_{ij}A^U_{ji}}\ \overline{A^U_{ii}}^2$. For the subleading, there are three non-crossing partitions with $\#\tau=2$, which are $(12)(34), (123)(4)$ and  $(124)(3)$. 
From the lattices, 
\begin{figure}[h]
\includegraphics[width=0.75\linewidth]{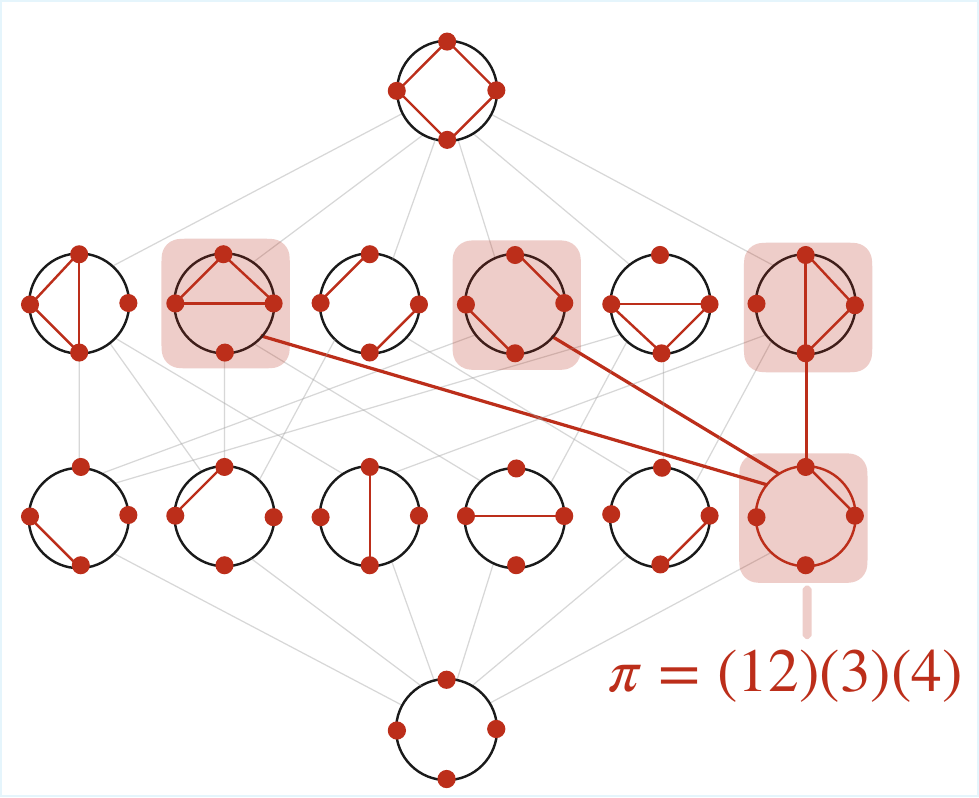} 
 \includegraphics[width=0.75\linewidth]{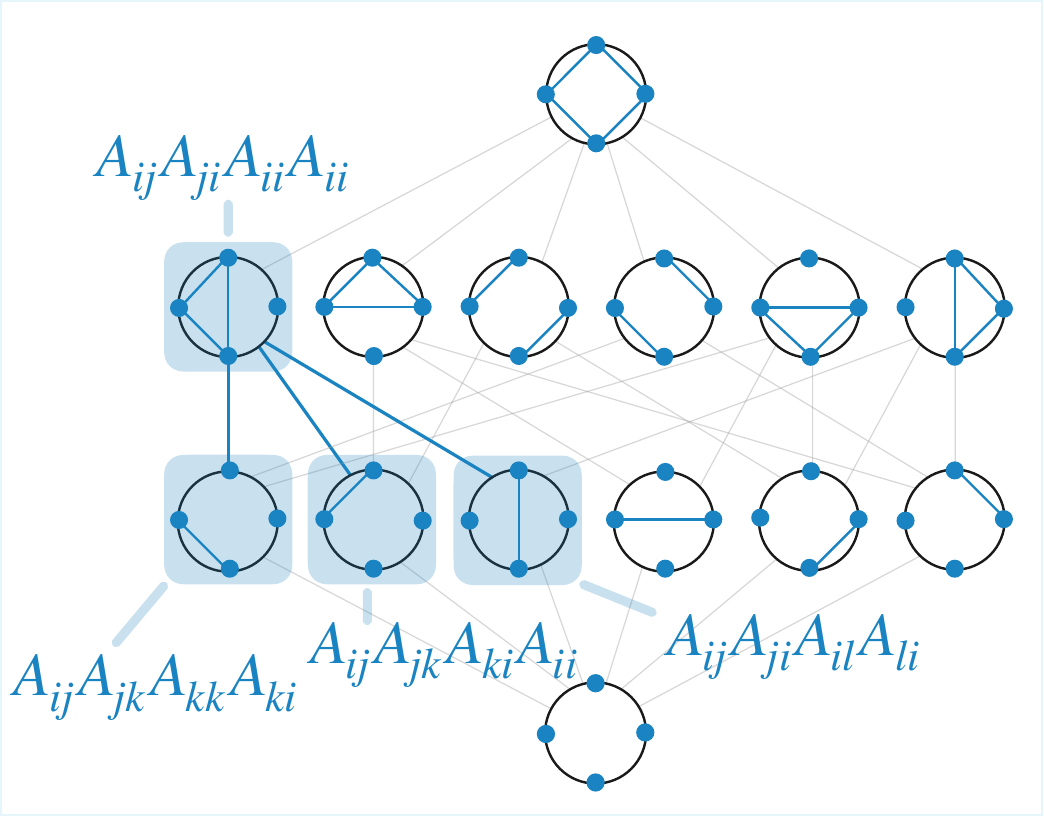}
\end{figure}
We represented both the lattice for partition associated to the repeated indices in a product of matrix elements (second lattice) and the lattice for its Kreweras complement (first lattice). Indeed, since the lattice is self-dual, one can read the result indifferently from one or the other, just reversing the order. 
From the lattice on the bottom, it appears more clearly that the partitions finer of one unity than $\pi$ correspond to products of matrix elements in which there is one additional distinct index, compared to the initial product. Graphically, this corresponds to \textit{breaking the lines} in the initial block one at a time, which effectively means considering one fewer repeated index and thus one fewer block. The subleading corrections are given by the leading contributions of these configurations. 
As a result:
\begin{align}
    \label{ex_ijii}
    &\overline{A^U_{ij}A^U_{ji}A^U_{ii}A^U_{ii}} = \frac{\kappa_2\kappa^2_1}{D} +\frac{\kappa^2_2+2\kappa_1\kappa_3}{D^2}+ \mathcal{O}\left( \frac{1}{D^3}\right) \\ &\simeq \overline{A^U_{ij}A^U_{ji}}\,\,\overline{A^U_{ii}}\,\,\overline{A^U_{ii}} + \overline{A^U_{ij}A^U_{ji}}\,\,\overline{A^U_{il}A^U_{li}}+ 2\overline{A^U_{ij}A^U_{jk}A^U_{ki}}\,\,\overline{A^U_{ii}}\nonumber\\
    \nonumber
    &\raisebox{-0.3\height}{\includegraphics[width=0.85\linewidth]{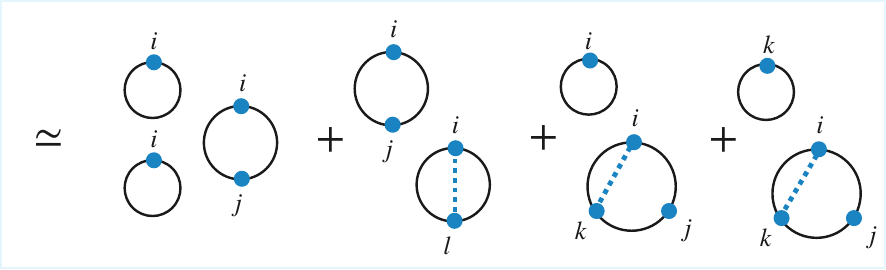}} \ .
\end{align}

The subleading is made by \textit{recombining} the factorized terms of the leading order in all possible ways, while treating indices that were equal as distinct; these indices are connected by a dashed line in the diagrams.

Consider as another example $\pi=(12)(34)$, the Kreweras complement of $\overline{A^U_{ij}A^U_{ji}A^U_{il}A^U_{li}}$: the leading order is given by $\kappa_2^2/D^2$, which corresponds to the factorization $\overline{A^U_{ij}A^U_{ji}}\,\overline{A^U_{il}A^U_{li}}$. For the subleading, there is just one non-crossing partitions with $\#\tau=1$, which is $\gamma_4$, the one at the top of the dual lattice.
This corresponds, indeed, to the product of matrix elements in which there is one additional distinct index, compared to the initial product.
As a result:
\begin{align}
    \nonumber
    &\overline{A^U_{ij}A^U_{ji}A^U_{il}A^U_{li}} = \frac{\kappa^2_2}{D^2}+\frac{\kappa_4}{D^3}
    \notag
    \qquad \raisebox{-0.3\height}{\includegraphics[width=0.35\linewidth]{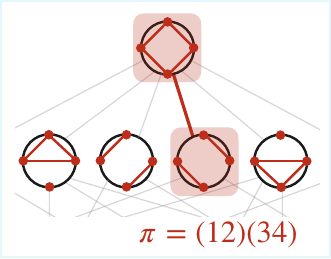}} \\ 
    &\qquad \qquad \quad\ \ \, \simeq \overline{A^U_{ij}A^U_{ji}}\,\,\overline{A^U_{il}A^U_{li}} + \overline{A^U_{ij}A^U_{jk}A^U_{kl}A^U_{li}}
    \nonumber \\
    \label{ex_ijil}
    &\raisebox{-0.3\height}{\includegraphics[width=0.7\linewidth]{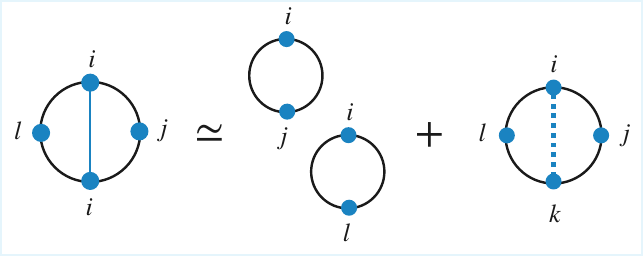}} \ .
\end{align}
\end{subequations}
Therefore, our finding is that \textit{each subleading contribution of cacti equals the leading contribution of a configuration with one more distinct index, previously coincident in the cacti}.
For non-cacti, the partition $\pi$ is not defined which reflects the fact that there is no unique way of factorizing these product of matrix elements, as it is better understood in the derivation in App.~\ref{app_GRIderivation}.

The expression in Eq.~\eqref{close_formula_RMT} contains no explicit energy dependence, indeed, it is well known that this simple RMT toy model is not sufficient to derive the ETH functions. In the following section, we extend this approach in order to incorporate the explicit energy dependence and overcome this limitation.

\section{Local Rotational Invariance}
\label{sec_local}

We now consider the second toy model, which takes into account that $A$ remains invariant under local basis transformations, i.e., reshufflings of its elements restricted to local neighborhoods, i.e. $A \sim A^{U_{\text{loc}}}$. We consider $U_{\rm loc}$ to have local structure in the energy of the Hamiltonian, as explained in Section~\ref{sec_change}. The Haar average for computing the matrix elements' $n-$th moments \eqref{mat_el_prod_local} is performed explicitly in App.~\ref{app_LRIderivation}, for some examples involving the first few $n$ orders. Although the derivations are made using the minimal model in Fig.\ref{fig_unitary}(b), the resulting expressions can be directly translated to the more refined model in Fig.~\ref{fig_unitary}(c). Here we present these results, separating the discussion between simple loops and cacti.

\subsection{Local Free Cumulants}
For simple loops, we show in App.~\ref{app_LRIderivation} that the average is given by a smooth function of the energies,
\begin{equation}
    \label{eq_result_LRI}
    \overline{A^U_{i_1 i_2} \dots A_{i_n i_1}^U}  \simeq K_n(E_{i_1},\dots,E_{i_n}) \ , 
\end{equation}
that we now write explicitly for the first few orders up to $n=4$.
 To specify the functions $K_n$, let us introduce a normalized delta operator, for each energy state 
$i$,
\begin{equation}
    \label{sigma}
    \hat \sigma_{E_i} \equiv \sum_\alpha 
    \frac {\delta_\Delta(E_i-E_\alpha)}{D_i} \ket \alpha \bra{\alpha} \ , \quad \text{Tr}(\hat \sigma_{E_i})=1\ ,
\end{equation}
where we recall that $\delta_\Delta(E_i-E_\alpha)$ is peaked around $E_i$ with a characteristic width $\Delta$, so that $\hat \sigma_{E_i}$ implements a smooth microcanonical projectors on energy $E_i$.
We denote with $\sigma(E_i-E_\alpha) = \delta_\Delta(E_i-E_\alpha)/D_i$ the diagonal matrix element of such operator.
This allows to write the smooth functions $K_n$ as
\begin{widetext}
\begin{subequations}
\label{CDEF}
\begin{align}
 K_1(E_i) & = \text{Tr}(\hat \sigma_{E_i} A) 
 \label{c_1_loc}
 \\
 \label{c_2_loc}
  K_2(E_i, E_j) & =  
\text{Tr}(\hat \sigma_{E_i} A\hat \sigma_{E_j} A) - [\text{Tr}(\hat \sigma_{E_i} A)]^2 \sigma(E_i-E_j)
\\
\begin{split}
    K_3(E_i, E_j, E_k) & = 
\text{Tr}(\hat \sigma_{E_i} A\hat \sigma_{E_j} A\hat \sigma_{E_k} A)  - \text{Tr}(\hat \sigma_{E_i} A)\sigma(E_i-E_j) \, 
  \text{Tr}(\hat \sigma_{E_i} A\hat \sigma_{E_k} A) + 2\rm{cycl.}
  \\ & \quad + 2 \text{Tr}(\hat \sigma_{E_i} A)^3\sigma(E_i-E_j)\sigma(E_j-E_k)
  \end{split} 
  \\
  \begin{split}
K_4(E_i, E_j, E_k, E_l) & =  \text{Tr}(\hat \sigma_{E_i} A\hat \sigma_{E_j} A\hat \sigma_{E_k} A\hat \sigma_{E_l} A)  \\ & \quad - \text{Tr}(\hat \sigma_{E_i} A)\sigma(E_i-E_j) \, 
  \text{Tr}(\hat \sigma_{E_i} A\hat \sigma_{E_k} A\sigma_{E_l} A) + 3\rm{cycl.}
  \\ & \quad - \text{Tr}(\hat \sigma_{E_i} A\sigma_{E_j} A)\sigma(E_i-E_k) \, \text{Tr}(\hat \sigma_{E_i} A\hat \sigma_{E_l} A) + \rm{cycl.}
  \\ & \quad + 2 \text{Tr}(\hat \sigma_{E_i} A)^2\sigma(E_i-E_j)\sigma(E_j-E_k)\text{Tr}(\hat \sigma_{E_i} A\hat \sigma_{E_l} A)+ 3\rm{cycl.}
  \\ & \quad + \text{Tr}(\hat \sigma_{E_i} A)\text{Tr}(\hat \sigma_{E_k} A)\sigma(E_i-E_j)\sigma(E_k-E_l)\text{Tr}(\hat \sigma_{E_i} A\hat \sigma_{E_k} A)+ \rm{cycl.}
  \\ & \quad - 5\text{Tr}(\hat \sigma_{E_i} A)^4\sigma(E_i-E_j)\sigma(E_j-E_k)\sigma(E_k-E_i) \ .
  \end{split}
\end{align}
\end{subequations}
It is straightforward to see that for a flat $\delta_\Delta()=1$ over the entire energy spectrum, one recovers the results of global rotational invariance \eqref{simple_diagram}, discussed in Sec.~\ref{sec_global}. In this limit, these correlators correspond to rescaled free cumulants, $K_n(E_{i_1}, \dots, E_{i_n}) = \kappa_n/ D^{n-1}$, where $\kappa_n$ were defined from normalized traces of powers of the entire matrix, cf. Eq.~\eqref{ex_momcum}.

We assume that, when the smooth functions are multiplied by the energy projectors, then they can be evaluated at the same energy, i.e. $K_1(E_i) \sigma(E_i-E_j)=K_1(E_j) \sigma(E_i-E_j)$. This implies the continuity of the smooth functions when the indices are the same (zero frequencies). 

By inverting the \eqref{CDEF} we find:

\begin{subequations}
\label{localfc_explicit}
\begin{align}
     \text{Tr}(\hat \sigma_{E_i} A) & = K_1(E_i) 
     \\
     \text{Tr}(\hat \sigma_{E_i} A\hat \sigma_{E_j} A) & = K_2(E_i, E_j ) + K_1(E_i) K_1(E_j) \sigma(E_i-E_j)
     \\
     \begin{split}
      \text{Tr}(\hat \sigma_{E_i} A\hat \sigma_{E_j} A\hat \sigma_{E_k} A)  & =
      K_3(E_i, E_j, E_k) 
      + K_1(E_i) K_2(E_j, E_k) \sigma(E_i-E_j) + 2{\rm cycl.} 
       \\ & \quad + K_1(E_i) K_1(E_j) K_1(E_k) \sigma(E_i-E_j) \sigma(E_j-E_k)
      \end{split} \\
        \begin{split}
      \text{Tr}(\hat \sigma_{E_i} A\hat \sigma_{E_j} A\hat \sigma_{E_k} A\hat \sigma_{E_l} A)  & =
      K_4(E_i, E_j, E_k, E_l) 
      + K_1(E_i)K_3(E_j,E_k,E_l)\sigma(E_i-E_j) + 3{\rm cycl.} \\ & \quad  + K_2(E_i,E_j) K_2(E_k, E_l) \sigma(E_i-E_k) + {\rm cycl.} 
       \\ & \quad + K_1(E_i) K_1(E_j) K_2(E_k,E_l) \sigma(E_i-E_j) \sigma(E_j-E_k)+ 3{\rm cycl.} \\ & \quad + K_1(E_i) K_1(E_k) K_2(E_j,E_l) \sigma(E_i-E_j) \sigma(E_k-E_l)+ {\rm cycl.}  
       \\ & \quad + K_1(E_i) K_1(E_j) K_1(E_k)K_1(E_l) \sigma(E_i-E_j) \sigma(E_j-E_k)\sigma(E_k-E_l) \ .
      \end{split}
\end{align}
\end{subequations}
\end{widetext}

One can notice the similarity between these expressions and Eq.~\eqref{ex_momcum}, in fact, $K_n$ can be defined implicitly from a generalized moments-cumulants formula as

\begin{subequations}
\label{localfc_main}
\begin{align}
     \text{Tr}(\hat \sigma_{E_{i_1}} A\dots\hat \sigma_{E_{i_n}} A) &= \\
     = \sum_{\mu\in NC(n)} &K_\mu(E_{i_1}, \dots E_{i_n}) \sigma_{\mu^*}(E_{i_1}, \dots, E_{i_n})\ , \nonumber
\end{align}

where $\mu^*$ is the dual of the non-crossing partition $\mu$, and the function $K_\mu(E_{i_1},\dots,E_{i_n})$ 
is defined as:

\begin{equation}
K_\mu (E_{i_1},\dots, E_{i_n} )= \prod_{m=1}^{\#\mu} K_{l_m}(E_{i_{B_m(1)}},\dots,E_{i_{B_m(l_m)}})\ .
\label{free_cum_local}
\end{equation}
\end{subequations}

The function $\sigma_{\mu^*}(E_{i_1},\dots,E_{i_n})$ is a product of rescaled delta functions $\sigma()$ that equate indices belonging to the same block of $\mu^*$;
note that if multiple indices belong to the same block we should choose the minimum number of $\sigma()$ functions. By inspection of the first few orders, one immediately finds Eqs.~\eqref{localfc_explicit}. 

As it is clear from Eqs.~\eqref{CDEF}, when the energies are well separated, these quantities can be rewritten as traces with appropriate normalization, so that their scaling reads:
\label{eq_scaling_cn}
\begin{equation}
\label{eq_scaling_cn_a}
    K_n(E_1, \dots, E_n) \simeq {D_{E_+}^{1-n}} \ ,
\end{equation}
 where on the right-hand side, we take the large-size limit $L\gg1$ and use that for a finite-rank observable, the trace scales as the density of states at the average energy $E_+ = (E_1+\dots+E_n)/n$. \\

We note that, besides dimensional factors, the same expression \eqref{localfc_main} can be found in Refs.~\cite{hruza2023coherent, bernard2024structured, barraquand2025introduction}, where the corresponding quantities are defined as \textit{local free cumulants}. Therefore, in this work we will likewise refer to $K_n$ as local free cumulants. The main difference is that our $K_n$, derived within the local rotational invariant model, possess the different dimensionality associated with the energies $D_i, D_j, \dots$. In \cite{bernard2024structured} the density of states of the matrices is flat: there is no energy dependency and all the $D$ are the same, as for Floquet models, so one can define properly normalized functions.
Moreover, Ref.~\cite{bernard2024structured}, shows that the local free cumulants introduced there can be related to operator-valued free cumulants. \\
Since Eqs.~\eqref{localfc_main} are the generalization of Eq.~\eqref{free_cum_NC} to matrices with an energy structure, this holds also in our case: we show in App.~\ref{app_linkFB} how Eqs.~\eqref{localfc_main} naturally appear when generalizing scalar free probability to operator-valued free probability. Consequently, local free cumulants can be identified with the diagonal components of specific operator-valued free cumulants, namely, those weighted by projectors onto energy windows.

\subsection{Cacti Diagrams}
 
Finally, let us discuss the implications for the scaling of matrix elements with repeated indices. For these diagrams, one can characterise not only the leading-order, but also the subleading scaling, as we did for the model with global rotational invariance.

We illustrate here examples; for $n=2$, the expectation value with repeated indices reads
\begin{subequations}
    \label{eq_result_LRI_cacti}
    \begin{align}
    \label{ex_ii_LRI}
    \overline{A^U_{ii}A^U_{ii}}
        & \simeq [K_1(E_i)]^2 + K_2(E_i, E_i) \\
        & \simeq  \overline{A^U_{ii}}^2(E_i) + \overline{A^U_{ij}A^U_{ji}}(E_i,E_i) \notag\ .
    \end{align}
This corresponds to the usual factorization with correction given by the second local free cumulant evaluated at equal energies. 
In Eqs.~\eqref{eq_result_LRI_cacti} and in the following, we use the notation $\overline{A^U_{ij}A^U_{ji}}(E_i,E_i)$ to highlight the energy dependencies. 
This result for the diagonal matrix elements matches the well known ETH two-point function result \cite{mondaini2017eigenstate}. \\

For $n=4$, two examples are:
\begin{align}
\label{4p_1_ene}
&\overline{A^U_{ij}A^U_{ji}A^U_{ii}A^U_{ii}}
    \simeq 
K_2(E_i, E_j) [K_1(E_i)]^2 +\\ 
&   + K_2(E_i, E_j) K_2(E_i, E_i) + 2 K_1(E_i) K_3(E_i, E_j, E_i) \nonumber \\ \nonumber 
& \simeq \overline{A^U_{ij}A^U_{ji}}(E_i,E_j)\, \overline{A^U_{ii}}^2(E_i) + \overline{A^U_{ij}A^U_{ji}}(E_i,E_j)\overline{A^U_{il}A^U_{li}}(E_i,E_i)  \\ &+ 2 \overline{A^U_{ii}}(E_i)\overline{A^U_{ij}A^U_{jk}A^U_{ki}}(E_i,E_j,E_i) \ ,
\nonumber
\end{align}

and 
\begin{align}
\nonumber
&\overline{A^U_{ij}A^U_{ji}A^U_{il}A^U_{li}}
   \simeq 
K_2(E_i, E_j) K_2(E_i, E_l) + K_4(E_i, E_j, E_i, E_l) \nonumber  \\ 
& \qquad\qquad \nonumber \simeq \overline{A^U_{ij}A^U_{ji}}(E_i,E_j)\,\,\overline{A^U_{il}A^U_{li}}(E_i,E_l)
\\ &
\qquad\qquad+ \overline{A^U_{ij}A^U_{jk}A^U_{kl}A^U_{li}}(E_i,E_j,E_i,E_l) \ ,
\label{4p_2_ene}
\end{align} 
\end{subequations}
from which it is clear that, at the leading order, the usual factorization happens, while the corrections arise from higher-order local free cumulants, corresponding to products of matrix elements with more different indices, but evaluated at equal mesoscopic energies. \\

These results generalize those obtained for global rotational invariance in Sec.~\ref{sec_global} to the case of structured matrices. In particular, Eqs.~\eqref{4p_1_ene}--\eqref{4p_2_ene} are analogous to Eqs.~\eqref{ex_ijii}--\eqref{ex_ijil}, but now incorporating explicit energy dependence. One could notice that the dashed lines in the diagrams above are, in the case of local rotational invariance, connecting equal mesoscopic energies. 

Therefore, the contributions can be read again from the lattice of non-crossing partitions, and the general expression for the leading and subleading behavior reads: 

\begin{align}
\label{close_formula_LRI}
     \overline{A^U_{i_1 i_{\pi(1)}} \dots A^U_{i_n i_{\pi(n)}}} 
    &\simeq K_{\pi}(E_{i_1},\dots, E_{i_n} ) \,+ \\&+ \sum_{\substack{\tau\in NC(n)\,  \text{with}\\ \tau > \pi, \  \#\tau= \#\pi-1}} K_{\tau}(E_{i_1},\dots, E_{i_n}) \ ,
    \notag
\end{align}

where, in the set $\{E_{i_1},\dots, E_{i_n}\}$ the energies are repeated whenever indices are repeated in \textbf{i}; the function $K_\mu$ was defined in Eq.~ (\ref{free_cum_local}), as a product of local free cumulants.
For $K_\pi$, even if in $\{E_{i_1},\dots, E_{i_n}\}$ there can be repetitions, in the dependencies of its local free cumulants $K_{l_m}(\pi)$ only distinct energies are present, since, by definition, each block of $\pi$ accounts only for different indices. 
For $K_\tau$, instead, since in the products associated there is one additional distinct index compared to the initial product, $K_{l_m}(\tau)$ is calculated at equal energies whenever indices that were previously identical becomes distinct.

As a result, \textit{each subleading contribution of cacti equals the leading contribution of a configuration with one more distinct index, calculated at equal energies for the indices previously coincident.}

Eq.~\eqref{close_formula_LRI} represents one of the main results of this paper, and it directly generalizes the result of full RMT in Eq.~\eqref{close_formula_RMT} to structured matrices. The same behavior is expected to hold for matrix elements of non-integrable Hamiltonians, hence providing a refinement of the full ETH. In Sec.~\ref{sec_numerical}, we shall verify that the same also holds for a non-integrable Floquet model.

\section{Empirical averages \\ over small energy windows}
\label{sec_ETH}

After the local rotational invariant toy model, we are now in position to make a connection with the standard ETH ansatz.
Firstly, the scaling in Eq.~\eqref{eq_scaling_cn_a} of the 
local free cumulants $K_n$ suggests defining some order-one functions
\begin{align}
    F^{(n)}_{e_+}(\omega_{1}, \dots \omega_{n-1}) = \lim_{L\to \infty} e^{(n-1) L s(e_+)} K_n(E_1, \dots E_n)\ ,
\end{align}
commonly known as \emph{ETH smooth functions}, which are generally written as functions of the mean energy density $e_+=E_+/N$ and of all the other $n-1$ independent frequencies $\omega_i = E_i-E_{i+1}$.

In order to make a connection with the ETH ansatz in Eq.~\eqref{eth}, an important step forward that we make in this work is to link the previously defined average of matrix elements over an ensemble of banded unitaries to the \emph{empirical averages} of matrix elements of a given instance of the operator under study, in the basis of the particular Hamiltonian that one has chosen.
We can derive, for diagrams with distinct indices:
\begin{align}
\label{eq_cn_diffInd}
\nonumber
&K_n(E_{i_1}, \dots, E_{i_n}) 
 \\
&  =  {\overline{A_{i_1 i_2}^U\dots A_{i_n i_1}^U}}
\nonumber
\\ 
\nonumber
& \simeq\sum_{\alpha_1 \neq \dots \neq \alpha_n} 
     A_{\alpha_1\alpha_2} \frac{\delta_\Delta(E_{i_1}-E_{\alpha_1})}{D_{i_1}}
     \dots 
     A_{\alpha_n\alpha_1} \frac{\delta_\Delta(E_{i_n}-E_{\alpha_n})}{D_{i_n}}
     \\ & \equiv
     {\overline{A_{i_1 i_2}\dots A_{i_n i_1}}}\quad \text{for $i_1\neq \dots \neq i_n$}\ . 
\end{align}
In order to derive this relation we have assumed that the statistical properties of $A$ are the same as those of $A^U$
so that in the \textit{generalized moments} in (\ref{CDEF}) or in the generalized sums in (\ref{eq_cn_diffInd}) we could arbitrarily use $A$ or $A^U$. We have then assumed that these quantities are self-averaging, which is typically the case in RMT.
This expression corresponds just to a definition of average over smoothed different indices for a single realization of the matrix; this is what we define as empirical average.
The nice aspect of this equation is that \emph{it corresponds to the numerical recipe for deterministic matrix elements}, as commonly used in the literature \cite{pappalardi2025full}.  Furthermore, this expression is continuous for coinciding energies, i.e. when $E_{i}\to E_j$ or $\omega_{ij}=E_i-E_j \to 0$.

Similarly, one can define empirical averages of diagrams with repeated indices. The strategy is to average the matrix elements over small windows of energies, keeping the sum and the smoothing only for the distinct indices. As an example we consider:
\begin{equation}
\overline{A_{ii}A_{ii}}(E_i) \equiv \sum_{\alpha} A_{\alpha\alpha}A_{\alpha\alpha}
\frac{\delta_\Delta(E_i-E_\alpha)}{D_{i}} \ . 
\end{equation}
Under the assumptions made above one can recognize these empirical averages as the averages over the ensemble of unitaries, and obtain the factorization and its corrections.
Other examples of empirical averages can be found in \eqref{app_eq_num}.

\section{Numerical Test  \\ in Floquet Models}
\label{sec_numerical}

This section presents a numerical test of both the well established ETH and the new subleading corrections derived analytically in the previous sections. 
We consider a non-integrable one-dimensional spin chain under a stroboscopic periodic drive, a setup similar to those commonly studied in the literature on Floquet many-body systems \cite{bukov2015universal, fleckenstein2021thermalization} and in recent works on quantum chaos \cite{akila2016particle, bertini2018exact, braun2020transition, pappalardi2025full}. We introduce the model below and subsequently present the simulation methods and results. All numerical codes and datasets used in this work are openly available on Zenodo \cite{zenodo2025}.

\subsection{The Model}
We consider the following time-dependent Hamiltonian
\begin{equation}
H(t) = H_0 + \text{sgn}[\cos(\Omega t)] V_0 \ ,
\label{KIM}
\end{equation}
with $H_0$ and $V_0$ accounting for (i) Ising interactions on a chain of $L$ spin 1/2, (ii) longitudinal and transverse fields, and (iii) mixed interactions between nearest neighbors in the transverse directions, similar to the Dzyaloshinskii–Moriya interactions. Specifically:
\begin{align}
H_0 &= H_Z + H_X+H_{M} \ , \qquad V_0 = H_Z - H_X -H_{M} \ , \notag \\
\notag
H_Z &= J \sum^L_{i} Z_i Z_{i+1} + h \sum^L_i Z_i, \qquad
H_X = g \sum^L_i X_i, \\
\label{Hamiltonian_num}
H_{M} &= d\sum^L_{i} \left(X_i Y_{i+1}+Y_i X_{i+1}\right)\ .
\end{align}
We consider periodic boundary conditions. The time evolution operator for \eqref{KIM}, over one period $T$, is
\begin{align}
\notag
U_F &= e^{-i (H_0+V_0) \frac{T}{4}} \, e^{-i (H_0-V_0) \frac{T}{2}} \, e^{-i (H_0+V_0) \frac{T}{4}}  \\ &=e^{-i H_Z \frac{T}{2}} \, e^{-i (H_X+H_{M}) T} \, e^{-i H_Z \frac{T}{2}}
\label{Floquet_operator}
\\
& \equiv \sum_\alpha e^{-i\nu_\alpha} \ket{\nu_\alpha}\bra{\nu_\alpha}\,, 
\notag
\end{align}
whose spectrum is given by the Floquet quasienergies $\nu_\alpha$, which are $2\pi$ periodic; the choice $\nu_\alpha \in [-\pi,\pi)$, $\forall \alpha$ is made.
The frequency of the drive $\Omega$ is connected to the period as \( T = 2\pi/\Omega \). For definiteness, we fix $\Omega=2 \pi$, and the parameters
$J=1,\, h=0.905,\, g=0.809,\, d=1$.

This ergodic model exhibits translational invariance and parity symmetry, but not time-reversal symmetry, which is broken by the $H_M$ interaction\footnote{These interactions, similarly to those considered in \cite{bouverotdupuis2025random}, were specifically chosen to ensure the system belongs to the unitary symmetry class.}. 
It therefore belongs to the unitary symmetry class, meaning that the level-spacing statistics of the Floquet operator eigenvalues follow those of a random Hamiltonian drawn from the GUE \cite{haake2010quantum}. This is precisely the case for which we derived the analytical results: the observable $A$ was required to be invariant under unitary rotations, reflecting that the Floquet eigenvectors belong to the unitary class. As observable, we choose the total magnetization, rescaled by the number of spins, in order to preserve translational invariance.

In App.~\ref{app_num} we present results for slightly different Hamiltonians. Firstly, we consider a Hamiltonian without translational invariance, in order to check our assumptions for different observables, for instance, the magnetization on a single site. Then, we consider the case for which the time reversal symmetry is restored, and the Hamiltonian belongs to the orthogonal symmetry class; we will briefly discuss the discrepancies observed with respect to the unitary case.

\subsection{Methods}

Our aim is to prove numerically Eq.~\eqref{close_formula_LRI}, in particular we will look at two specific matrix elements' $4-$th moments that are the ones in the examples~\eqref{4p_1_ene}--\eqref{4p_2_ene}. We want to verify the validity of the factorization and its subleading corrections, when calculating numerically the product of matrix elements through the empirical averages. Moreover, we want to confirm the scaling with the Hilbert (sub)space dimension of the different contributions. 

We begin by specializing our findings to Floquet models. 
In this case, the ETH smooth functions do not depend on the absolute energies, but only on the energy differences\footnote{As these differences are also modulo $2\pi$, we restrict to the interval $\omega_{\alpha\beta}\in[-\pi,\pi)$.} $\omega_{\alpha\beta}=\nu_\alpha-\nu_\beta$. This reflects the fact that in periodically driven systems there is no conserved total energy, and thus the statistical properties of matrix elements can only depend on relative quasienergies separations. Moreover, the density of states is flat so that $D_{E^+} = e^{Ls(e_+)} = D$, Hilbert (sub)space dimension. Therefore, for simple loops, calling  $\omega_{\alpha} \equiv \omega_{i_\alpha i_{\alpha+1}}$,
\begin{equation}
\overline{A_{i_1i_2}\dots A_{i_ni_1}}(\omega_1,\dots,\omega_{n-1})  \simeq  F^{(n)}(\omega_1,\dots,\omega_{n-1})  / D^{n-1} \ .
\end{equation}
The relation is similar to the one of global rotational invariance, while still keeps the dependence on the difference of quasienergies.  

Let us write the expressions for the cacti in \eqref{4p_1_ene}--\eqref{4p_2_ene}, specialized for the Floquet case. 
\begin{subequations}
\begin{align}
\label{4p_1_ene_Floquet}
&\overline{A_{ij}A_{ji}A_{ii}A_{ii}} (\omega)
\simeq \overline{A_{ij}A_{ji}}(\omega)\, \overline{A_{ii}}^2 \\ & + \overline{A_{ij}A_{ji}}(\omega)\overline{A_{il}A_{li}}(0) + 2 \overline{A_{ii}}\,\,\overline{A_{ij}A_{jk}A_{ki}}(\omega,-\omega) \ ,
\nonumber
\end{align}
with $\omega\equiv \omega_{ij}=\nu_i-\nu_j$, and 
\begin{align}
\nonumber
&\overline{A_{ij}A_{ji}A_{il}A_{li}}(\omega_1,\omega_2) \simeq \overline{A_{ij}A_{ji}}(\omega_{1})\,\,\overline{A_{il}A_{li}}(\omega_{2}) +\\ &+ \overline{A_{ij}A_{jk}A_{kl}A_{li}}(\omega_{1},-\omega_{1},\omega_{2}) \ ,
\label{4p_2_ene_Floquet}
\end{align} 
with $\omega_1:=\omega_{ij}=\nu_i-\nu_j,\ \omega_2:=\omega_{il}=\nu_i-\nu_l$. \\
\end{subequations}
The numerical calculation of each product of matrix elements is made by employing empirical averages, where the summing and the smoothing are performed directly on the quasienergy differences, and involve only distinct indices. 
It is important to account for the intrinsic periodicity of the quasienergies, ensuring that the smoothing procedure is formulated in a correspondingly periodic manner. Specifically, since $\nu_\alpha $ is identified with $\nu_\alpha\pm 2\pi$, we employ:
\begin{align}
&\delta^{\,\text{per.}}_{\omega}(\nu_\alpha,\nu_\beta) = \delta_{\Delta}(\omega-(\nu_\alpha-\nu_\beta))+   \\ &\delta_{\Delta}(\omega+2\pi-(\nu_\alpha-\nu_\beta)) + \delta_{\Delta}(\omega-2\pi-(\nu_\alpha-\nu_\beta))\, .
\notag
\end{align}
Therefore, for simple loops the empirical average reads:

\begin{align}
\notag
&{\overline{A_{i_1 i_2}\dots A_{i_n i_1}}} (\omega_{1},\dots \omega_{n-1}) \simeq \label{emp_av_Floquet}
\\ &\frac{1}{D}\sum_{\alpha_1 \ne \dots \ne \alpha_n} 
A_{\alpha_1\alpha_2} \frac{\delta^{\,\text{per.}}_{\omega_1}(\nu_{\alpha_1},\nu_{\alpha_2})}{D_{\omega_1}}
\dots \\ &\qquad \dots A_{\alpha_{n-1}\alpha_n}  \frac{\delta^{\,\text{per.}}_{\omega_{n-1}}(\nu_{\alpha_{n-1}},\nu_{\alpha_n})}{D_{\omega_{n-1}}} A_{\alpha_n\alpha_1}  \ , 
\notag
\end{align}
with $D_{\omega} = \sum_{\alpha,\beta} \delta^{\,\text{per.}}_{\omega}(\nu_{\alpha},\nu_{\beta})/D$. Consequently the formulation follows for the cacti. 
All the expressions numerically implemented are explicitly reported in App.~\ref{app_num_supp}.

\begin{figure*}[t]
\centering
\includegraphics[width=1 \textwidth]{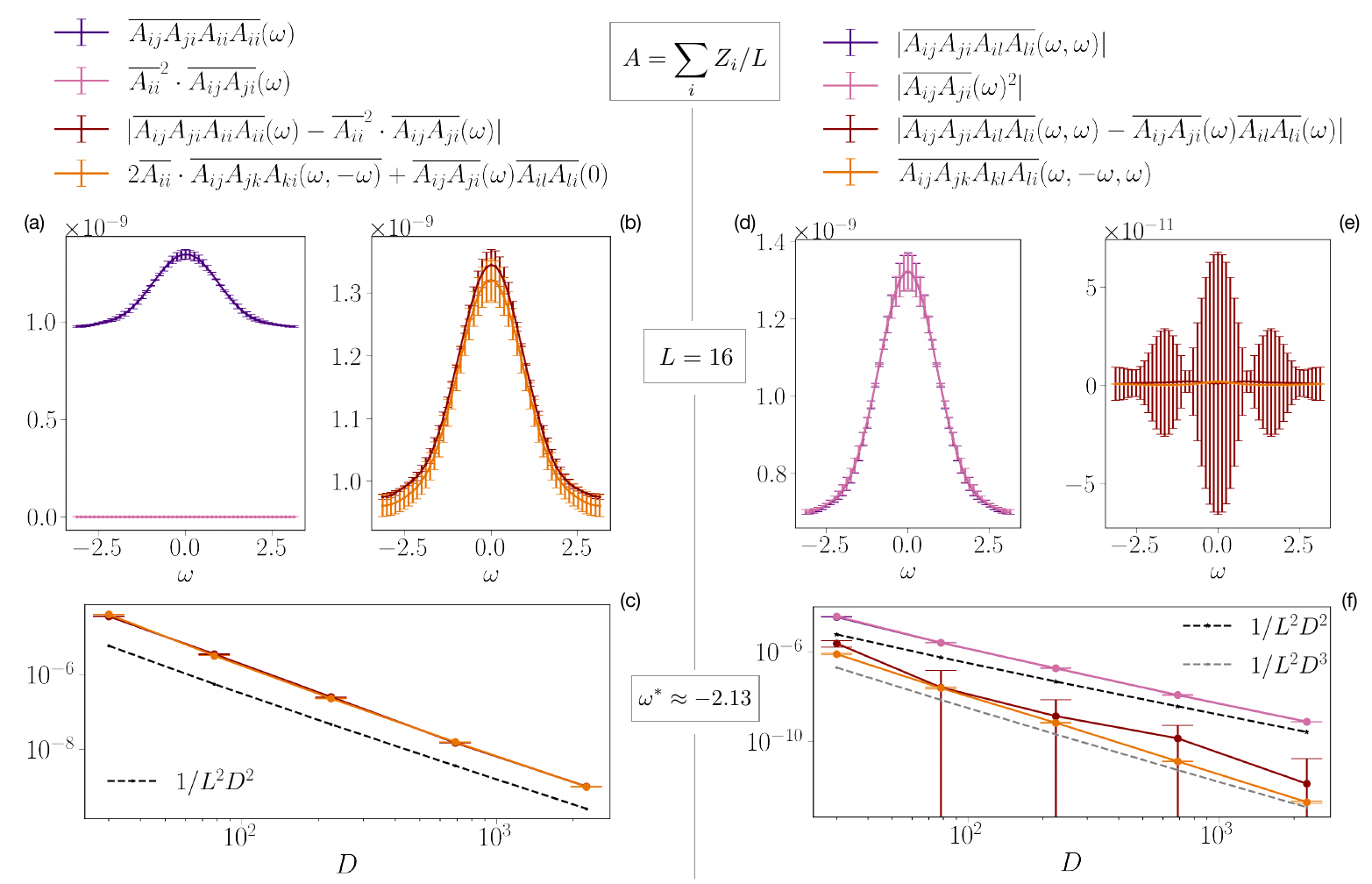}
\caption{
Numerical results for the Floquet system in \eqref{Floquet_operator}, and for the observable $A=\sum_i Z_i/L$. We numerically reproduce Eqs.~\eqref{4p_1_ene_Floquet}--\eqref{4p_2_ene_Floquet}, for $\omega_1=\omega_2=\omega$ chosen in $[-\pi,\pi]$. The error bars are given by averaging each product, for each $\omega$, over different values of the smoothing parameter $\Delta$, as explained in the text. (a),(b),(c) refer to \eqref{4p_1_ene_Floquet}, while (d),(e),(f) to \eqref{4p_2_ene_Floquet}.  \\
In (a),(d) we compare the right-hand side (purple curve) with the respective leading order (pink curve) for fixed $L=16$; for (a) the leading order is zero, while in (d) the data nicely reproduce the expected leading order factorization.
In (b),(e) we compare the difference between the right-hand side and the leading contribution (brown curve) with the respective subleading correction (orange curve) for fixed $L=16$. The data verify the subleading correction to the factorization.
In (c),(f) we represent the scalings of all the curves as a function of $D$. The scalings were studied for values $L\in[8,10,12,14,16]$, and for a fixed value of $\omega^*\approx-2.13$. While all the curves in (c) scale as $1/L^2D^2$, since the leading order is zero, in (f) we observe the different scalings between the leading (purple and pink curves), as $1/L^2D^2$ and the subleading (brown and orange curves), as $1/L^2D^3$. }
\label{fig_numerics}
\end{figure*}

\subsection{Numerical Results}
\label{subsec_num}
We perform exact diagonalization of the Floquet operator \eqref{Floquet_operator}, using the QuSpin package \cite{weinberg2017quspin, quspin} for writing the different terms in \eqref{Hamiltonian_num}. In particular, we restrict the Hilbert space to the symmetry block of positive parity and zero momentum; we indicate with $D$ the dimension of this subspace. 
As observable, we choose the total magnetization along the $z$ direction, rescaled by the number of spins, meaning $ A = \sum_i Z_i/L$. After evaluating the matrix elements in the Floquet basis $A_{\alpha\beta} = \langle \nu_\alpha|A|\nu_\beta \rangle$, we calculate the averages of the products of matrix elements involved in Eqs.~\eqref{4p_1_ene_Floquet}--\eqref{4p_2_ene_Floquet}, through the expressions \eqref{app_eq_num}, fixing, in the second example, $\omega_2=\omega_1\equiv\omega$. We choose 51 values of $\omega$ in $[-\pi,\pi)$.
Numerically, we implement the smoothing through a Gaussian function
\begin{equation}
    \delta_\Delta(\omega-(\nu_\alpha-\nu_\beta)) = \frac{1}{\sqrt{\pi\Delta}} \exp\left({-\frac{(\omega-(\nu_\alpha-\nu_\beta))^2}{\Delta}}\right)\ ,
\end{equation}

choosing $\Delta$ so that a sufficient number of states is satisfying the condition, still much smaller than $D$. Moreover, we average each matrix elements' moment over multiple values of $\Delta$s, in order to have a final function with the respective error bars, for each value of $\omega$. In App.~\ref{app_num_supp} we explicitly write the intervals chosen for the parameter $\Delta$. 

All results are plotted in Figure~\ref{fig_numerics}. For a fixed number of spin $L=16$, we compare the right-hand side (purple curve) of Eqs.~\eqref{4p_1_ene_Floquet}--\eqref{4p_2_ene_Floquet} with the respective leading order (pink curve), and then we subtract the leading to the right-hand side, in order to compare this difference (brown curve) with the prediction for the subleading (orange curve). All curves are plotted as a function of $\omega$. Moreover, in panels (c)(f), we study the scaling with the subspace dimension $D$ for each of these curves, for a fixed $\omega^*\approx -2.13$. This value was chosen arbitrarily as a representative parameter value; the qualitative behavior of the plots is not sensitive to variations of $\omega^*$ which confirms the robustness of the results. The numerical results nicely agree the predictions, 
which follow from the fact that, for distinct indices, simple loops scale as $\overline{A_{i_1i_2}\dots A_{i_ni_1}} \propto 1/D^{(n-1)}$. 

Since $\braket{M_z}=0$, we notice that the leading order of $\overline{A_{ij}A_{ji}A_{ii}A_{ii}}(\omega)$ is zero, therefore the average follows directly the subleading contribution $\overline{A_{ij}A_{ji}}(\omega)\overline{A_{il}A_{li}}(0)$, and scales as $1/L^2D^2$, as expected. On the contrary, $\overline{A_{ij}A_{ji}A_{il}A_{li}}(\omega)$ is very well factorized in $\overline{A_{ij}A_{ji}}(\omega)\overline{A_{il}A_{li}}(\omega)$, scaling as $1/L^2D^2$, and the subleading well follows the expected curve $\overline{A_{ij}A_{jk}A_{kl}A_{li}}(\omega,-\omega,\omega)$, scaling as $1/L^2D^3$.
The rescaling with $L$ are required since the magnetization has been rescaled with the total number of spins $L$, for being a local observable.

\section{Conclusions and Perspectives}
\label{sec_conclusions}

In this work we have revisited the general ideas behind the development of 
full ETH. 
In particular, we have discussed how the properties of matrix elements can be studied under random basis transformations which are associated to local perturbations of the Hamiltonian.
Different levels of refinement of the model led us to consider (i) a completely random basis defined by the application of a Haar distributed random unitary matrix, (ii) a block diagonal unitary where each block is independent and Haar distributed, (iii) a smoothed version of this last model which can be associated to a banded diagonal unitary matrix.

The natural mathematical framework of our study is provided by free probability theory.
In fact, for all these ensembles, we derived analytical expressions of moments of the matrix elements of the observable in terms of (local) free cumulants.
As it was noticed before, the products of interest, for the purposes of computing correlation functions, can be classified as diagrams such as "simple loops", when indices concatenate but do not repeat, and ``cacti", which are non-crossing partitions of the indices with some repetitions.
It was postulated that the expectation of cacti factorizes and in our work we provide a detailed characterization of the next to leading term beyond this assumption, which can be expressed in terms of non-crossing partitions at higher order.

In addition, we show how, under the assumption of (local) rotational invariance, 
the averages of the matrix elements under the application of random unitaries are compatible with the empirical averages performed over a single instance of the matrix ensemble, upon averaging over energy windows, the method that is usually performed in numerical simulations to extract the statistics of such entries.\\

As emphasized in the Introduction, deriving ETH and its subleading corrections from the principle of local rotational invariance provides a deeper understanding of the physical meaning of matrix-element correlations, of the averages entering the ETH ansatz, and of the ETH functions themselves in terms of global properties of the operator. At the same time, the characterization of subleading corrections, already present in standard ETH, fills a structural gap in the full ETH framework.\\
More generally, these results open several promising directions for future research, which we discuss in the remainder of this section.

\subsection*{Applications of our results}
We first discuss applications of the subleading corrections we have computed, which play a quantitative role in quantum dynamics.
\begin{itemize}[leftmargin=1.2em, itemsep=0.5em, topsep=0.5em]
\item \emph{Higher-order correlation functions: late-time plateaus and freeness.} The subleading corrections discussed in this paper determine the late time plateau of \emph{thermal free cumulants}~\cite{pappalardi2022eigenstate} at long-times.
In App.~\ref{app_corr_fun} we compute higher-order thermal correlation functions and explicitly identify the corresponding subleading terms; these can be incorporated into thermal free cumulants, yielding a refinement of the ETH free cumulants introduced in Ref.~\cite{pappalardi2022eigenstate}.
Since the leading ETH free cumulants average to zero at late times due to the non-resonance condition~\cite{fava2025designs}, these corrections control the late-time plateau and the fluctuations around it. 
For the second and fourth thermal free cumulants, we find
\begin{align}
    [k_2(t,0)]_\infty &\simeq  {K_2(\omega =0)}\ , \\
    [k_4(t,0,t,0)]_\infty
    \simeq & \ 2
    \sum_{i\neq j}
    K_4(\omega,-\omega,-\omega)\ , \notag
\end{align}
of order $\mathcal{O}(D^{-1})$, with plateau fluctuations of order $\mathcal{O}(D^{-2})$. The subleading corrections thus set the exponentially small late-time plateau, providing a quantitative characterization of long-time freeness in chaotic dynamics. \\
We note that for non-Floquet systems, one should consider further corrections coming from the dependency of observables on the average energy, see e.g. Ref.~\cite{zhang2026finite}.

\item \emph{Breakdown of ETH.}
Our subleading corrections may also offer a sharp diagnostic of chaos. At $k=2$ they predict that the variance of the diagonal matrix elements equals the off-diagonal two-point function at zero frequency [Eq.~\eqref{ex2b_summary}], decaying as $1/D$. In integrable systems, instead, the diagonal matrix elements fluctuate much more strongly, with a variance that decays only \emph{polynomially} in the system size, as $1/L^{a}$~\cite{vidmar2016generalized}, a violation of this $k=2$ prediction. The all-order corrections derived here could thus provide a precise way to distinguish integrable from non-integrable systems. 
Relatedly, recent works on the \emph{breaking of ergodicity}, e.g.\ the ``fading ergodicity'' near the many-body-localized regime~\cite{kliczkowski2024fading}, address the anomalous low-frequency scaling of the ETH functions, where our framework could find natural applications.

\item \emph{Different symmetry classes.}
The subleading corrections also depend on the symmetry class. Already at $k=2$, the diagonal variance is fixed by the off-diagonal two-point function at zero frequency up to a class-dependent factor, one for the unitary class and two for the orthogonal one~\cite{dalessio2016from, mondaini2017eigenstate, foini2019eigenstate2} [Eqs.~\eqref{GOE_eq}--\eqref{GOE_eq2}], as we verify numerically in App.~\ref{app_num_GOE}. The difference stems from rotational invariance: in the orthogonal class the observable is invariant under local \emph{orthogonal} transformations, so the permutation Weingarten calculus used here no longer applies. A full prediction of the higher-$k$ discrepancies between symmetry classes remains an open problem.
\end{itemize}

\subsection*{New directions}
Beyond these applications, the principle of local rotational invariance suggests several directions where it could be applied more broadly.
\begin{itemize}

\item \emph{Characterization of an ETH ensemble.}
Much recent attention has been devoted to characterizing ensembles of unitaries $\mathcal{E}$ and their ability to reproduce the Haar ensemble. A major open challenge is to identify the ensemble appropriate to chaotic Hamiltonian time evolution,
\begin{equation}
\mathcal{E}_H = \{\, e^{-iHt},\ t\sim\mathcal{P} \,\}\ ,
\end{equation}
where $\mathcal{P}$ is a suitable distribution of times. Since such an evolution conserves energy, and, in fact, every eigenstate, the Haar ensemble must be endowed with additional structure in order to capture the randomness of chaotic dynamics. Most efforts so far have focused either on ensembles of states, in the context of deep thermalization and Hilbert-space ergodicity~\cite{mark2024maximum, mcginley2025scrooge, gosh2025late, tirrito2025anticoncentration}, or on random Hamiltonians~\cite{cui2025random}. Local rotational invariance, through the associated block unitary $U_{\rm loc}$, provides a first constructive definition of such an \emph{ETH ensemble}: an ensemble of unitaries that, applied to the matrix elements of an observable, reproduces by construction the correlations dictated by full ETH, including both the smooth energy dependence and the subleading corrections. Making the relation between $U_{\rm loc}$ and the ensembles above precise remains an important open problem.

\item \emph{Connection to eigenstate correlations.}
Here we have applied local rotational invariance to the matrix elements of \emph{observables}. There is, however, great interest in the correlations between the \emph{eigenstates} themselves~\cite{hosur2016chaos, hahn2024eigenstate}, which control information-theoretic quantities and the spreading of operators. Since such quantities can be written as traces of suitable projector operators, local rotational invariance could in principle be applied to those projectors as well, providing a route to derive eigenstate-correlation results within the same framework.

\item \emph{Block and banded unitary averages.}
On the technical side, our work introduces \emph{a new methodological tool}: the average over block-structured random unitaries, representing a first step towards a systematic Weingarten calculus for unitaries with an energy structure. While we have shown how to treat block-diagonal Haar unitaries, the main open challenge is to extend this framework to genuinely smooth banded unitaries, such as the one shown in Fig.~\ref{fig_unitary}(c). Beyond the present setting, we expect this methodology to be of independent interest and to find applications in other communities where structured random unitaries naturally arise, for instance in the study of quantum dynamics with conserved quantities.

\end{itemize}

\begin{acknowledgments}
We are grateful to R. Speicher and A. Wendel for their valuable and insightful feedback on free probability.
We thank J. Kurchan for continuous collaborations on this topic.
E.V. and S.P. are supported by the Deutsche Forschungsgemeinschaft (DFG, German Research Foundation) under Germany’s Excellence Strategy - Cluster of Excellence Matter and Light for Quantum Computing (ML4Q) EXC 2004/1 -390534769, and DFG Collaborative Research Center (CRC) 183 Project No. 277101999 - project B02.
L.F. acknowledges support by the French government through the France 2030 program (PhOM – Graduate School of Physics), under reference ANR-11-IDEX-0003 (Project Mascotte, L. Foini).\\
The data that support the findings of this article are openly available \cite{zenodo2025}.
\end{acknowledgments}

\bibliographystyle{apsrev4-2}
\bibliography{biblio}

\newpage
\appendix

\section{More on Free Probability}
\label{app_FB}
In Sec.~\ref{sec_free_prob} we introduced the basic notions of free probability theory, with particular emphasis on the combinatorial structure of partitions. In this section, we instead follow a more standard mathematical presentation, highlighting directly why free probability provides a natural framework for studying quantum operators in chaotic many-body quantum systems and their statistical properties. We then introduce operator-valued free probability, since some of the quantities introduced in this work can be naturally identified with objects already appearing in this framework. Finally, we discuss the isomorphism between partitions and permutations in more detail, as it will be essential for the derivations in the following appendices.
\subsection{Standard introduction to free probability}
\paragraph{Non-commutative probability spaces.}
The fundamental object in probability theory is a probability space $(\mathcal{X},\varphi)$, consisting of a unital $*$-algebra $\mathcal{X}$ and a linear functional $\varphi:\mathcal{X}\to\mathbb{C}$ satisfying
$$
\varphi(\mathbb{1})=1,\qquad
\varphi(x^\dagger x)\ge 0,\qquad
\varphi(x^\dagger)=\varphi^*(x)\ .
$$
When $\mathcal{X}$ is commutative, one recovers classical probability theory. A standard example is
$\mathcal{X}=L^\infty(\Omega,P)$,
the algebra of bounded random variables on a probability space $(\Omega,P)$, with expectation functional
$\displaystyle \varphi(x)=\int_\Omega x(\omega),dP(\omega)$.\\

Free probability extends this framework by allowing $\mathcal{X}$ to be non-commutative. Two important examples are: 
(i) the algebra $M_D(\mathbb{C})$ of $D\times D$ matrices equipped with the normalized trace
$\varphi(x)=\mathrm{Tr}(x)/D$;
(ii) a von Neumann algebra of observables acting on a Hilbert space $\mathcal{H}$, together with a vector state
$
\varphi_\psi(x)=\langle \psi|x|\psi\rangle \ .
$
These two examples are closely connected. Indeed, free probability originated in the study of operator algebras, and, since large random matrices often provide finite-dimensional models for operator-algebraic structures, it was strongly influenced by random matrix theory; subsequently it supplied powerful tools for instance for studying the asymptotic eigenvalue distributions of large random matrices. \\
In the main text (and in the following sections), for a generic quantum observable $A$ seen as a $D\times D$ matrix, we adopt the convention $ \langle A \rangle\equiv  \varphi(A) := \text{Tr}(A)/D$.\\
When $\mathcal{X}$ is non-commutative, many classical probabilistic notions require a non trivial reformulation. The most important example is the concept of independence. For classically independent random variables $x$ and $y$, mixed moments factorize. For instance,
$$
\varphi(xyxy)=\varphi(x^2y^2)
=\varphi(x^2)\varphi(y^2).
$$
The question is how this property should be generalized in a non-commutative setting.\\
\paragraph{Freeness.}
The non-commutative analogue of independence is free independence, or \emph{freeness}. Let $X$ and $Y$ be unital subalgebras of $\mathcal{X}$. They are said to be freely independent if
\begin{equation}
    \varphi(c_1c_2\cdots c_n)=0
\end{equation}
whenever:
\begin{itemize}[leftmargin=1.2em, itemsep=0.5em, topsep=0.5em]
\item each $c_k$ belongs either to $X$ or to $Y$;
\item consecutive elements come from different subalgebras;
\item $\varphi(c_k)=0$ for all $k$.
\end{itemize}
As an example, if $x\in X$ and $y\in Y$ are free, then
$$
\varphi(xyxy) =
\varphi(x^2)\varphi(y)^2
+
\varphi(x)^2\varphi(y^2)
\varphi(x)^2\varphi(y)^2\ .
$$
Freeness and classical independence are distinct notions. In general, freely independent variables are not classically independent, and conversely.\\
A fundamental source of free random variables arises in random matrix theory: independent random matrices from ensembles such as the Gaussian Unitary Ensemble (GUE) become asymptotically free in the limit of large matrix dimension.
\paragraph{Further developments.}
Many central constructions of classical probability admit free analogs, including the free central limit theorem, free convolution, free cumulants, and their associated generating functions. A distinctive feature of free probability is its deep connection with combinatorial structures, most notably non-crossing partitions, as highlighted in Sec.~\ref{sec_free_prob}.
\subsection{Operator-valued free probability}
\label{app_OVFB}
Operator-valued free probability generalizes standard free probability by replacing the scalar-valued expectation $\varphi : \mathcal{X} \rightarrow \mathbb{C}$ with a \textit{conditional} expectation
$\varphi_{\mathcal{D}} : \mathcal{X} \rightarrow \mathcal{D}$
where $\mathcal{D}$ is a distinguished subalgebra of $\mathcal{X}$, satisfying the condition $\varphi_{\mathcal{D}}(\hat d_1 x \hat d_2) = \hat d_1\varphi_{\mathcal{D}}(x)\hat d_2$, for $\hat d_{1,2}\in \mathcal{D}$.
Moments retain information encoded in the chosen subalgebra, rather than collapsing all variables to scalar quantities: the choice of $\varphi_\mathcal{D}$ determines which degrees of freedom are averaged over and which are preserved. This additional structure makes operator-valued free probability particularly well suited for systems with internal degrees of freedom. \\
In the context of random matrices, this framework is particularly useful for describing ensembles with non trivial internal structure, so that the resulting moments preserve information about the underlying structure that would be lost under a global trace. 
We call \textit{operator-valued moments} the matrices 
$\langle A\hat d_1 A\hat d_2 A\dots \hat d_n A \rangle_{\mathcal{D}}\in \mathcal{D}$, with $\hat d_k\in \mathcal{D}, \forall k$.
The corresponding \textit{operator-valued free cumulants} are defined implicitly through the same moments–cumulants relation as in Eq.~\eqref{free_cum_NC}, that is 
\begin{equation}
\langle A \hat d_1A\dots \hat d_{n} A \rangle_{\mathcal{D}}  = \sum_{\mu \in NC(n+1)} \kappa^{\mathcal{D}}_\mu(A \hat d_1,\dots, A \hat d_{n}, A) \ .
\label{op_val}
\end{equation}
The difference with respect to scalar free probability is that 
the matrices $K^{\mathcal{D}}_\mu$ are not factorized in a product of block cumulants as in \eqref{free_cum_NC}. Instead, the cumulants associated with the blocks of $\mu$ are recursively composed according to the nesting structure of the non-crossing partition, since they take values in $\mathcal{D}$.\\

For the purposes of this work, we restrict our attention to a specific example, that is the one of block matrices with the convenient choice of $\mathcal{D}$ given by a block-diagonal matrix with diagonal blocks. We focus on this particular setting because it admits a direct connection with the results presented in this work (see App.~\ref{app_linkFB}).\\
Concretely, let us consider four blocks:\\
\begin{align}
&A= 
\begin{pmatrix}
    \mathcal{A}_{11} & \mathcal{A}_{12} \\
    \mathcal{A}_{21} & \mathcal{A}_{22}
\end{pmatrix} \ , \qquad 
\langle A \rangle_{\mathcal{D}} := \begin{pmatrix}
    \langle \mathcal{A}_{11}\rangle \mathbb{I} & 0 \notag \\
    0 &  \langle \mathcal{A}_{22}\rangle \mathbb{I}
    \end{pmatrix} &\, ,\\ \notag \\ 
& \label{example_ovfb} \mathcal{D} =  \bigg\{ \begin{pmatrix}
    d_1\mathbb{I} & 0 \\
    0 & d_2\mathbb{I}
    \end{pmatrix} \bigg\}_{d_1,d_2 \in \mathbb{R}} =  \\ & \qquad \qquad 
    = \mathrm{span} \bigg\{\hat \Pi_1 =  \begin{pmatrix}
    \mathbb{I} &0\\
    0 &0
    \end{pmatrix} ,\hat \Pi_2 = \begin{pmatrix}
    0 & 0 \\
    0 & \mathbb{I}
    \notag
    \end{pmatrix} \bigg\}
\notag
\end{align}

Using Eq.~\eqref{op_val}, we have 
\begin{equation}
\label{example_ovfb_k1}
    \kappa^{\mathcal{D}}_1(A) = \mathrm{diag}(\langle \mathcal{A}_{11}\rangle \mathbb{I},  \langle \mathcal{A}_{22} \rangle \mathbb{I} )\ .
\end{equation}
We can then calculate some examples of second order operator-valued moments and free cumulants. We are interested in the ones weighted with the projectors on the diagonal blocks $\hat \Pi_{I=1,2}$:
\begin{align}
\langle A\hat \Pi_1 A\rangle_{\mathcal{D}} =
\mathrm{diag}(\langle \mathcal{A}^2_{11}\rangle \mathbb{I},  \langle \mathcal{A}_{21}\mathcal{A}_{12}\rangle \mathbb{I})\ , \notag \\
\langle A\hat \Pi_2 A\rangle_{\mathcal{D}} = \mathrm{diag}(
\langle \mathcal{A}_{12}\mathcal{A}_{21}\rangle \mathbb{I},  \langle \mathcal{A}^2_{22}\rangle \mathbb{I})\ .
\label{example_ovfb_m2}
\end{align}
The moments-cumulants formula reads
\begin{align}
\langle A\hat \Pi_{I}A \rangle_{\mathcal{D}}  = 
\kappa^{\mathcal{D}}_2(A\hat \Pi_{I}, A) + \kappa_1 (A \hat \Pi_{I}) \kappa_1(A) \ ,
\end{align}
and one can check that $\kappa_1 (A \hat \Pi_1) = \kappa_1 (A)\hat \Pi_1$.
Therefore, \\ 
\begin{equation}
\begin{aligned}
\kappa^{\mathcal{D}}_2(A\hat \Pi_{1}, A)
&= \mathrm{diag}\left(
(\langle \mathcal{A}^2_{11}\rangle-\langle \mathcal{A}_{11}\rangle^2)\mathbb{I},
\langle \mathcal{A}_{21}\mathcal{A}_{12}\rangle\mathbb{I}
\right),\\
\kappa^{\mathcal{D}}_2(A\hat \Pi_{2}, A)
&= \mathrm{diag}\left(
\langle \mathcal{A}_{12}\mathcal{A}_{21}\rangle\mathbb{I},
(\langle \mathcal{A}^2_{22}\rangle-\langle \mathcal{A}_{22}\rangle^2)\mathbb{I}
\right).
\end{aligned}
\label{example_ovfb_k2}
\end{equation}\\
This can be generalized to higher order and for a generic number of blocks. 

\newpage

\subsection{More on the isomorphism \\ between partitions and permutations}
\label{app_permutations}
In Sec.~\ref{sec_free_prob} we briefly mentioned the isomorphism between partitions and permutations, without entering into its details. Here we clarify and formalize this correspondence, as it will be essential for the derivation of Eq.~\eqref{close_formula_RMT} in App.~\ref{app_GRIderivation}. 

\paragraph{Permutations.} 
A permutation of $n$ elements is a bijective rearrangement of the set 
$\{1, \dots, n\}$. Using the cyclic notation, we represent it as a sequence of disjoint cycles 
$\alpha = (C_1) (C_2) \dots (C_{\#\alpha})$, where $\#\alpha$ is the number of cycles and each cycle $(C_m)$ lists the elements that are mapped cyclically. Two special permutations are the identity $e$ ($n$ cycles of 1 element) and the maximal cycle $\gamma_n = (1\,2\,\dots\,n)$ (a single normal-ordered $n$-cycle).

\paragraph{Lattice.}
A useful notion is the one of rank or Cayley weight of a permutation $\alpha \in S_n$: $|\alpha|= n-\# \alpha$. The rank represents the minimal number of transpositions needed to generate $\alpha$: it is minimized by $e$ (rank $0$) and maximized by $\gamma_n$ (rank $n-1$).
The Cayley weight induces a natural metric on the symmetric group $S_n$,
\begin{equation}
    d(\alpha,\beta)=|\alpha^{-1}\beta| = n-\#(\alpha^{-1}\beta)\ ,
    \label{Cayley_dist}
\end{equation}
known as the Cayley distance. This distance measures the minimal number of transpositions required to transform $\alpha$ into $\beta$.
With this metric, we consider the set of all permutations $\eta$ that saturate the triangle inequality between $\alpha$ and $\beta$, that is, all permutations lying on some shortest path between $\alpha$ and $\beta$:
\begin{equation}
[\alpha,\beta] = \bigl\{\, \eta \in S_n \; : \;
d(\alpha,\beta) = d(\alpha,\eta) + d(\eta,\beta) \,\bigr\}\ .
\end{equation}
Between $\alpha$ and $\beta$ there may exist multiple distinct shortest paths, corresponding to different sequences of transpositions, each defining a geodesic sequence
\[
\alpha = \eta_0, \; \eta_1, \; \dots, \; \eta_k = \beta,
\]
where each consecutive pair differs by a single transposition, $d(\eta_{i-1},\eta_i) = 1$, and the total number of steps is minimal, $k = d(\alpha,\beta)$. 
This metric naturally defines the Cayley graph of $S_n$, whose vertices are all permutations and whose edges connect permutations differing by a single transposition. In this graph, the Cayley distance corresponds to the length of the shortest path between two vertices. Geodesics in the Cayley graph are therefore sequences of permutations connected by transpositions that realize this minimal distance.
Moreover, this notion of geodesic naturally induces a partial order on the set of permutations lying along geodesics between two fixed endpoints. Permutations on different single geodesics paths are, in general, not comparable under this order. However, restricted to a single geodesic, for instance one starting at the identity $e$, an order is defined as 
\begin{equation}
\beta \leq \alpha \quad \text{if and only if} \quad d(e,\beta) \leq d(e,\alpha) \ ,
\end{equation}
that is, $\beta$ lies closer to the identity than $\alpha$, along the same geodesic. 
Under this order, permutations along geodesics between fixed endpoints form a lattice which is a subset of the Cayley graph\footnote{This is not the case for the entire symmetric group; indeed, the entire Cayley graph itself does not form a lattice.}.
A special role is played by the lattice of permutations in $[e,\gamma_n]$, as we will see below.

\paragraph{Dual.}
One can define a natural dual by
\begin{equation}
\label{normal_dual}
\alpha^* = \alpha^{-1} \gamma_n\ .
\end{equation}
While the dual can be defined for any permutation in $S_n$, it is an order-reversing bijection only when considering permutations lying along geodesics between two fixed endpoints. Therefore, if $\beta \leq \alpha$ along the geodesic, then $\beta^* \geq \alpha^*$. When considering permutations in $[e,\gamma_n]$, intuitively, the dual exchanges positions relative to the two endpoints, mapping permutations close to $e$ to permutations close to $\gamma_n$ and vice versa. If $\alpha \in [e,\gamma_n]$, then also $\alpha^* \in [e,\gamma_n]$, and $|\alpha| + |\alpha^*|=n+1$.

\paragraph{Embedding partitions-permutations.}
As said in the main text, every permutation $\alpha \in S_n$ induces a partition $\mu_\alpha$ by grouping together the elements belonging to the same cycle: $\mu_{\alpha} = \{\text{cycles of } \alpha\}$. 
The converse is not true: a partition does not specify a unique permutation, because one must choose an order of the elements within each block to define a cycle. However, when restricting the symmetric group to a geodesic interval from the identity to a maximal cycle, the correspondence becomes bijective. We focus in particular on $[e,\gamma_n]$; the cycles of a permutation can be represented as non-crossing blocks when the elements are arranged on a circle in the natural order, that is, when they lie on the interval $[e,\gamma_n]$.
Conversely, any non-crossing partition can be uniquely realized by a permutation lying on $[e,\gamma_n]$, since the circular order of elements is fixed.
Not only does this establish an isomorphism between the geodesic interval in the symmetric group and the lattice of non-crossing partitions, but it also preserves the partial order\footnote{This is not true if considering the geodesic interval from the identity to a generic maximal cycle: in this case, it is an isomorphism, but not an embedding.}: if one permutation is smaller than another along the geodesic in $[e,\gamma_n]$, the corresponding non-crossing partitions are ordered in the same way according to the refinement order. In this sense, the isomorphism respects the lattice structure. Moreover, it turns out
that the dual permutation, when viewed as a partition, is the Kreweras complement.
Therefore, we have the well known embedding of $NC(n)$ into $S_n$: 
\begin{equation}
(\,[e,\gamma_n], \le_{\mathrm{geod}}\,) \;\cong\; (NC(n), \le_{\mathrm{ref}})\,.
\end{equation}

\begin{widetext}
\section{Global rotational invariance derivations}

\label{app_GRIderivation}
In this Appendix we derive Eq.~\eqref{close_formula_RMT} given in the main. 
The first step consists in inserting Eq.~\eqref{Haar_av_1} into Eq.~\eqref{mat_el_prod}, meaning inserting the averages over the Haar ensemble in the expression for the matrix elements' moments. 
As a first step, we obtain:

\begin{align}
    \overline{A^U_{i_1j_1}\dots A^U_{i_nj_n}} &= 
    \sum_{\alpha\in S_n} \sum_{\beta\in S_n} 
    \text{Wg}_{\alpha,\beta}(D) 
    \sum_{\substack{ \bar{i}_1 \dots \bar{i}_n}}
    A_{\bar{i}_1\bar{i}_{\beta(1)}}\dots A_{\bar{i}_n\bar{i}_{\beta(n)}}\, \delta_{i_{\alpha(1)},j_1}\dots \delta_{i_{\alpha(n)},j_n}\,  \notag \\
    &\equiv \sum_{\alpha\in S_n} \frac{\mathcal{K}_\alpha}{D^{n-\#\alpha}}\, \delta_{i_{\alpha(1)},j_1}\dots \delta_{i_{\alpha(n)},j_n}\, ,
\label{ex_expr}
\end{align}
where we defined, with prefactor $D^{n-\#\alpha}$ for convenience: 
\begin{equation}
    \mathcal{K}_\alpha= D^{n-\#\alpha}\sum_{\beta\in S_n} 
    \text{Wg}_{\alpha,\beta}(D) \sum_{\substack{ \bar{i}_1 \dots \bar{i}_n}}
    A_{\bar{i}_1\bar{i}_{\beta(1)}}\dots A_{\bar{i}_n\bar{i}_{\beta(n)}}= \sum_{\beta\in S_n} \frac{D^{n-\#\alpha}}{D^{-\#\beta}} \,
    \text{Wg}_{\alpha,\beta}(D)  \prod_{m=1}^{\# \beta}  \left \langle A^{l_m}\right \rangle \ .
\label{Kappa_ex}
\end{equation}
\end{widetext}
The same expression can also be understood as the action of the corresponding Haar quantum channel on \(A^{\otimes n}\) in the energy basis, following the formulation of Ref.~\cite{kaneko2020characterizing}.

We now explain terms appearing in this expression:
\begin{itemize}[leftmargin=1.5em, itemsep=1em, topsep=1em]
    \item $\text{Wg}_{\alpha,\beta}(D)$ is the Weingarten function, a rational function of the Hilbert space dimension $D$. Viewed as a matrix, it is defined as the pseudo-inverse of the Gram matrix with entries $G_{\alpha,\beta} = D^{\#(\alpha^{-1}\beta)}$. The Weingarten function depends only on the cycle structure of $\alpha^{-1}\beta$. A well known result is the first order contribution of the Weingarten function when expanded for large $D$ :
\begin{align}
        \text{Wg}_{\alpha,\beta} (D) = \frac{1}{D^{2n-\#(\alpha^{-1}\beta)} }\left[\phi(\alpha,\beta) +\mathcal{O}
    \left(\frac{1}{D^2}\right)\right] \ .
    \label{Weing_approx}
\end{align}
 $\phi(\alpha,\beta)$ is a known multiplicative function, it factorizes according to the cycles of $ \alpha^{-1}\beta$ as $\displaystyle \prod^{\#(\alpha^{-1}\beta)}_{m=1} (-1)^{l_m-1} C_{l_m-1}$, with $C_n$ Catalan number;

\item in the last equality we defined $\langle\bullet\rangle=\text{Tr}(\bullet)/D$ as the normalized trace, and we used the equality 
\begin{equation}   
\sum_{\substack{ \bar{i}_1 \dots \bar{i}_n}}
    A_{\bar{i}_{\beta(1)}\bar{i}_1}\dots A_{\bar{i}_{\beta(n)}\bar{i}_n} = \prod_{m=1}^{\# \beta} \text{Tr}(A^{l_m})\,.
\end{equation}
\end{itemize}

Similarly to the calculation in Refs.~\cite{nica2006lectures, fava2025designs}, done at the level of the Haar quantum channel, we perform a systematic expansion in powers of $1/D$ of Eq.~\eqref{Kappa_ex}, for a fixed $\alpha$: inserting the expansion of the Weingarten function (Eq.~\eqref{Weing_approx}) we get
\begin{align}
    \mathcal{K}_\alpha  = \sum_{\beta \in S_n} \frac{D^{n-\#\alpha}}{D^{\Delta(\alpha,\beta)} }\left[\phi(\alpha,\beta) \prod_{m=1}^{\# \beta}  \left \langle A^{l_m}\right \rangle +\mathcal{O} \left(\frac{1}{D^2}\right) \right]\ ,
\label{kappa_beta}
\end{align}
where we substituted $\Delta(\alpha,\beta) = 2n-\#(\alpha^{-1}\beta)-\#\beta = d(e,\beta) + d(\beta,\alpha)$ with $d(,)$ Cayley distance, defined in Eq.~\eqref{Cayley_dist}. 

The leading order is given only by the $\beta$s that minimize $\Delta(\alpha,\beta)$. This happens when $\beta\in[e,\alpha]$, so that $\Delta(\alpha,\beta) = n-\#\alpha$, its minimum value. Moreover, the condition $\beta\in[e,\alpha]$ is equivalent to having $\beta\le\alpha$. 

Permutations that do not belong to the geodesic interval give subleading contributions of order $\mathcal{O}(1/D^{n-\#\alpha+2})$, and so must be taken into account along the Weingarten approximation. 
Indeed, one can show\footnote{This can be proven identifying $S_n$ with an orientable surface, and applying Euler’s formula: the only possible solutions to the triangle inequality are $d(e,\beta)+d(\beta,\alpha)=d(e,\alpha)+2g$,
for integer $g\ge0$ \cite{nica2006lectures}.} that $\cancel{\exists}\alpha, \beta$ so that $\Delta(\alpha,\beta)= n-\#\alpha+1$.

Therefore, the leading order is given by
\begin{equation}
\mathcal{K}_\alpha = \sum_{\beta \leq \alpha} \phi(\alpha, \beta) \prod_{m=1}^{\# \beta}  \left \langle A^{l_m}\right \rangle+\mathcal{O} \left(\frac{1}{D^{2}}\right) \, .
    \label{kappa_exact}
\end{equation}
We can interpret the result obtained in terms of partitions instead of permutations. Specifically, the leading order of $\mathcal{K}_\alpha$ is the free cumulant $\kappa_\alpha$ associated to the non-crossing partition defined by $\alpha$. In order to see the equivalence, we write explicitly the inverse relation of Eq.~\eqref{free_cum_NC} which expresses a generic cumulant $\kappa_\mu$, for $\mu\in NC(n)$ in terms of the moments, via the M\"obius function $\psi$:
\begin{equation}
\kappa_\mu = \sum_{\substack{\nu \leq \mu \\ \nu \in NC(n)}} \psi(\nu,\mu) 
\prod_{m=1}^{\# \nu} \langle A^{l_m} \rangle\ .
\label{free_cum}
\end{equation}

Considering $\alpha$ and $\beta$ as non-crossing partitions\footnote{The subtle point is that in general $\alpha$ is not forced to be in the interval $[e,\gamma_n]$, therefore one cannot identify it with a non-crossing partition; however, it is possible to construct another permutation $\alpha'=\lambda^{-1}\alpha\lambda$, with $\lambda\in S_n$, so that $\#\alpha=\#\alpha'$ and $\alpha' \in [e,\gamma_n]$. In this case $\mathcal{K}_\alpha = \mathcal{K}_{\alpha'}$: the term $\mathcal{K}$ is the same when taking $\alpha$ or $\alpha'$ and therefore can still be identified with the free cumulant associated to $\alpha'$; we will keep the notation $\alpha$ in the text, intending that when $\alpha$ is not in the interval $[e,\gamma_n]$, it has to be thought as $\alpha'$.\label{foot_rotated}}, the sum in Eq.~\eqref{kappa_exact} runs over $\beta \leq \alpha, \in NC(n)$, as in Eq.~\eqref{free_cum}. Moreover, one can identify $\phi$ with the M\"obius function. It is now straightforward to see that $\mathcal{K}_\alpha=\kappa_{\alpha}+\mathcal{O}(1/D^2)$.

For the matrix elements' moments \eqref{ex_expr} it follows 
\begin{align}
\label{appr_expr}
     &\overline{A^U_{i_1j_1}\dots A^U_{i_nj_n}}=\sum_{\alpha \in S_n} \frac{1}{D^{n-\#\alpha}}\,\ \cdot \\  & \qquad\quad  \cdot\,\ \left[ \kappa_{\alpha} \,\delta_{i_{\alpha(1)},j_1}\dots \delta_{i_{\alpha(n)},j_n} + \mathcal{O}\left( \frac{1}{D^{2}}\right) \right] \ .
\notag
\end{align}

This expression tells us that the matrix elements' moments are given as a sum over permutations, each coming with multiple $\delta$s. In other words, a product of matrix elements, averaged on the ensemble, is non zero only when the second set of indices $\{j_1\dots j_n\}$ is a permutation of the firsts $\{i_1\dots i_n\}$, or also, each index is repeated at least once in the product. This reflects the fact that the products in \eqref{Haar_av_1} are non zero only when a $U$ couples with a $U^*$. This is typical of the unitary group. 
As a consequence, the non-zero products of matrix elements can be written as
\begin{align}
\label{appr_expr_fin_perm}
&\overline{A^U_{i_1i_{\sigma(1)}}\dots A^U_{i_ni_{\sigma(n)}}}  =\\ &\qquad =\sum_{\alpha \in S_n} \frac{1}{D^{n-\#\alpha}}\left[ \kappa_{\alpha} \,\delta_{\alpha\sigma}(\textbf{i})  + \mathcal{O}\left( \frac{1}{D^{2}}\right) \right] \ ,
\notag
\end{align}
where we defined $\delta_{\alpha\sigma}(\textbf{i})=\delta_{i_{\alpha(1)},i_{\sigma(1)}}\dots \delta_{i_{\alpha(n)},i_{\sigma(n)}}$ to keep the notation shorter. We will also use the notation $\textbf{i}:=\{i_1,\dots,i_n\}$ and $\sigma(\textbf{i}):=\{i_{\sigma(1)},\dots,i_{\sigma(n)}\}$.
Differently from Eq.~\eqref{close_formula_RMT} and the surrounding discussion, here $\sigma$ is a generic permutation, since we are still considering all the possible product of matrix elements; however, it is non trivial to define the permutation $\sigma$, for a given product of matrix elements, since in $\textbf{i}$ there could be repeated indices.  
\begin{itemize}[leftmargin=*, itemsep=1em, topsep=1em]
\item When there are no repetitions, then $\sigma$ is uniquely defined: there is a bijective correspondence between a product of matrix elements with different indices and a permutation. For instance, $\sigma=(12)(34)$ defines uniquely $A_{ij}A_{ji}A_{kl}A_{lk}$, for $i\ne j\ne k\ne l$. 
\item When there are repetitions, instead, there are multiple permutations giving the same product of matrix elements; we define $\sigma$ as the finest, that is, the one with the biggest number of cycles.
For instance, one can use both $(12)(34)$ or $(1234)$ for defining $A_{ij}A_{ji}A_{il}A_{li}$, but we chose $\sigma$ as the first. We notice that, in this way, each cycle of $\sigma$ involves only different indices in \textbf{i}, meaning that $i_{\mathcal{C}_m(1)}\ne \dots \ne i_{\mathcal{C}_m(l_m)}$, for all the cycles $\mathcal{C}_m$ of $\sigma$ of length $l_m$.
However, the finest permutation might not be unique: there could be different $\sigma$s with the same number of cycles. This is the case for $A_{ij}A_{ji}A_{ij}A_{ji}$ which can be defined by two different finest permutations $(12)(34)$ and $(14)(23)$, both with two cycles. This happens only for the product of matrix elements that were called non-cacti.
\end{itemize}

It is clear that a permutation $\sigma$ will not uniquely define a product of matrix elements, since the same permutation could be describing a product with all different indices or with repeated indices, being the finest.  
This is the case, for instance, for $A_{ij}A_{ji}A_{ii}A_{ii}$ and $A_{ij}A_{ji}A_{ii}A_{jj}$, where $\sigma=(12)(3)(4)$. 
Therefore, together with $\sigma$, one needs to give the additional ordered set of indices $\textbf{i}$ on which the permutation acts, with the constraint that $\sigma$ is the finest, if there are repetitions. In the example, $\textbf{i}=\{ijii\}$ for the first, and $\textbf{i}=\{ijij\}$ for the second.
 
We highlight a special case that happens when $i_k =i_{\sigma(k-1)}$ mod $n$.
The condition depends both on $\sigma$ and $\textbf{i}$; of course is satisfied when $\sigma=\gamma_n$ and there are no repeated indices, but also, for instance, for  
$\sigma=(12)(3)(4)$ on \textbf{i}=$\{ijii\}$.
More specifically, this condition defines what we called ETH-cycles, for which the products of matrix elements are of the form $\overline{A^U_{i_1i_2} A^U_{i_2 i_3}\dots A^U_{i_ni_1}}$, where indices can be different or repeated. Excluding the non-cacti, as explained in the main, these products of matrix elements can be described uniquely by a non-crossing partition\footnote{The fact that a product of matrix elements can be described uniquely by a partition happens only in the case of ETH-cycles, which partitions the set \textbf{i} in block of equal indices, since one needs the indices to be on a circle for being in blocks of partitions.},
whose Kreweras complement, seen as a permutation, is exactly $\sigma$; moreover, $\sigma \in [e,\gamma_n]$ in those cases.

We will now focus the discussion on ETH-cycles, excluding the non-cacti, that is simple loops and cacti. For those, we can derive the closed formula Eq.~\eqref{close_formula_RMT}, while the destiny of the others will be discussed separately afterwards.

\subsection{Closed Formula for simple loops and cacti}
From expression \eqref{appr_expr_fin_perm}, we can deduce both the leading and the first subleading contributions.

The $\delta_{\alpha\sigma}({\textbf{i})}$ appearing in the equation is selecting only some $\alpha$s in the sum over the symmetric group, which are the ones that act as $\sigma$ on \textbf{i}. 
One could initially assume that the condition is satisfied only when $\alpha = \sigma$; this is indeed correct if $\sigma$ acts on a set of distinct indices, corresponding to the case of simple loops.
For cacti, instead, $\alpha$s are all the permutations giving the same result as $\sigma$, when applied to the set with repeated indices. As said above, with repeated indices $\sigma$ is not well defined, but we chose it to be the finest. Permutations $\alpha$s are all the others which could also define the same product of matrix elements, but have more cycles. 

\subsubsection{Leading Order}

Between all the $\alpha$s satisfying the $\delta_{\alpha\sigma}({\textbf{i})}$, the leading order is given by the one minimizing $n-\#\alpha$, the exponent of $1/D$. This happens for the permutation with the biggest number of cycles i.e. the finest which is $\alpha=\sigma$, for how we defined it. 
Therefore, 
\begin{align}
    \overline{A^U_{i_1i_{\sigma(1)}}\dots A^U_{i_ni_{\sigma(n)}}}=\frac{\kappa_{\sigma}}{D^{n-\#\sigma}} +\mathcal{O} \left(\frac{1}{D^{n-\#\sigma+1}}\right)\, ,
\label{leading_tot}
\end{align}
with 
\begin{align}
    \notag
   \mathcal{O} \left(\frac{1}{D^{n-\#\sigma+1}}\right) = &\sum_{\alpha \ne \sigma \in S_n} \frac{1}{D^{n-\#\alpha}}\left[ \kappa_{\alpha} \,\delta_{\alpha\sigma}(\textbf{i})  + \mathcal{O}\left( \frac{1}{D^{2}}\right) \right] \\ &+\mathcal{O}\left(\frac{1}{D^{n-\#\sigma+2}}\right) \, .
    \label{leading_corrections}
\end{align}

For simple loops and cacti we identify $\sigma$ with $\pi$, and we obtain the expression in the main Eq.~\eqref{close_formula_RMT}. In particular, for simple loops, $\pi=\gamma_n$; as a consequence, they are the smallest product of matrix elements, having the scaling $1/D^{n-1}$. 

\subsubsection{First Subleading Correction}
We want to read the first subleading term from \eqref{leading_corrections}. We immediately see that this is given by all the $\alpha$s satisfying the $\delta_{\alpha\sigma}({\textbf{i})}$ for which $\#\alpha=\#\sigma-1$, meaning the ones with one cycle less than $\sigma$. For convenience, we call them $\rho$s. Therefore, from \eqref{leading_corrections}:

\begin{align}
\notag
   \mathcal{O} \left(\frac{1}{D^{n-\#\sigma+1}}\right) =
   \sum_{\substack{\rho \in S_n \text{with}\\ \#\rho=\#\sigma-1 }}\frac{\kappa_{\rho}}{D^{n-\#\rho}}\delta_{\rho\sigma}({\textbf{i})} \\ +\mathcal{O} \left(\frac{1}{D^{n-\#\sigma+2}}\right)\, .
\label{subleading_generic}
\end{align}
We emphasize that $\rho$s are not forced to be in the geodesic interval $[e,\gamma_n]$, even if $\sigma$ is.
Before providing a geometric interpretation of the $\rho$s that appear in the expansion, we remark that only the first subleading correction admits a compact and explicit expression. Higher-order contributions require taking into account not only permutations $\alpha$ with progressively fewer cycles, but also subleading terms in $\mathcal{K}_\alpha$ arising from the higher-order structure of the Weingarten function and from permutations that do not belong to the geodesic, as indicated in Eq.~\eqref{kappa_beta}. Accordingly, contributions of order $\mathcal{O}(1/D^{n - \#\sigma + 2})$ lie beyond the analysis of this paper\footnote{Actually, the only reason why we can study the first order contribution is because of the power $1/D^2$ in Eq.~\eqref{kappa_exact}.}.

Let's then focus on the $\rho$s.
Firstly, we already stressed that for simple loops there are no other permutations acting as $\sigma$, apart from $\sigma$ itself, since in \textbf{i} there are no repetitions. In this case, the first order correction is zero. Since the leading order has scaling $1/D^{n-1}$, the error is of order $\mathcal{O}(1/D^{n+1})$.

When indices are repeated, we want to show that the condition given by $\delta_{\rho\sigma}(\textbf{i})$ restricts the sum for $\rho \in S_n$, with the additional condition $\#\rho=\#\sigma-1$, only over $\rho\in[e,\gamma_n]$.

We introduce the following two notations:
\begin{itemize}[leftmargin=1.5em, itemsep=1em, topsep=1em]
\item $\lambda:=\sigma\rho^{-1}$. The permutation $\lambda$ permits to obtain $\sigma$ when composed with $\rho$. The $\delta_{\rho\sigma}(\textbf{i})$ implies $i_k = i_{\lambda(k)}\,\, \forall k $, that is $\lambda$ acting as the identity on \textbf{i};
\item $ \hat \alpha := \gamma_n \alpha^{-1}$ for a generic permutation $\alpha\in S_n$, which is a sort of anti-dual, with the respect to Eq.~\eqref{normal_dual}. Indeed, it is still an order-reversing bijection, however it inverts the usual direction, bringing a dual permutation to the original one. For instance for permutations on the geodesic interval $[e,\gamma_n]$, one can see $\alpha$ as the Kreweras complement of $\hat \alpha$. As a consequence, $\hat \sigma$ is basically the non-crossing partition characterizing the repeated indices in \textbf{i}. Therefore, we will say that elements in the same cycle of $\hat  \sigma$ are associated to equal indices in \textbf{i}, meaning that $i_{\mathcal{\hat C}_m(1)}= \dots = i_{\mathcal{\hat C}_m(l_m)}$, for all the cycles $\mathcal{\hat C}_m$ of $\hat \sigma$ of length $l_m$.
\end{itemize}

The statement we want to prove is the following : under the assumptions (1) $\sigma$ satisfies the condition of being cyclic on the set \textbf{i}, $i_{k} = i_{\sigma(k-1)}$ mod $n$, and (2) $\#\rho=\#\sigma-1$, then 
\begin{equation}
\{\rho \in S_n \, /\, i_{\sigma(k)} = i_{\rho(k)}  \ \ \forall k \} \ \Longleftrightarrow \ \rho \in [e,\gamma_n] \ ,
\label{math_proof}
\end{equation}

that is, all and only the $\rho$s satisfying $\delta_{\rho\sigma}(\textbf{i)}$ are the ones on the geodesic interval $[e,\gamma_n]$.

We start by showing the left to right implication; as a first step, we show that if $\rho$ satisfies $\delta_{\rho\sigma}(\textbf{i})$ and $\#\rho=\#\sigma-1$, then $d(\rho,\sigma)=1$, that is 
$$
\#(\rho^{-1}\sigma) \equiv \#(\sigma\rho^{-1}) \equiv \#\lambda \,\,= \,\,n-1 \ .
$$
The reader should notice that the condition (2) is equivalent to 
$d(e,\rho)=d(e,\sigma)+1$, but does not imply directly that $\rho>\sigma$ which means $\sigma\in [e,\rho]$. This happens when the additional condition $d(\rho,\sigma)=1$ is satisfied. 
We want to show that $\lambda$ is a transposition; the permutation 
$\lambda$ needs to satisfy two conditions: it must unify two cycles of $\sigma$ which are distinct, so that $\rho$ has one cycle less than $\sigma$, and it must act as the identity on the set of indices \textbf{i}. This means that there should be a cycle of $\lambda$, let's say $C$, with elements belonging to two different cycles of $\sigma$ (not more than two), and so that $i_{C(1)}=\dots = i_{C(l)}$. Since in a single cycle of $\sigma$ there cannot be elements associated to the same index, by construction of $\sigma$, the cycle $C$ of $\lambda$ cannot contain more than two elements, that is $l=2$. We will label these elements as $x$ and $y$. Moreover, there can be only one cycle containing more than one element, all the other cycles of $\lambda$ need to have unit length. Let us admit this is not the case: let us imagine we have more than one cycle having more than one element. This could happen when more than two cycles of $\sigma$ are involved and so unified, which goes against $\#\rho=\#\sigma-1$, or when $\sigma$ is one of the permutations defining a non-cacti. Indeed, these are the only cases in which different cycles could be associated to more than one equal index, meaning that in two different cycles of $\sigma$, saying $C_1$ and $C_2$, more than one couple $(m,n)$ exists, so that $i_{C_1(m)}=i_{C_2(n)}$. These cases are excluded from this derivation, since we consider only simple loops and cacti. 
Therefore, $\lambda$ is a transposition, and $\rho>\sigma$ with one cycle less. \\
As a successive step, one can directly show that $\sigma\rho^{-1}=\hat{\sigma}^{-1} \hat \rho$, which implies $d(\hat \rho,\hat \sigma)=1$, for what we just showed. If two permutations are at distance one, they must have one cycle of difference, that is
\begin{equation}
\label{implication}
d(\hat \rho,\hat \sigma)=1  \ \ \Rightarrow \ \ \ 
 \#\hat \rho - \#\hat \sigma = \pm1 \ :
\end{equation}
$\hat \rho$ has one more cycle than $\hat \sigma$ or one less. \\
We also notice that, as a consequence of their definitions, $\hat{\sigma}^{-1} \hat \rho = \lambda$, which means that the same transposition $\lambda$ of above, composed with $\hat \sigma$ gives $\hat \rho$. Therefore, $\lambda$ is or dividing a cycle of $\hat \sigma$ in two cycles of $\hat \rho$ (+ in Eq.~\eqref{implication}), or vice versa unifying two cycles of $\hat \sigma$ in one of $\hat \rho$ (- in Eq.~\eqref{implication}). To decide the right direction, we have to look at the elements $x,y\in C$ and understand if they are in a same cycle of $\hat \sigma$ and different of $\hat \rho$, or vice versa; for construction, $i_x=i_y$, that is, $x$ and $y$ are constrained to be in the same cycle of $\hat \sigma$. Therefore, $\lambda$ is dividing the cycle of $\hat \sigma$ containing both $x$ and $y$ in two different cycles of $\hat \rho$, one with $x$ and another with $y$. As a consequence, $\#\hat \rho = \#\hat \sigma+1$ and $\hat \rho <\hat \sigma$. \\
Since $\sigma \in [e,\gamma_n]$, then also $\hat \sigma \in [e,\gamma_n]$, and since $\hat \rho <\hat \sigma$, the same is true for $\rho$. We notice that $\hat \rho<\hat \sigma$ is a stronger condition then $\rho>\sigma$, because it implies that $\hat \rho\in [e,\gamma_n] \Rightarrow \rho\in [e,\gamma_n]$. 
Indeed, there are $\rho$s bigger than $\sigma$ of one cycle, which do not satisfy the delta function, which are indeed all the ones not belonging to $[e,\gamma_n]$.\\

Let us now show the right to left implication in Eq.~\eqref{math_proof}; when $\rho>\sigma, \in[e,\gamma_n]$ with one cycle less, then  $\hat \rho<\hat \sigma, \in[e,\gamma_n]$ with one cycle more.  This means that each cycle of $\hat \rho$ is included in a cycle of $\hat \sigma$. Specifically, considering a generic cycle $\mathcal{\hat C}$ of $\hat \sigma$ of length $l$, it can be written as the disjoint union of all the cycles of $\hat \rho$ contained in $\mathcal{\hat C}$, that we indicate as $\mathcal{\hat R}_g$, for $g=1\dots r$, admitting they are $r$. 
Since it is true that $i_{\mathcal{\hat C}(1)}= \dots = i_{\mathcal{\hat C}(l)}$, it is also true that $i_{\mathcal{\hat R}_1(1)}= \dots = i_{\mathcal{\hat R}_1(l_1)}= i_{\mathcal{\hat R}_2(1)} \dots =i_{\mathcal{\hat R}_2(l_2)} = \dots = i_{\mathcal{\hat R}_g(1)}= \dots = i_{\mathcal{\hat R}_g(l_g)}$. Not only, this would be true even if instead of $\hat \sigma$ and $\hat \rho$ we consider their inverses, since the inverse operation is not mixing cycles, but just changing the order of the elements in the cycles.
This is valid for each cycle of $\hat \sigma$, therefore it is clear that $\forall k$:
\begin{align}
\notag
i_{\hat \rho(k)} = &i_{\hat \sigma(k)} \  \Leftrightarrow \  i_{\hat \rho^{-1}(k)} = i_{\hat \sigma^{-1}(k)} \ \Leftrightarrow   \ i_{ \hat \sigma^{-1}\hat \rho(k)} = i_{k} \\ &\Leftrightarrow   \ i_{ \sigma \rho^{-1}(k)} = i_{k} \ \Leftrightarrow   \ i_{ \sigma(k)} = i_{\rho (k)} \ , 
\end{align}
same condition imposed from the delta function $\delta_{\rho\sigma}(\textbf{i})$, as we wanted to prove.\\
 
This completes the proof.\\

An important remark is that the condition of $\alpha\in [e,\gamma_n]$ would not be necessary and sufficient for satisfying $\delta_{\alpha\sigma}(\textbf{i})$ when the assumption (2) is made for higher order corrections, that is, looking for a generic order $M$ for which $\# \alpha =\# \sigma-M$. This would still imply $d(\alpha,\sigma)=d(\hat \alpha,\hat \sigma)=M$, but it would not be sufficient for saying that $\#\hat \rho - \#\hat\sigma$ is equal to $\pm M$; the implication \eqref{implication} is true only for $M=1$.
Therefore, in all the other cases, taking only $\alpha\in[e,\gamma_n]$ would be too restrictive: there would be permutations $\alpha$s satisfying $\delta_{\alpha\sigma}(\textbf{i})$, but not on the geodesic interval. Since we already noticed that the higher corrections would be much more difficult to interpret, we do not go further in this direction.

\paragraph*{Example.} We consider useful to follow the derivation with a specific example. We choose \textbf{i}$=\{ijii\}$ and $\sigma=(12)(3)(4)$, that is the ETH-cycle $\overline{A^U_{ij}A^U_{ji}A^U_{ii}A^U_{ii}}$. We construct the permutations $\lambda$s so that they unify one single cycle of $\sigma$, with the constraint that elements in the same cycle are associated to equal indices.  
There are only three choices and they are all constrained to be transpositions;
with the respective $\rho = \lambda\sigma$, they are 
$\lambda_1 = (13)(2)(4) \rightarrow \rho_1=(123)(4)$, $\lambda_2 = (14)(2)(3) \rightarrow \rho_2=(124)(3)$, and $\lambda_3 = (34)(1)(2) \rightarrow \rho_3=(12)(34)$.
Clearly, for all of them $d(\rho,\sigma)=1$. However, we notice that there were other permutations with a single less cycle, but not at distance one, for instance $(142)(3)$. 

Calculating the anti-dual permutation,
$\hat \sigma = (134)(2)$ and $\hat \rho_1 = (14)(2)(3),\hat \rho_2 = (34)(1)(2),\hat \rho_3 = (13)(2)(4)$. 
We notice that $d(\hat \rho,\hat \sigma)=1$ in all the cases, and moreover $\#\hat \rho = \# \hat \sigma+1$ which is implied by the fact that the elements exchanged by the transpositions $\lambda$ are always belonging to the same cycle of $\hat \sigma$. Moreover, the relation $\hat \rho = \hat \sigma \lambda$ is valid. The $\rho$s satisfying the delta function, as reported in the main in Eq.~\eqref{ex_ijii}, are all and only the ones in the geodesic interval $[e,\gamma_4]$. 

Let us discuss quickly which permutations would be involved for $M=2$; in this case the permutations satisfying the delta function are $\alpha_1 = (1234)$ and $\alpha_2 = (2134) \cancel{\in}[e,\gamma_4]$. 
It is still true that $d(\alpha,\sigma)=2$ (even if it is evident that not all the permutations at distance 2 satisfy the delta function), however, in the dual the permutations are not forced to have two cycles more than $\hat \sigma$, for instance $\hat \alpha_2 = (132)(4)$ has the same number of cycles of $\hat \sigma$.

Therefore we proved that Eq.~\eqref{subleading_generic} is equivalent to: 
\begin{align}
\label{subleading_tot}
   O \left(\frac{1}{D^{n-\#\sigma+1}}\right)=\sum_{\substack{\rho\in [e,\gamma_n]\,  \text{with}\\ \rho > \sigma, \  \#\rho= \#\sigma-1}} +\mathcal{O} \left(\frac{1}{D^{n-\#\sigma+2}}\right) \ .
\end{align}
We remind the reader of the assumption that $\sigma$ is defining only ETH-cycles, excluding the non-cacti. 
In this case we can finally translate the formula to the language of partitions. Summing over $\rho\in [e,\gamma_n] >\sigma $ at distance 1 is equivalent to summing over the non-crossing partitions associated, at distance 1; in other words, over $\tau\in NC(n)>\pi $ at distance 1.
Therefore, Eqs.~\eqref{leading_tot} Eq.~\eqref{subleading_tot} give Eq.~\eqref{close_formula_RMT} in the main. Some examples were presented in Eqs.~\eqref{ex_ijkl}--\eqref{ex_ijil}.

\subsection{Destiny of non ETH-cycles and non-cacti}
Eq.~\eqref{appr_expr_fin_perm} is valid for all the products of matrix elements, and $\forall\sigma\in S_n$, however, Eq.~\eqref{leading_tot} and Eq.~\eqref{subleading_tot} were derived only for simple loops and cacti, that is, when $\sigma \in [e,\gamma_n]$, and it is cyclic on the set \textbf{i}. 
Let us examine where the derivation fails for the other cases of non-ETH-cycles and non-cacti.

\subsubsection{Non-cacti}
We said above that non-cacti cannot be described by a unique permutation $\sigma$. 
Indeed, the finest permutation is not unique: different $\sigma$s with the same number of cycles give the matrix product of non-cacti. As a consequence, the leading order will not be given by a single permutation, but by all the finest $\sigma$s, summed together. 
This reflects the non-existence of the Kreweras complement for the crossing partition and the fact that a crossing diagram cannot be factorized uniquely, as instead happens for all the other ETH-cycles. For instance, the unique non-cacti of $n=4$ is given by: 
\begin{equation}
\overline{A^U_{ij}A^U_{ji}A^U_{ij}A^U_{ji}} = \frac{2\kappa^2_2}{D^2} + \mathcal{O}
\left(\frac{1}{D^3}\right) \ .
\end{equation}
Since our interest regards simple loops and cacti, we don't discuss further the subleading of the non-cacti.

\subsubsection{Non-ETH-cycles}
Now we discuss the case of products of matrix elements which are not of the form $\overline{A^U_{i_1i_2} A^U_{i_2 i_3}\dots A^U_{i_ni_1}}$, that is, they do not satisfy $i_k =i_{\sigma(k-1)}$ mod $n$, and for which $\sigma$ is uniquely defined (we exclude all the cases similar to the cacti). The condition depends both on $\sigma$ and $\textbf{i}$.

When there are no repeated indices in \textbf{i}, whether $\sigma$ is cyclic or not, the leading order is given by Eq.~\eqref{leading_tot}, and the subleading is zero. This was the case for simple loops, and holds in general for all the permutations defined on \textbf{i} without repetitions, since no other permutations are satisfying $\delta_{\alpha\sigma}(\textbf{i})$. 

When there are repeated indices in \textbf{i}, the leading order is still given by Eq.~\eqref{leading_tot}, because one can still define a unique finest permutation; this contribution will be the same as the one of the ETH-cycles with the same $\sigma$, but different \textbf{i}.
The subleading, instead, is still given by Eq.~\eqref{subleading_generic}, but not by Eq.~\eqref{subleading_tot}. Indeed, its derivation relies on the assumption of $\sigma$ being cyclic on \textbf{i} and breaks down when this is not true. In particular what is important in the proof is that, when $\sigma$ has this property, $\hat \sigma$ is basically the non-crossing partition characterizing the repeated indices in \textbf{i}, so that, $i_{\mathcal{\hat C}_m(1)}= \dots = i_{\mathcal{\hat C}_m(l_m)}$, for all the cycles $\mathcal{\hat C}_m$ of $\hat \sigma$ of length $l_m$. This was essential for proving that Eq.~\eqref{implication} is satisfied with the sign +; for non-ETH-cycles, $x,y$, elements of the transposition $\lambda$, are not forced to be in the same cycle of $\hat \sigma$, therefore the implication can be valid both for $\pm$. As a consequence, the condition $\hat \rho < \hat \sigma$ is not implied and can happen that $\rho \cancel{\in}[e,\gamma_n]$ still satisfies the $\delta_{\rho\sigma}(\textbf{i})$. 

As an example, let us consider $\overline{A^U_{ij}A^U_{ji}A^U_{ii}A^U_{jj}}$, in order to compare it with $\overline{A^U_{ij}A^U_{ji}A^U_{ii}A^U_{ii}}$ that we discussed above. First of all, $\sigma=(12)(3)(4)$ in both the cases and the leading order is exactly the same for both the two averages.
Focusing on the non-ETH cycle, we  construct the permutations $\lambda$s so that they unify one single cycle of $\sigma$, with the constraint that elements in the same cycle are associated to equal indices.  
There are only two choices and they are all constrained to be transpositions;
with the respective $\rho = \lambda\sigma$, they are 
$\lambda_1 = (13)(2)(4) \rightarrow \rho_1=(123)(4)$, and $\lambda_2 = (24)(1)(3) \rightarrow \rho_2=(142)(3)$.
Clearly, for all of them $d(\rho,\sigma)=1$. Indeed, this part of the proof relied only on the definition of $\sigma$ being the finest. Problems come when calculating the anti-dual permutations, $\hat \rho_1 = (14)(2)(3), \hat \rho_2 = (1342)$. 
We still have $d(\hat \rho,\hat \sigma)=1$ in both cases, but $\#\hat \rho_2 = \# \hat \sigma-1$. Indeed the elements exchanged by the transposition $\lambda_2$ are not belonging to the same cycle of $\hat \sigma$, since $\hat \sigma =(123)(4)$ is mixing the positions of equal indices. 

Again, since our interest regards the simple loops and cacti, we don't discuss further the subleading corrections of non-ETH-cycles.

\bigskip
 
To summarize, the leading order given as Eq.~\eqref{leading_tot} is valid for all the matrix elements' moments apart from when $\sigma$ cannot be uniquely defined. 
Eq.~\eqref{subleading_tot}, instead, is not valid for non-ETH-cycles with repeated indices, and, again, for the non-cacti.

\section{Local rotational invariance derivations}
\label{app_LRIderivation}
In this Appendix we evaluate Eq.~\eqref{mat_el_prod_local}. In contrast to the globally invariant case, the matrices here behave as random unitaries only within their respective index subsets. Consequently, Eq.~\eqref{Haar_av_1} and Weingarten calculus cannot be applied straightforwardly. We explicitly evaluate the first few $n$ orders, while showing that the Weingarten formalism can still be employed by properly coupling matrices acting on different intervals. Using the same approximations as in the global case will lead to similar results, now encoding the energy dependencies. For the computation, we employ the minimal block model in Fig.~\ref{fig_unitary}(b), and we show afterwards how to translate the expressions to the more refined model in Fig.~\ref{fig_unitary}(c). We will consider, since the beginning, ETH-cycles only, starting from simple loops in order to derive Eqs.~\eqref{eq_result_LRI},\eqref{CDEF} given in the main, and then evaluating explicitly some examples of cacti in order to show how the factorization and the first subleading correction appear in the case of rotational invariance.

\subsubsection{$n=1$}
Since $U^{(I)}$ and $U^{(J)}$ have two independent unitaries, respectively only to the set of indices $I$ and $J$, the average in Eq.~\eqref{loc_1pt} is non zero only when $I=J$:
\begin{align}
    \overline{A^U_{ij}} = \sum_{\substack{\bar i\in I \\ \bar j \in J}} \overline{U^{(I)\,*}_{\bar{i}i}U^{(J)}_{\bar{j}j}}A_{\bar{i}\,\bar{j}} = \delta_{I,J} \sum_{\substack{\bar i, \bar j \in I}} \overline{U^{(I)\,*}_{\bar{i}i}U^{(I)}_{\bar{j}j}}A_{\bar{i}\,\bar{j}} \ .
\label{loc_1pt}
\end{align}
After fixing this condition, the product is a rotation that involves only the submatrix $A_{\bar i\, \bar j}$ for $\bar i,\bar j\in I$; then, the average can be calculated through the Weingarten calculus for the dimension of the submatrix, that is $D_i$. From Eq.~\eqref{Haar_av_1}: 
$$\overline{U^{(I)\,*}_{\bar{i}i}U^{(I)}_{\bar{j}j}} = \text{Wg}_{e,e}(D_i) \delta_{ij}\delta_{\bar i\bar j}\ .$$ Therefore, since $\delta_{ij}$ directly implies $\delta_{I,J}$,
\begin{align}
    \overline{A^U_{ij}} = \delta_{ij} \frac{1}{D_i} \sum_{\bar i \in I} A_{\bar i\,\bar i } \equiv \delta_{ij} \, K_1(E_i)\ ,
    \notag
\end{align}
where $c_1(E_i)$ depends on the mesoscopic energy $E_i$ of the interval $I$.

It is straightforward to translate this result to the smoothed version, just explicitly writing 
\begin{equation}
    K_1(E_i)= \frac{1}{D_i}\sum_{\bar i \in I} A_{\bar i\,\bar i } =\sum_{\alpha}  \frac{\delta_{\Delta}(E_{i}-E_{\alpha})}{D_i} A_{\alpha\alpha }\ . 
\end{equation}
When $\delta_\Delta()$ is peaked around $E_i$, one obtains the first line of Eq.~\eqref{CDEF}. 
This expression reproduces the result of global rotational invariance, but taking into account the energy dependence of the matrix elements; for a flat $\delta_{\Delta}()$ one recovers, indeed, the normalized trace of the entire matrix.

\subsubsection{$n=2$}

We can distinguish two cases in the expression
\begin{align}
    \overline{A^U_{ij}A^U_{ji}} = \sum_{\substack{\bar i,\bar l \in I \\ \bar j,\bar k \in J} } \overline{U^{(I)\,*}_{\bar{i}i} U^{(J)\,*}_{\bar{k}j}U^{(J)}_{\bar{j}j} U^{(I)}_{\bar{l}i}}
    A_{\bar{i}\,\bar{j}}\,A_{\bar{k}\,\bar{l}} \, .
\label{loc_2pt}
\end{align}

When $I=J$, we can use Weingarten calculus for dimension $D_i$. 
Expanding $\text{Wg}_{\alpha,\beta}(D_i)$ at the leading order in the dimension, one obtains:

\begin{align}
&\delta_{ij}\left(\frac{1}{D_i} \sum_{\bar i\in I} A_{\bar i\, \bar i}\right)^2 +\\ &+\frac{\delta_{I,J}}{D_i} \left(  \frac{1}{D_i}\sum_{\substack{\bar{i}, \bar{k}\in I }} A_{\bar{i}\,\bar{k}}A_{\bar{k}\,\bar{i}} - \left(\frac{1}{D_i} \sum_{\bar i\in I} A_{\bar i\, \bar i}\right)^2\right) \ .  \nonumber
\end{align}

When $I\ne J$, the average in Eq.~\eqref{loc_2pt} factorizes in two $$\overline{U^{(I)\,*}_{\bar i i}U^{(I)}_{\bar l i}} \,\, \,\overline{{U^{(J)\,*}_{\bar k j}U^{(J)}_{\bar j j }}}\ .$$ Again, one can use Eq.~\eqref{Haar_av_1} separately for the two averages, one with $D_i$ and the other with $D_j$: 
$$
\frac{1-\delta_{I,J}}{D_iD_j} \sum_{\substack{\bar{i}\in I \\ \bar{k}\in J }} A_{\bar{i}\,\bar{k}}A_{\bar{k}\,\bar{i}} \ .
$$
Summing these two cases together: 
\begin{align}
\label{app_LRI_ex2}
    \overline{A^U_{ij}A^U_{ji}} & \simeq \delta_{ij} \left[K_1(E_i)\right]^2 + \\  + \frac{1}{D_{i}D_{j}}&\sum_{\substack{\bar{i}\in I 
    \notag
    \\ \bar{k}\in J }} A_{\bar{i}\,\bar{k}}A_{\bar{k}\,\bar{i}}
- \left ( \frac 1{D_i}\sum_{\bar i\in I} A_{\bar i\, \bar i} \right )^2 \frac{\delta_{I,J}}{D_i} \\
 &\equiv \delta_{ij} \left[K_1(E_i)\right]^2 + K_2(E_i,E_j) \ .
\notag
\end{align}
Again, one can use equivalently smooth delta functions, so that
\begin{align}
\notag
K_{2}(E_i, E_j)  = &\sum_{\alpha \beta} A_{\alpha\beta} \frac{\delta_\Delta(E_i-E_\alpha)}{D_i}A_{\beta \alpha} \frac{\delta_\Delta(E_j-E_\beta)}{D_j} \\&- [K_1(E_i)]^2 \sigma(E_i-E_j) \ ,
\end{align}
where $\sigma(E_i-E_j)$ accounts for the smoothed version of $\delta_{I,J}/D_i$ on the mesoscopic energies. This expression shows that the simple loop is given by 
the second line of Eq.~\eqref{CDEF}.

For the cactus $\overline{A^U_{ii}A^U_{ii}}$ we recover, from Eq.~\eqref{app_LRI_ex2}, the factorization and the first subleading term with the right energy dependencies, as in Eq.~\eqref{ex_ii_LRI}. 

\begin{widetext}
\subsubsection{$n=3$}
For the expression 
\begin{align}
    \overline{A^U_{ij}A^U_{jk}A^U_{ki}}= \sum_{\substack{\bar{i},\bar n \in I \\ \bar{j},\bar k \in J\\
    \bar{l},\bar{m} \in K}} \overline{U^{(I)\,*}_{\bar{i}i} U^{(J)\,*}_{\bar{k} j} U^{(K)\,*}_{\bar{m} k} U^{(J)}_{\bar{j}j} U^{(K)}_{\bar{l}k} U^{(I)}_{\bar{n}i}}
    A_{\bar{i}\,\bar{j}}A_{\bar{k}\bar{l}} A_{\bar{m}\bar{n}} \, ,
    \label{loc_3pt}
\end{align}
one has now to distinguish multiple cases, each computable with Eq.~\eqref{Haar_av_1} choosing the right dimension, and expanding the Weingarten function. 

We begin with the case of simple loop $i\ne j \ne k$. Eq.~\eqref{loc_3pt} needs to be considered in three different cases:
\begin{description}
    \item[$ \bullet  \ \ I=J=K$] the Weingarten function has to be evaluated only for $\alpha=(123)$, since it is the only one respecting $i\ne j\ne k$, and expanded at the first order in $1/D_i$. The result is 
\begin{align}
    \frac{\delta_{I,J}\delta_{I,K}}{D^2_i} \left( \frac{1}{D_i}\sum_{\bar{i},\bar{k},\bar{m} \in I} A_{\bar i \bar k }A_{\bar k \bar m}A_{\bar m \bar i}  -\frac{3}{D^2_i}\sum_{\bar i \in I}A_{\bar i \bar i} \sum_{\bar k, \bar m \in I} A_{\bar k, \bar m \in I} A_{\bar m \bar k} + 2\frac{\left(\sum_{\bar i \in I}A_{\bar i \bar i}\right)^3}{D^3_i}\right) \ .
\end{align}
    \item[$\bullet \ \ I=J\ne K + \ 2 \,cycl.$] the average factorizes in two parts; one is easily computable directly with $\text{Wg}_{e,e}(D_k)$, and the other has to be computed only for $\alpha=(12)$. The result reads: 
\begin{align}
    \frac{\delta_{I,J}(1-\delta_{J,K})(1-\delta_{K,I})}{D_i} \left( \frac{1}{D_iD_k}\sum_{\bar{i},\bar{k}\in I,\bar{m} \in K} A_{\bar i \bar k }A_{\bar k \bar m}A_{\bar m \bar i}  -\frac{1}{D^2_iD_k}\sum_{\bar i \in I}A_{\bar i \bar i} \sum_{\bar k\in I, \bar m \in K} A_{\bar k \bar m} A_{\bar m \bar k}\right) \ .
\end{align}
    \item[$\bullet \ \ I \ne J \ne K$] the average factorizes in three parts, each of them computable with $\text{Wg}_{e,e}$ for dimension $D_i$, $D_j$, $D_k$, respectively. The result is given by : 
\begin{align}
    (1-\delta_{I,J})(1-\delta_{J,K})(1-\delta_{K,I})\frac{1}{D_iD_jD_k}\sum_{\bar{i}\in I,\bar{k}\in J,\bar{m}\in K} A_{\bar i \bar k }A_{\bar k \bar m}A_{\bar m \bar i} \ .
\end{align}
\end{description}
The different cases have to be summed together to get the final result:
\begin{align}
    &\overline{A^U_{ij}A^U_{jk}A^U_{ki}} \simeq K_3(E_i, E_j, E_k)  = \sum_{\alpha\in I, \beta\in J, \gamma\in K} A_{\alpha\beta} A_{\beta \gamma}A_{\gamma\alpha} - [K_1(E_i)]\frac{\delta_{I,J}}{D_i}
     \sum_{\alpha\in I, \beta\in K} 
     A_{\alpha\beta} 
     A_{\beta \alpha}+  2{\rm cycl.} + 2 [K_1(E_i)]^3 \frac{\delta_{I,J}\delta_{J,K}}{D^2_i} \ .
     \notag
\end{align}
The smoothing of this expression, done in the same way as explained above, gives the third line of Eq.~\eqref{CDEF}.

\medskip

We move to the cactus $\overline{A^U_{ij}A^U_{ji}A^U_{ii}}$;
Eq.~\eqref{loc_3pt} becomes 
\begin{align}
    \overline{A^U_{ij}A^U_{ji}A^U_{ii}}= \sum_{\substack{\bar{i},\bar n,\bar{l},\bar{m} \in I \\ \bar{j},\bar k \in J}} \overline{U^{(I)\,*}_{\bar{i}i} U^{(J)\,*}_{\bar{k} j} U^{(I)\,*}_{\bar{m} i} U^{(J)}_{\bar{j}j} U^{(I)}_{\bar{l}i} U^{(I)}_{\bar{n}i}}
    A_{\bar{i}\,\bar{j}}A_{\bar{k}\bar{l}} A_{\bar{m}\bar{n}} \, ,
\end{align}
and needs to be considered in two different cases:
\begin{description}
    \item[$\bullet \ \ I=J$] the Weingarten function has to be evaluated for $\alpha=(12)(3)$ and $(123)$, and expanded at the first order in $1/D_i$; not only, for each $\alpha$ one has to consider only the permutations $\beta<\alpha$ in order to have the first subleading contribution. We highlight that the other $\alpha$s are not respecting the condition on the initial set of indices \textbf{i}. It is remarkable to notice that the permutations to be considered are the ones on the geodesic between $e$ and $\gamma_3$, equal or bigger than $(12)(3)$, exactly as in the case of global rotational invariance. For instance, the permutation $(132)\cancel{\in} [e,\gamma_3]$ and, indeed, does not respect the repetitions in the set of indices \textbf{i}, even if with one cycle more than $(12)(3)$. The computation leads to 
    
\begin{align}
    \frac{\delta_{I,J}}{D_i}  \left( \frac{1}{D^2_i}\sum_{\bar i \in I}A_{\bar i \bar i} \sum_{\bar k, \bar m \in I} A_{\bar k \bar m} A_{\bar m \bar k}-\frac{\left(\sum_{\bar i}A_{\bar i \bar i}\right)^3}{D^3_i}+\frac{1}{D^2_i}  \sum_{\bar{i},\bar{k},\bar{m} \in I} A_{\bar i \bar k }A_{\bar k \bar m}A_{\bar m \bar i}
     + 2\frac{\left(\sum_{\bar i \in I}A_{\bar i \bar i}\right)^3}{D^4_i} -\frac{3}{D^3_i}\sum_{\bar i \in I}A_{\bar i \bar i} \sum_{\bar k, \bar m \in I} A_{\bar k \bar m} A_{\bar m \bar k}\right) \ .
\end{align}

    \item[$\bullet \ \ I \ne J$] the average factorizes in two parts; one easily computable directly with $\text{Wg}_{e,e}(D_j)$, and the other to be evaluated for both $\alpha=(12)$ and $\alpha=e$, always considering only $\beta<\alpha$. The result is: 
    
\begin{align}
    (1-\delta_{I,J}) \left( \frac{1}{D_jD^2_i}\sum_{\bar i \in I}A_{\bar i \bar i} \sum_{\bar k\in J, \bar m\in I} A_{\bar k \bar m} A_{\bar m \bar k} +\frac{1}{D_jD^2_i}  \sum_{\bar{i},\bar{m} \in I,\bar{k} \in J} A_{\bar i \bar k }A_{\bar k \bar m}A_{\bar m \bar i}
      -\frac{1}{D_jD^3_i}\sum_{\bar i \in I}A_{\bar i \bar i} \sum_{\bar k \in J,\bar m\in I} A_{\bar k \bar m} A_{\bar m \bar k}\right) \ .
\end{align}
\end{description}

Summing all together, one gets:

\begin{align}
&\overline{A^U_{ij}A^U_{ji}A^U_{ii}}\simeq    \frac{1}{D_jD^2_i}\sum_{\bar i \in I}A_{\bar i \bar i} \sum_{\bar k\in J, \bar m \in I} A_{\bar k \bar m} A_{\bar m \bar k}
-\frac{\delta_{I,J}}{D_i}\frac{\left(\sum_{\bar i}A_{\bar i \bar i}\right)^3}{D^3_i} \\ &+ \frac{1}{D_jD^2_i}\sum_{\bar{i},\bar{m}\in I,\bar{k} \in J} A_{\bar i \bar k }A_{\bar k \bar m}A_{\bar m \bar i} +2\frac{\delta_{I,J}}{D_I} \frac{\left(\sum_{\bar i}A_{\bar i \bar i}\right)^3}{D^4_i}
-\frac{1}{D_jD^3_i} \sum_{\bar i \in I}A_{\bar i \bar i} \sum_{\bar k\in J, \bar m\in I} A_{\bar k \bar m} A_{\bar m \bar k} -2\frac{\delta_{I,J}}{D_i} \frac{1}{D^3_i} \sum_{\bar i \in I}A_{\bar i \bar i} \sum_{\bar k, \bar m \in I} A_{\bar k \bar m} A_{\bar m \bar k}
\notag \\ \\
& \qquad \qquad \quad \equiv K_2(E_i,E_j)K_1(E_i) + K_3(E_i,E_j,E_i) \ , \notag
\end{align}

which reproduces the right factorization and the first subleading contribution given by $K_3$ evaluated at equal mesoscopic energies, for coincident initial indices.\\

The case of the cactus $\overline{A^U_{ii}A^U_{ii}A^U_{ii}}$ is the easiest, but significant. 
There is only one average in Eq.~\eqref{loc_3pt} to be considered, for which Weingarten calculus for dimension $D_i$ can be employed, exactly as in the case of global rotational invariance.  
The Weingarten function has to be evaluated for all $\alpha$s, since all of them respect the condition on the initial set of indices \textbf{i}, and have more cycles than $e$, but two considerations have to be made: firstly, $(123)$ and $(132)$ would give the second subleading correction, so they have to be excluded, for being coherent with the approximation at the leading order in the dimension made on the Weingarten function. Interestingly, $(132)$ is respecting the repetitions of the indices in \textbf{i}, even if it is not on the geodesic $[e,\gamma_3]$; indeed, as said in App.~\ref{app_GRIderivation}, only for the first subleading correction it is sufficient to restrict to non-crossing partitions, but for higher orders this is not the case. The final result, in this case, is:

\begin{align}  
\overline{A^U_{ii}A^U_{ii}A^U_{ii}}\simeq 
&\frac{(\sum_{\bar i \in I}A_{\bar i \bar i})^3}{D^3_i}+ \frac{3}{D^3_i}\sum_{\bar i \in I}A_{\bar i \bar i} \sum_{\bar k, \bar m \in I} A_{\bar k \bar m} A_{\bar m \bar k} - 3\frac{\left(\sum_{\bar i \in I}A_{\bar i \bar i}\right)^3}{D^4_i} \notag\\
 \equiv &[K_1(E_i)]^3 + 3K_1(E_i)K_2(E_i,E_i) \ .
 \notag
\end{align}

Again, this reproduces the right factorization and the first subleading contribution.

\subsubsection{$n=4$}
The same logic as outlined for the previous cases applies here, but the explicit derivation is much longer and therefore omitted. It is important to separate the different cases of macroscopic energies and, within each case, to factorize the averages before applying the Weingarten formalism. Only the relevant $\alpha$s should be considered, the function should be approximated to leading order, and only terms with $\beta < \alpha$ retained.
\newpage
\end{widetext}

\section{Link between local free cumulants \\ and operator-valued free cumulants}
\label{app_linkFB}

We now show how Eqs.~\eqref{localfc_main} emerge within the framework of operator-valued free probability. To this end, let us reconsider the example introduced in App.~\ref{app_OVFB}, namely a four-block matrix and the choice of $\mathcal{D}$ as the subalgebra of block-diagonal matrices generated by the projectors $\hat \Pi_{I=1,2}$ onto the individual blocks, see Eq.~\eqref{example_ovfb}. Generalizing this construction to an arbitrary number of blocks provides the natural setting in which operator-valued free cumulants, valued by these projectors, can be related to local free cumulants.
Indeed, local free cumulants arise when a physical observable $A$ in the energy basis is regarded as a $D\times D$ matrix partitioned into blocks according to the energies of the states, with each block being independently reshuffled by a local rotation. However, unlike operator-valued free cumulants, which are matrix-valued, local free cumulants are scalar quantities. In fact, we now show explicitly that they correspond to the diagonal entries of the blocks of the operator-valued free cumulants. Furthermore, a suitable rescaling of the projectors is required to recover Eqs.~\eqref{localfc_main}.\\
We define the normalized projectors on the two diagonal blocks as $\hat \sigma_{I} := \hat \Pi_{I}/D_{I}$ being $D_{I}$ the dimension of $\mathcal{A}_{II}$. \\
For the first-order free cumulant, we have (Eq.~\eqref{example_ovfb_k1}):
\begin{equation}
\langle A \rangle_{\mathcal{D}} = \kappa^{\mathcal{D}}_1(A) = \mathrm{diag}\big(\langle \mathcal{A}_{11}\rangle \mathbb{I}, \langle \mathcal{A}_{22} \rangle \mathbb{I} \big)\, .
\end{equation}

The value on the diagonal of each block can be expressed as the global trace of the matrix projected onto the corresponding diagonal block via the projector:
\begin{equation}
\langle \mathcal{A}_{II}\rangle = \mathrm{Tr}( A \hat{\sigma}_{I}) \, .
\label{first_gen_mom}
\end{equation}
Indeed, this corresponds simply to the normalized trace of the diagonal block of the original matrix.
We call the r.h.s. of Eq.~\eqref{first_gen_mom} as \textit{generalized moment}, and we introduce the first \textit{generalized free cumulant} as 
\begin{equation}
    K_1(\mathcal{A}_{II})= \mathrm{Tr}( A\hat \sigma_{I}) \ .
\end{equation}
These generalized quantities are scalars and explicitly retain the dependence on the specific diagonal block. \\
We can proceed analogously for the second-order case, for which we report Eqs.~\eqref{example_ovfb_m2},\eqref{example_ovfb_k2}:
\begin{align}
\notag
\langle A\hat \sigma_1 A\rangle_{\mathcal{D}} &=
\mathrm{diag}(\langle \mathcal{A}^2_{11}\rangle/D_1 \mathbb{I},  \langle \mathcal{A}_{21}\mathcal{A}_{12}\rangle/D_{1} \mathbb{I})\ , \\
\langle A\hat \sigma_2 A\rangle_{\mathcal{D}} &= \mathrm{diag}(
\langle \mathcal{A}_{12}\mathcal{A}_{21}\rangle/D_2 \mathbb{I},  \langle \mathcal{A}^2_{22}\rangle/D_2 \mathbb{I})\ ,
\notag \\ 
\notag
\kappa^{\mathcal{D}}_2(A\hat \sigma_1, A)&= \mathrm{diag}\left((\langle \mathcal{A}^2_{11}\rangle - \langle \mathcal{A}_{11}\rangle^2) /D_1\mathbb{I},  \langle \mathcal{A}_{21}\mathcal{A}_{12}\rangle /D_1 \mathbb{I}\right ) ,\\
\notag
\kappa^{\mathcal{D}}_2(A\hat \sigma_2, A) &= 
\mathrm{diag}\left (
\langle \mathcal{A}_{12}\mathcal{A}_{21}\rangle/D_2 \mathbb{I} , ( \langle \mathcal{A}^2_{22}\rangle - \langle \mathcal{A}_{22}\rangle^2)/D_2\mathbb{I} \right).
\end{align}
We now define the second-order generalized moments as 
\begin{align}
    \langle \mathcal{A}^2_{II}\rangle/D_{I} &= \mathrm{Tr}( A\hat \sigma_{I} A \hat \sigma_{I})\\ \notag \langle \mathcal{A}_{IJ}\mathcal{A}_{JI}\rangle/D_J &= \mathrm{Tr}( A\hat \sigma_I A \hat \sigma_J) \ ,
\end{align}
which are again the diagonal values of the operator-valued free cumulants.
The generalized second-order free cumulants follow as  
\begin{align}
\notag
K_2(\mathcal{A}_{II},\mathcal{A}_{II}) & =\mathrm{Tr}( A\hat \sigma_{I} A \hat \sigma_{I}) - \mathrm{Tr}( A\hat \sigma_{I})^2 \\ \notag &\equiv \frac{1}{D_{I}}\left(\langle \mathcal{A}^2_{II}\rangle - \langle \mathcal{A}_{II}\rangle^2\right) \ , \\
K_2(\mathcal{A}_{II},\mathcal{A}_{JJ}) & =  \mathrm{Tr}( A\hat \sigma_I A \hat \sigma_J)\\ & \equiv \frac{1}{D_J}\langle \mathcal{A}_{IJ}\mathcal{A}_{JI}\rangle\ ,
\notag
\end{align}
which compactly is
\begin{align}
K_2(\mathcal{A}_{II},\mathcal{A}_{JJ}) =  \mathrm{Tr}(A\hat \sigma_{I} A \hat \sigma_{J}) -  \frac{\delta_{I,J}}{D_{I}}\mathrm{Tr}( A\hat \sigma_{I})^2 \ . 
\end{align}
One should note that these formulas extend straightforwardly to a generic number of blocks in the initial matrix. The resulting free cumulants are scalars, but they explicitly encode correlations between different diagonal blocks. Moreover, an additional Kronecker $\delta$ structure is inherited from operator-valued free probability. \\
In the case of an observable $A$ expressed in the energy basis, the generalized free cumulants capture correlations between different diagonal blocks, each associated with a given mesoscopic energy. In this setting, the projector
$ \displaystyle 
\hat \sigma_I = \sum_\alpha |\alpha\rangle\langle\alpha| \, \frac{\delta_\Delta(E_i - E_\alpha)}{D_I}
$
selects the set of states with energy $E_i$, where $i$ labels a representative state within the block. We may therefore replace the block index $I$ with the corresponding energy $E_i$, and identify $\delta_{I,J}/D_I$ with
$\displaystyle
\sigma(E_i - E_j) = \frac{\delta_\Delta(E_i - E_j)}{D_I}\, .
$
In this representation, the first two cumulants read
\begin{align}
    K_1(E_i) &= \mathrm{Tr}\big( \hat \sigma_{E_i} A \big) \\
    K_2(E_i,E_j) &= \mathrm{Tr}\big(\hat \sigma_{E_i} A \hat \sigma_{E_j} A \big)
    - \big[\mathrm{Tr}( \hat \sigma_{E_i} A )\big]^2 \, \sigma(E_i - E_j)\, .
    \notag
\end{align}
These expressions correspond to Eqs.~\eqref{c_1_loc}--\eqref{c_2_loc}. Upon computing higher orders, one obtains the \textit{moments–cumulants formula} in Eq.~\eqref{localfc_main}. This establishes the identification of the local free cumulants with the generalized free cumulants here introduced, namely the diagonal-block components of the operator-valued free cumulants. \\
The equivalence has been shown explicitly for the toy model with blocks. To obtain smooth functions of the energy, one must smooth each block around its associated energy: the factor $\delta_{\Delta}(E_i - E_\alpha)$ is then understood as a function peaked around $E_i$, rather than a sharp box. In this regime, the connection to operator-valued free probability becomes more subtle. In particular, one must appropriately integrate the local cumulants over energy in order to recover operator-valued cumulants. This was carried out in \cite{bernard2024structured}, even if in a different setting.

\section{A different approach for global rotational invariance: low rank HCIZ integral}\label{App_HCIZ}

The Harish-Chandra-Itzykson-Zuber (HCIZ)
integral \cite{harish1957differential,IZ1980} ${\cal I}_{\beta_{\text{RMT}}}(A,B)$ is defined as:
\begin{equation}\label{eq:HCIZ-def}
{\cal I}_{\beta_{\text{RMT}}}(A,B) = \int_{G(D)} {\mathcal D} \Omega \,\, e^{\frac{\beta_{\text{RMT}} D}{2} \text{Tr}  B \Omega A  \Omega^{^\dagger}},
\end{equation}
where the integral is over the Haar measure of the compact group $\Omega \in  G(D)= O(D)$ (orthogonal) or 
$U(D)$ (unitary)  in $D$ dimensions   and $A,B$ are arbitrary 
$D \times D$ symmetric or hermitian matrices \cite{bun2017cleaning}. 
The parameter $\beta_{\text{RMT}}$ is the usual Dyson ``inverse temperature'', with $\beta_{\text{RMT}}=1$ or $2$, respectively for the two groups.

Several results are known for this integral but in the following we will be interested in a simple yet beautiful explicit prediction which is available when one of the matrices has lower rank $n \ll D$. Precisely, let us assume that $B$ has $n$ eigenvalues $\lambda^{B}_1,\dots,\lambda^{B}_n$ and $D-n$ zero eigenvalues. Then we have \cite{marinari1994replica,guionnet2005fourier}:
\begin{equation}\label{Eq_low_rank}
	{\cal I}_{\beta_{\text{RMT}}}(A,B) = \exp\left[ \frac{D\beta_{\text{RMT}}}{2} \sum_{i=1}^{n} {{\cal W}_{A}}(\lambda^{B}_i) \right] \, 
\end{equation}
where ${\cal W}_{A}$ is the primitive of the ${\mathcal{R}}$-transform of $A$:
\begin{equation}
\label{Def_W}
{{\cal W}_{A}}(z) = \int_0^z \, {\rm d} x  \, {\mathcal{R}}_{A}(x) \ .
\end{equation}
with $\mathcal{R}_A(z)$ defined as the generating function of free cumulants:
\begin{align}\label{eq:expansion_R_freecum}
\mathcal{R}_A(z) = \sum_{k=0}^\infty \kappa_{k+1}^A \, z^k.
\end{align} 
We note that $\sum_{i=1}^{n} {{\cal W}_{A}}(\lambda^{B}_i)$ can also be written as $\text{Tr} \, {{\cal W}_{A}}(B)$ and the exponent in Eq. (\ref{Eq_low_rank}) can be written as
\begin{equation}
\label{Def_W2}
{{\cal W}_{A}}(B) = \sum_{k=1}^\infty \frac{1}{k} \kappa_{k}^A \, \text{Tr}(B^{k}) .
\end{equation}
For an intuitive derivation of Eq. (\ref{Eq_low_rank}) we refer to \cite{bun2017cleaning}.\\

We now sketch how the predictions for the leading terms derived in the main
section in the unitary ensemble can be generalized also to the orthogonal case.
These results and the proof that we present are discussed in
\cite{maillard2019high}.
Generalizing the previous notation, let us now consider a matrix $A^O = O^\dagger A O$ with $O$ a Haar distributed orthogonal matrix (but we could consider different universality classes), and $A$ a matrix with limiting spectral function $\rho^A$.
The result (\ref{Eq_low_rank}) can be used to show that for any $n \geq 1$ and any set of pairwise distinct indices $i_1,\cdots,i_n$, one has:
 \begin{align}\label{eq:freecum_expectation_nonsummed}
 \lim_{D \to \infty} \left[ L_n^{(D)} \equiv D^{n-1} \overline{ A_{i_1 i_2}^O A_{i_2 i_3}^O \cdots A_{i_{n-1} i_n}^O A_{i_n i_1}^O  } \right] &= \kappa_n^{A}
 \end{align}
where the average is over the orthogonal matrix.
Note that the free cumulants of $A$ are the same as those of $A^O$ as the two matrices share the same spectral properties.

In order to simplify the following calculation, we assume that $(i_1,\cdots,i_n) = (1,\cdots,n)$, since the desired result does not depend on the particular choice of indices (as is clear by rotational invariance).

We note that we can use (\ref{eq:HCIZ-def}), and in particular its low rank limit, as a generating function for the matrix elements of $A^O$.
The cases $n=1$ and $n=2$ have to be treated separately, but 
one can generalize the case we consider
here for $n \geq 3$ where we can write  $L_n^{(D)}$ as:
 \begin{align}\label{Generating_function}
 L_n^{(D)} &= \frac{1}{D} \prod_{l=1}^n \frac{\partial}{\partial b_l} \left[\int_{O(N)} \, {\mathcal D}O \, e^{\frac{D}{2} \mathrm{Tr}\, \left[B(\bb) O  A O^\dagger \right]}\right]_{\bb=0},
 \end{align}
 in which we denoted $\bb \equiv (b_1,\cdots,b_n)$ and the following symmetric block matrix of rank $n$:
 \begin{align}
 B(\bb) \equiv \begin{pmatrix}
 B_1(\bb) & (0) \\
 (0) & (0) \\
 \end{pmatrix},
 \end{align}
 in which $B_1(\bb)$ is:
 \begin{align}\label{cycle_matrix}
 B_1(\bb) \equiv  \begin{pmatrix}
 0 & b_1 & 0 & \cdots & 0 & b_n \\
 b_1  & 0 & b_2 & \cdots & 0& 0 \\
 0 & b_2 & 0 & \cdots & 0 &0 \\
 \vdots & \vdots & \vdots & \ddots & \vdots & \vdots \\
 0 & 0 & 0 & \cdots & 0 & b_{n-1} \\
 b_n & 0 & 0 & \cdots & b_{n-1}  & 0
 \end{pmatrix}.
 \end{align}
Upon taking derivatives in (\ref{Generating_function}), using the 
low rank result (\ref{Def_W2}), 
and the form (\ref{cycle_matrix}), 
one can recover Eq. (\ref{eq:freecum_expectation_nonsummed}).

With the same strategy one can construct a $B(\bb)$ matrix which is used to compute averages of cactus diagrams.
As deriving the exponent in (\ref{Def_W2}) allows to gain a factor $D$ at the leading order it is clearly convenient to decompose
the cactus in as many leaves (cycles) as possible, noting that this decomposition is unique, and then the argument follows the same lines as for simple loops.

The same approach can be generalized to the unitary case (see Ref.~\cite{maillard2019high}) and the leading terms of the expectations turns out to be the same for real and complex matrices. 

It turns out that the low rank integral captures also the corrections with scaling $1/{D}$ smaller than the leading term for the cactus diagrams. At this order, however, one might expect some differences in the two universality classes.
In fact, different
factors appear in front of the expressions of the subleading corrections, beyond the one provided by the free cumulants, for products involving diagonal matrix elements in the case of orthogonal matrices.
One example is the cactus diagram which describes the fluctuations of diagonal matrix elements:
\begin{equation}
\label{GOE_eq}
\overline{A^{\Omega}_{ii}A^{\Omega}_{ii}}   \simeq  \overline{A_{ii}^{\Omega}}^2 + \frac{2}{\beta_{\text{RMT}}} \overline{A_{ij}^{\Omega}A_{ji}^{\Omega}} \ ,
\end{equation}
for $\Omega=O$ or $\Omega=U$,
as it was noticed for instance in \cite{mondaini2017eigenstate,foini2019eigenstate2}.

\section{Additional Numerical Results}
\label{app_num}
In the first section of this Appendix, we report some additional details on the implementations of the numerics; we remind the reader that all the codes and data can be found in \cite{zenodo2025}.
Successively, we present numerical results for slightly different Hamiltonians than the one in the main text: we still consider a non-integrable periodically driven spin chain, described by Hamiltonian in \eqref{KIM}, and by the Floquet operator in \eqref{Floquet_operator}, but we slightly change the form of $H_0$ or $V_0$ in \eqref{Hamiltonian_num}. Firstly, we break translational invariance, in order to check our assumptions for different observables, then, we restore the time reversal symmetry, so that the symmetry class of the Hamiltonian is orthogonal instead of unitary.

\subsection{Different observables}
\label{app_obs}
One can break the translational invariance, considering, for instance, a different amplitude for the longitudinal magnetic field on each site in Eq.~\eqref{Hamiltonian_num}: 
$$
H^{\text{disorder}}_z = J\sum_i Z_i Z_{i+1}+\sum_i h_i Z_i \ .
$$
Each $h_i$ is independently sampled from a uniform distribution centered at $\bar{h} = 0.905$ and spanning the interval $[\bar{h} - \delta_h,\, \bar{h} + \delta_h]$, with width $\delta_h = 0.1\sqrt{3}$.
In this case, we can consider as observable the magnetization on a single site; we choose $A=X_{L/2}$. 
We perform the same analysis as in Sec.~\ref{subsec_num} of the main, evaluating the averages of the products of matrix elements involved in Eqs.~\eqref{4p_1_ene_Floquet}--\eqref{4p_2_ene_Floquet}, through the expressions \eqref{app_eq_num}.

\begin{figure*}[h]
\centering
\includegraphics[width=1 \textwidth]{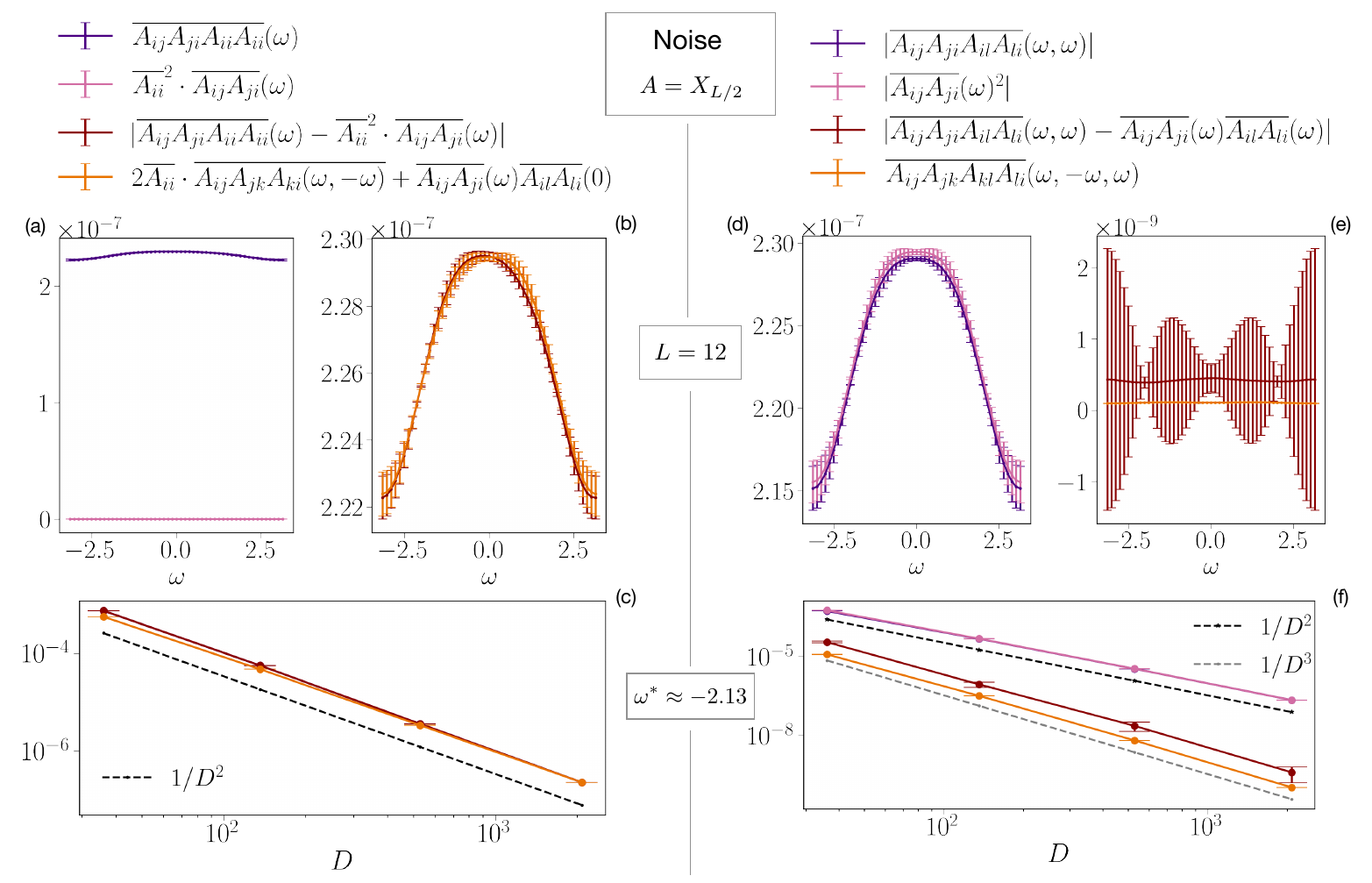}
\caption{Numerical results for the Floquet system with disorder presented in App.~\ref{app_obs}, and for the observable $A=X_{L/2}$. Similarly to Fig.~\ref{fig_numerics}, we numerically reproduce equations Eqs.~\eqref{4p_1_ene_Floquet}--\eqref{4p_2_ene_Floquet}, for $\omega_1=\omega_2=\omega$ chosen in $[-\pi,\pi]$. The error bar is given by averaging each product, for each $\omega$, over different values of the smoothing parameter $\Delta$, as explained in the text. (a),(b),(c) refer to \eqref{4p_1_ene_Floquet}, while (d),(e),(f) to \eqref{4p_2_ene_Floquet}.
In (a),(d) we compare the right-hand side (purple curve) with the respective leading order (pink curve) for fixed $L=12$; for (a) the leading order is zero, while in (d) the data very well reproduce the expected leading order factorization.
In (b),(e) we compare the difference between the right-hand side and the leading contribution (brown curve) with the respective subleading correction (orange curve) for fixed $L=12$. The data verify the subleading correction to the factorization. 
In (c),(f) we represent the scalings of all the curves as a function of $D$. The scalings were studied for values $L\in[6,8,10,12]$, and for a fixed value of $\omega^*\approx-2.13$. While all the curves in (c) scale as $1/D^2$, since the leading order is zero, in (f) we observe the different scalings between the leading (purple and pink curves), as $1/D^2$ and the subleading (brown and orange curves), as $1/D^3$. }
\label{fig_numerics_noise}
\end{figure*}

All results are plotted in Figure~\ref{fig_numerics_noise}. For a fixed number of spin $L=12$, we compare the right-hand side (purple curve) of Eqs.~\eqref{4p_1_ene_Floquet}--\eqref{4p_2_ene_Floquet} with the respective leading order (pink curve), and then we substract the leading to the right-hand side, in order to compare this difference (brown curve) with the prediction for the subleading (orange curve). Moreover, we study the scaling with the subspace dimension $D$ for each of these curves, for a fixed $\omega^*\approx -2.13$
Since $\braket{X_{L/2}}=0$, the leading order of $\overline{A_{ij}A_{ji}A_{ii}A_{ii}}(\omega)$ is zero, therefore the average follows directly the subleading contribution $\overline{A_{ij}A_{ji}}(\omega)\overline{A_{il}A_{li}}(0)$, and scales as $1/D^2$, as expected. On the contrary, $\overline{A_{ij}A_{ji}A_{il}A_{li}}(\omega,\omega)$ is very well factorized in $\overline{A_{ij}A_{ji}}(\omega)\overline{A_{il}A_{li}}(\omega)$, scaling as $1/D^2$, and the subleading well follows the expected curve $\overline{A_{ij}A_{jk}A_{kl}A_{li}}(\omega,-\omega,\omega)$, scaling as $1/D^3$.

\begin{figure}[h]
\centering
\includegraphics[width=0.5 \textwidth]{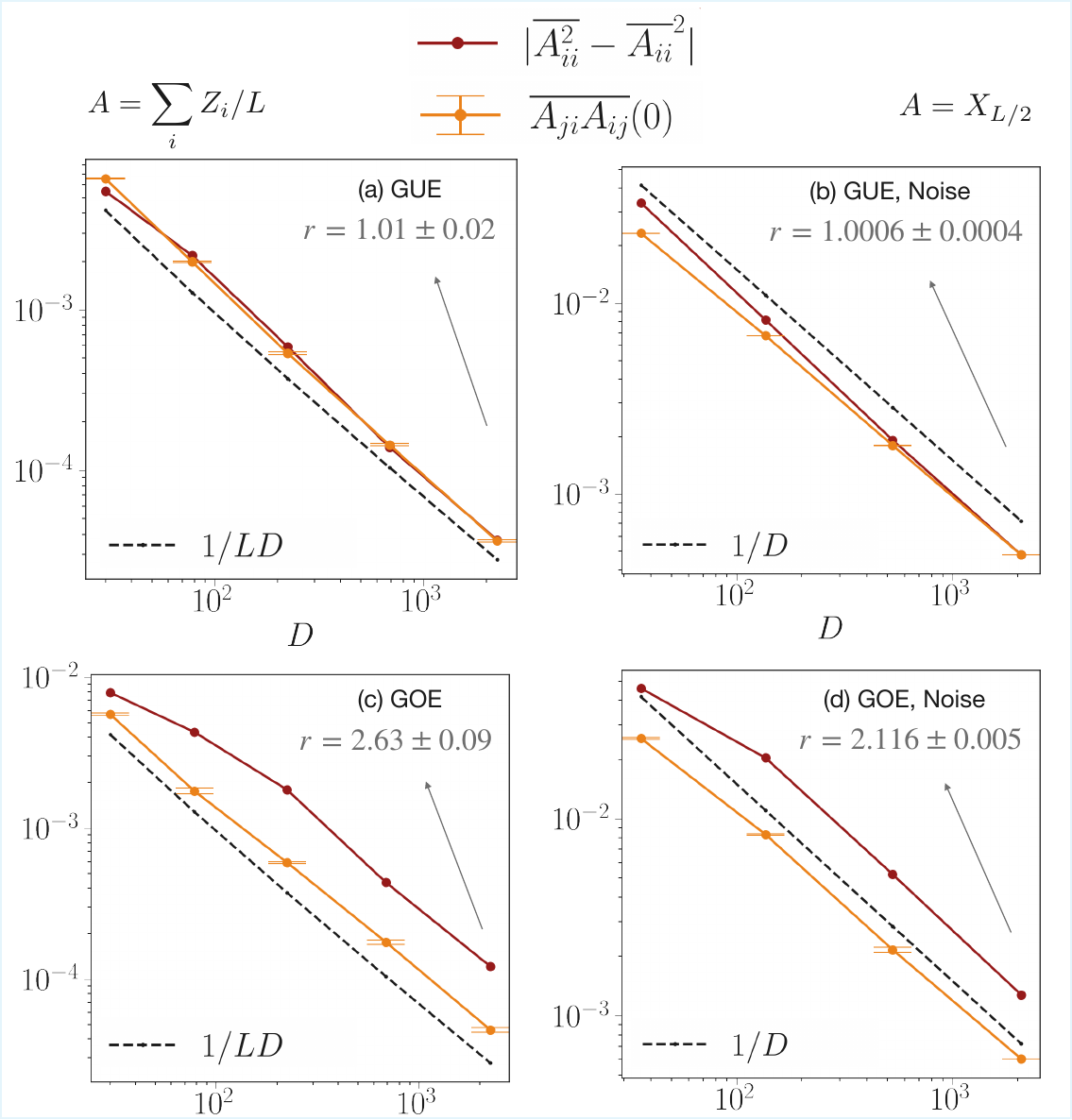}
\caption{Numerical comparison between the Floquet systems with and without time reversal symmetry, that is, belonging to the orthogonal or the unitary symmetry class, with and without disorder, presented in App.~\ref{app_num_GOE}. 
We numerically reproduce equation Eq.~\eqref{GOE_eq}. The error bar is given by averaging $\overline{A_{ij}A_{ji}}(0)$ over different values of the smoothing parameter $\Delta$, as explained in the text. For the first column $L\in [8,10,12,14,16]$, while for the second $L \in [6,8,10,12]$. 
We compare $|\overline{A_{ii}^2}-\overline{A_{ii}}^2|$ (brown curve) with the respective subleading correction $|\overline{A_{ij}A_{ji}}(0)|$ (orange curve), as functions of $D$, and we calculate the ratio $r=|\overline{A_{ii}^2}-\overline{A_{ii}}^2|/|\overline{A_{ij}A_{ji}}(0)|$ for the biggest system size for each figure. 
All the curves well reproduce the expected scaling of $1/LD$ for the system without disorder, and $1/D$ for the system with disorder. 
In (a) and (b) we notice that the two curves almost coincide, reproducing the result for the unitary symmetry class \eqref{ex_ii_LRI}. In (c) and (d), instead, $|\overline{A_{ii}^2}-\overline{A_{ii}}^2|$ is approximately twice as large as the subleading one, as expected for the orthogonal symmetry class \eqref{GOE_eq2}. }
\label{fig_numerics_2pt}
\end{figure}

\subsection{Orthogonal symmetry class}
\label{app_num_GOE}
\begin{figure*}[h]
\centering
\includegraphics[width=1 \textwidth]{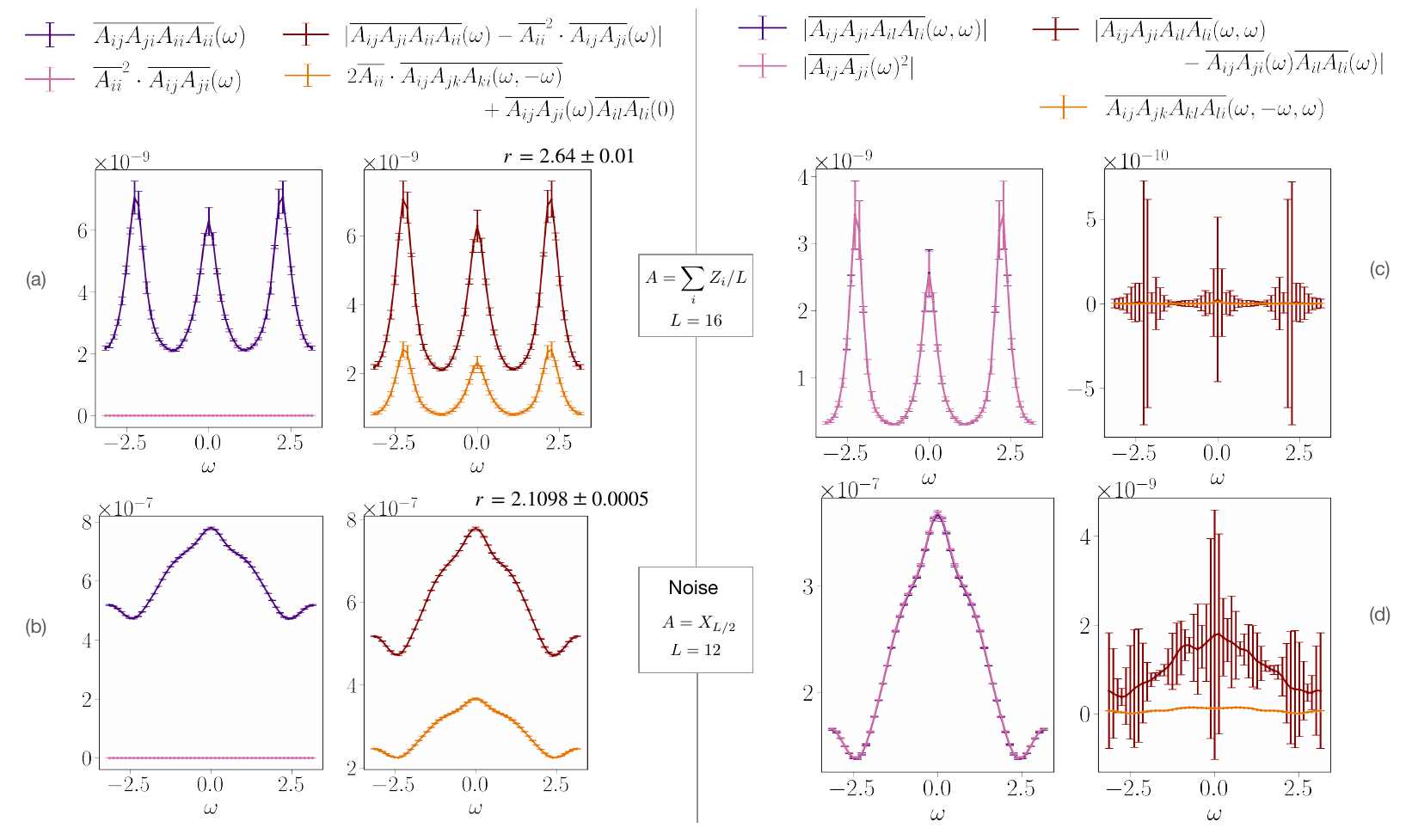}
\caption{Numerical results for the Floquet system belonging to the orthogonal symmetry class presented in App.~\ref{app_num_GOE}. In particular, (a) and (c) refer to the system in Eq.~\eqref{Hamiltonian_num}, while (b) and (d) to the variation of the system with disorder.
We numerically reproduce equations Eqs.~\eqref{4p_1_ene_Floquet}--\eqref{4p_2_ene_Floquet}, for $\omega_1=\omega_2=\omega$ chosen in $[-\pi,\pi]$. The error bar is given by averaging each product, for each $\omega$, over different values of the smoothing parameter $\Delta$, as explained in the text. (a),(b) refer to \eqref{4p_1_ene_Floquet}, while (c),(d) to \eqref{4p_2_ene_Floquet}.
We compare the right-hand side (purple curve) with the respective leading order (pink curve) for fixed system size; the data very well reproduce the expected leading order factorization.
Then, we compare the difference between the right-hand side and the leading contribution (brown curve) with the respective subleading correction (orange curve) for fixed system size. In (a) and (b) we notice that the first curve is approximately twice as large as the subleading one, in contrast to the unitary case (see Figs.~\ref{fig_numerics},\ref{fig_numerics_noise}). The ratio $r$ between these two curves, averaged over $\omega$, is respectively $r=2.66\pm 0.01$ and $r=2.095\pm0.007$ for the two systems considered, in accordance to Fig.~\ref{fig_numerics_2pt}.  
In (c) and (d), instead, we notice no difference between the unitary and the orthogonal case.}
\label{fig_numerics_confronto}
\end{figure*}

One can restore the time reversal symmetry considering no $H_M$ interactions, i.e. $d=0$ in Eq.~\eqref{Hamiltonian_num}. In this case, the system belongs to the orthogonal symmetry class, instead of the unitary \cite{haake2010quantum}. 
As mentioned in App.~\ref{App_HCIZ}, it is well known that the predictions of standard ETH for the subleading of the cacti for $n=2$ depend on the symmetry class \cite{dalessio2016from,mondaini2017eigenstate}, see Eq.~\eqref{GOE_eq}. For the orthogonal:
\begin{equation}
\overline{A_{ii}A_{ii}} - \overline{A_{ii}}^2 \simeq 2\overline{A_{ij}A_{ji}}(0) \ . 
\label{GOE_eq2}
\end{equation}
Therefore, we expect to have discrepancies with respect to the unitary case for the subleading of higher $n$, in particular when diagonal matrix elements are involved.
The origin of the differences is in the rotational invariance: the observable $A$ would be invariant under local orthogonal transformations, as a consequence of the pseudorandomness of real eigenvectors of the Hamiltonian. Therefore, the computation of averages of products of matrix elements could not be carried out via the Weingarten calculus with permutations (Eq.~\eqref{Haar_av_1}), as we did instead in the unitary case. Providing an analytical and general prediction of how the subleading corrections would change for the orthogonal case goes beyond the scope of this paper, but numerically we can perform again the same analysis as in Sec.~\ref{subsec_num} of the main in order to observe the discrepancies. 
Firstly, in Fig.~\ref{fig_numerics_2pt} for both the model given by the~\eqref{Hamiltonian_num} and the variation with disorder presented in App.~\ref{app_obs}, we study the factorization in Eq.~\eqref{GOE_eq2}. Since we are considering a Floquet model, the products of matrix elements on the left-hand side are constant for a fixed $L$, therefore the averages are computed as normal means over all the diagonal elements and plotted as a function of the dimension $D$. We observe the difference between the unitary $(d=1)$ and the orthogonal case $(d=0)$ in the proportionality factor $r= |\overline{A_{ii}A_{ii}} - \overline{A_{ii}}^2 |/ |\overline{A_{ij}A_{ji}}(0)|$ between the left-hand side (brown curve) and the right-hand side (orange curve),  that is respectively close to one and close to two. The proportionality factors are explicitly reported in Fig.~\ref{fig_numerics_2pt}, for the biggest system size of each case.\\
We then perform the same analysis as in Sec.~\ref{subsec_num} and App.~\ref{app_obs}, plotting the result in Fig.~\ref{fig_numerics_confronto} for $d=0$. 
Specifically, we compare the right-hand side (purple curve) of Eqs.~\eqref{4p_1_ene_Floquet}--\eqref{4p_2_ene_Floquet} with the respective leading order (pink curve), and then we subtract the leading to the right-hand side, in order to compare this difference (brown curve) with the prediction for the subleading (orange curve). We notice that the factorizations are always very well reproduced, while for the sublaeding contributions, the same difference observed in Fig.~\ref{fig_numerics_confronto} persists while considering the case of Eq.~\eqref{4p_1_ene_Floquet}. The proportionality factors are explicitly reported in Fig.~\ref{fig_numerics_confronto}, and are obtained averaging over all the $\omega$.
This discrepancy is not present in the case of Eq.~\eqref{4p_2_ene_Floquet} which, indeed, does not involve the diagonal.
\begin{widetext}
\subsection{Numerical Implementation}
\label{app_num_supp}
Regarding more specifically the numerical implementation, we write the empirical averages used for the numerical calculation of Eqs.~\eqref{4p_1_ene_Floquet}--\eqref{4p_2_ene_Floquet}, in the case of a Floquet system.
\begin{subequations}
\label{app_eq_num}
\begin{align}
    \overline{ A_{ij} A_{ji}A_{ii}A_{ii}}(\omega)= &\frac {\sum_{i\neq j} A_{ij}  \delta^{\,\text{per.}}_{\omega}(\nu_i,\nu_j)A_{ji}A_{ii}A_{ii} } {\sum_{i, j} \delta^{\,\text{per.}}_{\omega}(\nu_i,\nu_j)} \ \ ,
\end{align}
\begin{align}
    \overline{ A_{ij} A_{ji}A_{il}A_{li}}(\omega_1,\omega_2)= D \frac{\sum_{i\neq j \neq l} A_{ij}  \delta^{\,\text{per.}}_{\omega_1}(\nu_i,\nu_j) A_{ji}A_{il}\delta^{\,\text{per.}}_{\omega_2}(\nu_i,\nu_l)A_{li}}{\sum_{i, j} \delta^{\,\text{per.}}_{\omega_1}(\nu_i,\nu_j) \sum_{i, l }\delta^{\,\text{per.}}_{\omega_2}(\nu_i,\nu_l)} \ \ ,
\end{align}
\begin{align}
 \overline{ A_{ii}} = \frac{1}{D} \sum_i A_{ii} \ \ , \qquad \qquad 
    \overline{ A_{ij} A_{ji}}(\omega) = \frac{\sum_{i\neq j} A_{ij}  \delta^{\,\text{per.}}_{\omega}(\nu_i,\nu_j) A_{ji}} { \sum_{i, j} \delta^{\,\text{per.}}_{\omega}(\nu_i,\nu_j) } \ \ ,
\end{align}
\begin{align}
    \overline{ A_{ij} A_{jk}A_{ki}}(\omega,-\omega) = D \frac {\sum_{i\neq j \neq k} A_{ij}  \delta^{\,\text{per.}}_{\omega}(\nu_i,\nu_j)A_{jk}\delta^{\,\text{per.}}_{-\omega}(\nu_j,\nu_k)A_{ki}}{\sum_{i, j} \delta^{\,\text{per.}}_{\omega}(\nu_i,\nu_j)\sum_{j,k } \delta^{\,\text{per.}}_{-\omega}(\nu_j,\nu_k)} \ \ ,
\end{align}
\begin{align}
    \overline{ A_{ij} A_{jk}A_{kl}A_{li}}(\omega_1,-\omega_1,\omega_2) = D^2 \frac{\sum_{i\neq j \neq k \neq l} A_{ij}  \delta^{\,\text{per.}}_{\omega_1}(\nu_i,\nu_j) A_{jk}\delta^{\,\text{per.}}_{-\omega_1}(\nu_j,\nu_k)A_{kl}  \delta^{\,\text{per.}}_{\omega_2}(\nu_k,\nu_l) A_{li}}{\sum_{i, j} \delta^{\,\text{per.}}_{\omega_1}(\nu_i,\nu_j) \sum_{j, k }\delta^{\,\text{per.}}_{-\omega_1}(\nu_j,\nu_k)  \sum_{k, l }  \delta^{\,\text{per.}}_{\omega_2}(\nu_k,\nu_l)}\ \ .
\end{align}
\end{subequations}

Moreover, when we implemented the smoothing through the Gaussian function
$$
\delta_\Delta(\omega-(\nu_\alpha-\nu_\beta)) = \frac{1}{\sqrt{\pi\Delta}} \exp\left({-\frac{(\omega-(\nu_\alpha-\nu_\beta))^2}{\Delta}}\right)\ ,
$$
we calculated each matrix elements' moment for multiple values of $\Delta$s. In particular, we fixed intervals of $\Delta$, based on the model considered (unitary or orthogonal symmetry class) and on the number of spins $L$, as shown in the Table below. 
Inside each interval we chose 3 different values of $\Delta$, we calculated each matrix elements' moment for each $\Delta$, and we averaged them in order to have a final function, with the respective error bars. 
\newpage 
\begin{table}[h!]
\centering
\renewcommand{\arraystretch}{1.2}
\setlength{\tabcolsep}{8pt}
\begin{tabular}{l|c c c c}
\toprule
$\Delta$'s intervals & $L=6,8$ & $L=10$ & $L=12$ & $L=14,16$ \\
\midrule
Unitary & [2, 4] & [0.5, 1.5] & [0.5, 1.5] & [0.1, 0.7] \\
Orthogonal  & [0.3, 0.7] & [0.1, 0.5] & [0.07, 0.1] & [0.01, 0.07]  \\
\bottomrule
\end{tabular}
\end{table}
\end{widetext}

\section{Thermal free cumulants and subleading corrections}
\label{app_corr_fun}
In this Appendix we focus on thermal correlation functions [cf. Eq.~\eqref{ther_corr_fun}] of Floquet dynamics\footnote{For non-Floquet systems one should consider further corrections coming from saddle-point fluctuations.}, for which  $\langle \bullet\rangle_{\beta = 0} = \text{Tr}(\bullet)/D$ indicates the thermal average. In the following we will drop the subscription $\beta=0$.
Moreover, we focus here explicitly on the case of the correlation function $\langle A(t)A\rangle$ and of the OTOC $\langle A(t)AA(t)A\rangle$; other correlations can be treated similarly. For the thermal moments one can define thermal free cumulants between observables at different times $\vec t =(t_1, t_2, \dots t_n)$ through the moments-cumulants formula (see Eq.~\eqref{free_cum_NC}) as:
\begin{equation}
    \langle A(t_1) A(t_2) \dots A(t_n) \rangle = \sum_{\pi \in NC(n)} k_\pi( A(t_1), \dots A(t_n))\ ,
\end{equation}
where $k_\pi$ is a product of free cumulants over the blocks of partitions $\pi$, as in Eq.~\eqref{free_cum_NC}. 
We will denote the thermal free cumulants as $k_n(\vec{t})$.\\
These free cumulants are connected correlation functions, but in the ETH setting they were proven \emph{at the leading order} to be given only by sums over over simple loops \cite{pappalardi2022eigenstate}
\begin{align}
\label{eq_free_cumu_ETH}
    k_n(\vec{t}) \approx k^{\text{ETH}}_n(\vec{t}) &= \frac{1}{D} \sum_{i_1\ne \dots \ne i_n} A_{i_1i_2}\dots A_{i_ni_1} e^{i\vec{\omega} \vec{t}} \\ 
    & \simeq \frac{1}{D} \sum_{i_1\ne \dots \ne i_n} K_n(\vec{\omega}) e^{i\vec{\omega} \vec{t}}\ .
\end{align}
In the second line, we have also written the expression through the local free cumulants, which depend only on the energy differences $\vec\omega = (\omega_1,\dots,\omega_n)$ with $\omega_{\alpha} \equiv \omega_{i_\alpha i_{\alpha+1}} = E_{i_\alpha}-E_{i_{\alpha+1}}$. 
Eq.~\eqref{eq_free_cumu_ETH} is valid because, via ETH, crossing contributions are shown to be suppressed and non-crossing diagrams factorize at the leading order.  \\ \\
\emph{We can now add correction to this statement, using the subleading corrections that we calculated.} \\ \\ 
Firstly, we have that the first-order thermal free cumulant is time independent and equal exactly to the ETH free cumulant, as well as to the local free cumulant: 
\begin{equation}
    k_1\equiv \langle A\rangle  = k^{\text{ETH}}_1 = K_1\ .
\end{equation}
For the two-times correlation function,
\begin{align}
    &\langle A(t) A\rangle = \frac 1D\sum_{i,j }A_{ij}A_{ji} e^{i\omega_{ij}t} \\
    &= \frac 1D\sum_{i \neq j }A_{ij}A_{ji} e^{i\omega_{ij}t} + \frac 1D \sum_i A_{ii}A_{ii} \notag \\
    &=\frac 1D\sum_{i \neq j }K_2(\omega_{ij}) e^{i\omega_{ij}t}  + K_1^2 +K_2(\omega_{ij}= 0) +\mathcal{O}(D^{-2})  \notag\\
    & = k^{\text{ETH}}_2(t,0) + (k_1^{\text{ETH}})^2 + K_2(\omega_{ij}= 0)  +\mathcal{O}(D^{-2}) \ , \notag
\end{align}
and therefore for the second-order thermal free cumulant 
\begin{align}
\begin{split}
    k_2(t,0)& \equiv \langle A(t) A\rangle-\langle A\rangle^2 \\
    &=k^{\text{ETH}}_2(t,0) + K_2(\omega_{ij}= 0) +\mathcal{O}(D^{-2}) \ .    
\end{split}
\label{thermal_2}
\end{align} \\ 
For the OTOC we take $A$ to be traceless, for convenience. We compute
\begin{align}
   & \langle A(t) AA(t)A\rangle = \frac 1D\sum_{i,j,k,l }A_{ij}A_{jk}A_{kl}A_{li}e^{i\omega_{ij}t+i\omega_{kl}t} \\
    \notag &= \frac 1D\sum_{i \neq j \ne k \ne l } A_{ij}A_{jk}A_{kl}A_{li}e^{i\omega_{ij}t+i\omega_{kl}t} 
    \\ \notag&+ \frac 2D \sum_{i \neq j \ne l} A_{ij}A_{ji}A_{il}A_{li}e^{i\omega_{ij}t+i\omega_{il}t} 
    \\ \notag&+ \frac 1D \sum_{i \neq j} A_{ij}A_{ji}A_{ij}A_{ji}e^{2i\omega_{ij}t}=\\ \notag
    &=k^{\text{ETH}}_4(t,0,t,0)  + 2(k^{\text{ETH}}_2(t,0))^2 \\
     \\ \notag &+ \frac 2D \sum_{i \neq j \ne l} K_4(\omega_{ij},\omega_{ji},\omega_{il})e^{i\omega_{ij}t+i\omega_{il}t} 
    \\\notag &+ \frac 1D \sum_{i \neq j} A_{ij}A_{ji}A_{ij}A_{ji}e^{2i\omega_{ij}t} +\mathcal{O}(D^{-2})
\end{align}
Finally, the fourth-order out-of-time-order thermal free cumulants is 
\begin{align}
\label{thermal_4}
   &k_4(t,0,t,0) \equiv \langle A(t) A A(t) A\rangle - 2\langle A(t) A\rangle^2 
   \\ \notag \\ \notag
   & = k^{\text{ETH}}_4(t,0,t,0)  + 2(k^{\text{ETH}}_2(t,0))^2
     \\ \notag &+ \frac 
     2D \sum_{i \neq j \ne l} K_4(\omega_{ij},\omega_{ji},\omega_{il})e^{i\omega_{ij}t+i\omega_{il}t} 
    \\ \notag&+ \frac 1D \sum_{i \neq j} A_{ij}A_{ji}A_{ij}A_{ji}e^{2i\omega_{ij}t} \\
    & \notag - 2\left(k^{\text{ETH}}_2(t,0)+K_2(\omega_{ij}=0)\right)^2+\mathcal{O}(D^{-2}) \\  \notag \\ 
    & = k^{\text{ETH}}_4(t,0,t,0) \notag \\
    & \notag + \frac 2D \sum_{i \neq j \ne l} K_4(\omega_{ij},\omega_{ji},\omega_{il})e^{i\omega_{ij}t+i\omega_{il}t} 
    \\&+ \notag\frac 1D \sum_{i \neq j} A_{ij}A_{ji}A_{ij}A_{ji}e^{2i\omega_{ij}t} \\ \notag & - 4k^{\text{ETH}}_2(t,0)K_2(\omega_{ij}=0)+\mathcal{O}(D^{-2}) \ .
    \notag
\end{align}
The corrections to thermal correlations and free cumulants may become relevant for quantities in which the leading-order terms cancel, but mostly they are particularly important because they determine the finite-size scaling of the long-time plateau of the thermal free cumulants, as we now show.
\subsection{Long-time plateau of freeness}
We now take the infinite-time limit of the thermal free cumulants.  The ETH free cumulants, being oscillatory contribution, vanishes due to the non-resonance condition, as shown in \cite{fava2025designs}. Therefore, the plateau is given by the subleading corrections. We identify $\omega \equiv \omega_{ij}$ in the following equations.\\ 
For the second-order, from Eq.~\eqref{thermal_2}, the time-independent correction gives
\begin{align}
    \label{long_time_K2}
    [k_2(t,0)]_\infty
    \simeq
    {K_2(\omega=0)}\ .
\end{align}
Similarly, taking the infinite-time limit of Eq.~\eqref{thermal_4}, the leading contributions and the crossing term average to zero, whereas the subleading correction remains finite:
\begin{align}
\label{long_time_K4}
    [k_4(t,0,t,0)]_\infty
    &\simeq 2 \sum_{i\neq j}
    K_4(\omega,-\omega,-\omega) \ .
\end{align}
Since $K_2\propto D^{-1}$ and $K_4\propto D^{-3}$, Eqs.~\eqref{long_time_K2}--\eqref{long_time_K4} scale as $D^{-1}$. Similar arguments show that the fluctuations around these plateaus are of order $\mathcal{O}(D^{-2})$, see e.g. Ref.~\cite{vallini2024longtime}. Therefore, the subleading corrections derived in this work are responsible for the finite-size late-time plateaus, which are exponentially suppressed in the system size. This provides a quantitative characterization of long-time freeness in chaotic dynamics.

\end{document}